\documentclass[3p,times]{elsarticle}

\usepackage{ecrc}
\usepackage{amsmath}
\usepackage{amsfonts}
\usepackage{nomencl}
\usepackage{placeins}
\usepackage{multirow}
\usepackage{graphicx,subfigure}
\usepackage{epstopdf}
\usepackage{setspace}

\volume{00}

\firstpage{1}

\journalname{Acta Astronautica}

\runauth{M. Zebenay T. Boge, and D. Choukroun}

\begin{document}

\begin{frontmatter}
 \author{M. Zebenay\corref{cor1}\fnref{a}}
 \ead{melak.zebenay@uniklinik-Freiburg.de}
 \cortext[cor1]{{This work was presented at the AIAA Guidance, Navigation, and Control Conference that was held in Boston, August 19-22, 2013.}}
\title{Modeling, Stability Analysis, and Testing \\of  a Hybrid Docking Simulator}

\author[b]{T. Boge}
\author[c,d]{D. Choukroun}
\address[a]{University of Freiburg, 79106 Freiburg, Germany}
\address[b]{German Aerospace Center, German Space Operation Center, 82234 Wessling, Germany}
\address[c]{Ben-Gurion University of the Negev, POB 653, 84105, Beer-Sheva, Israel}
\address[d]{Delft University of Technology, Faculty of Aerospace Engineering, Space Systems Engineering, 2629 HS Delft, The Netherlands}

\begin{abstract}
A hybrid docking simulator is a hardware-in-the-loop (HIL) simulator that includes a hardware element within a numerical simulation loop. One of the goals of performing a HIL simulation at the European Proximity Operation Simulator (EPOS) is the verification and validation of the docking phase in an on-orbit servicing mission.  A key feature of the HIL docking simulator set-up is a feedback loop that is closed on the real force sensed at the docking interface during the contact with the probe. This force signal is used as input to the numerical simulation of the free-floating bodies in contact. The resulting relative 3D trajectory serves as a position command for the two robots end-effectors holding the probe and the docking interface. The high stiffness of the robots causes the contact duration to be shorter than the time delay in the robots' dynamics. This can lead to inconsistencies in the simulation results, to instability of the closed-loop system, and eventually to damages in the HIL system. This work presents a novel mitigation strategy to the given challenge, accompanied with stability analysis and validating experiments. The high-stiffness compliance issue is addressed by combining virtual and real compliances in the software and hardware, respectively. The method is presented here for six degrees of freedom. A linear stability analysis is provided for a 2D case. Experimental results are presented for a translational linear motion and for a 3D motion. This hybrid contact dynamics model and the accompanying analysis is envisioned to provide a safe and flexible  docking simulator tool. This tool shall allow reproduction of the desired impact dynamics for any stiffness and damping characteristics within a desired stability, and thus safe, domain of operation.
 \end{abstract} 
\end{frontmatter}
 \newcommand{\eqnref}[1]{(\ref{#1})}
 \newcommand{\class}[1]{\texttt{#1}}
 \newcommand{\package}[1]{\texttt{#1}}
 \newcommand{\file}[1]{\texttt{#1}}
 \newcommand{\BibTeX}{\textsc{Bib}\TeX}
 \renewcommand{\include}{\input}
\newcommand{\beq}{\begin{equation}}
\newcommand{\eeq}{\end{equation}}
\newcommand{\bdm}{\begin{displaymath}}
\newcommand{\edm}{\end{displaymath}}

\newcommand{\bqmat}{\left\{ \begin{matrix} }
\newcommand{\eqmat}{\right\{\end{matrix} }

\newcommand{\ever}{\ensuremath{\widetilde{\bf \theta}}}
\newcommand{\mever}{\ensuremath{\overline{\ever}}}
\newcommand{\sever}{\ensuremath{\sigma\ever}}

\newcommand{\mlmx}{\ensuremath{\overline{\lambda_{max}}}}
\newcommand{\slmx}{\ensuremath{\sigma_{\lmx}}}

\newcommand{\ger}{\ensuremath{\tilde{\bf{y}}}}

\newcommand{\Zer}{\ensuremath{\widetilde{\bf{Z}}}}

\newcommand{\sa}{\ensuremath{\sin{\alpha}}}
\newcommand{\ca}{\ensuremath{\cos{\alpha}}}
\newcommand{\sib}{\ensuremath{\sin{\beta}}}
\newcommand{\cb}{\ensuremath{\cos{\beta}}}

\newcommand{\Btfr}{\ensuremath{\Bfr{_{t}}}}
\newcommand{\Befr}{\ensuremath{  \widehat{\Bfr} }}
\newcommand{\Bfrk}{\ensuremath{ \Bfr_{_k} }}
\newcommand{\Bfri}{\ensuremath{ \Bfr_{_i} }}

\newcommand{\eqdef}{\ensuremath{{\;\stackrel{\triangle}{=}\;}}}

\newtheorem{lem}{Lemma}
\newtheorem{defi}{Definition}

\newcommand{\adjoint}{\ensuremath{ \,\mbox{adj} }}
\newcommand{\const}{\ensuremath{ \,\mbox{const}   }}
\newcommand{\cov}{\ensuremath{ \,\mbox{cov}   }}
\newcommand{\degree}{\ensuremath{ \,\mbox{deg} }}
\newcommand{\dghr}{\ensuremath{ \,\mbox{$\frac{deg}{hr}$} }}
\newcommand{\dt}{\ensuremath{ \,\Delta t   }}
\newcommand{\dst}{\ensuremath{ \, dt   }}
\newcommand{\eqr}{Eq.~\eqref }
\newcommand{\expm}{\ensuremath{ \mbox{\emph{\Large e}}} }
\newcommand{\half}{\ensuremath{ \,\frac{1}{2}\,   }}
\newcommand{\smallhalf}{\ensuremath{ \texttt{\small\half} }}
\newcommand{\quarter}{\ensuremath{ \,\frac{1}{4}\,   }}
\newcommand{\smallquarter}{\ensuremath{ \texttt{\small\quarter} }}
\newcommand{\halfdt}{\ensuremath{ \,\frac{\dt}{2}\,   }}
\newcommand{\hertz}{\ensuremath{ \,\mbox{Hz} }}
\newcommand{\kernel}{\ensuremath{ \,\mbox{Ker} }}
\newcommand{\mdghr}{\ensuremath{ \,\mbox{mdeg/hr} }}
\newcommand{\Order}{\ensuremath{ \,\mathcal{O} }}
\newcommand{\radian}{\ensuremath{ \,\mbox{rad} }}
\newcommand{\rank}{\ensuremath{ \,\mbox{rank} }}
\newcommand{\real}{\ensuremath{ \mathbb{R} }}
\newcommand{\realthree}{\ensuremath{ \real^{3} }}
\newcommand{\realfour}{\ensuremath{ \real^{4} }}
\newcommand{\realsixteen}{\ensuremath{ \real^{16} }}
\newcommand{\realn}{\ensuremath{ \real^{n} }}
\newcommand{\realm}{\ensuremath{ \real^{m} }}
\newcommand{\realfourbyfour}{\ensuremath{ \real^{4\times 4} }}
\newcommand{\realnbym}{\ensuremath{ \real^{n\times m} }}
\newcommand{\remo}{ \emph{Remark 1}: }
\newcommand{\remtw}{ \emph{Remark 2}: }
\newcommand{\remth}{ \emph{Remark 3}: }
\newcommand{\remf}{ \emph{Remark 4}: }
\newcommand{\Span}{ \mbox{Span} }
\newcommand{\second}{\ensuremath{ \,\mbox{sec} }}
\newcommand{\sigbyfor}{\ensuremath{ \frac{\sig^2}{4} }}
\newcommand{\smallsigbyfor}{\ensuremath{ \texttt{\small\sigbyfor} }}
\newcommand{\sigepsbyfor}{\ensuremath{ \frac{\sigeps^2}{4} }}
\newcommand{\smallsigepsbyfor}{\ensuremath{ \texttt{\small\sigepsbyfor} }}
\newcommand{\spectrum}{\ensuremath{ \,\mbox{Sp} }}
\newcommand{\thrsigbyfor}{\ensuremath{ \frac{3\sig^2}{4} }}
\newcommand{\smallthrsigbyfor}{\ensuremath{ \texttt{\small\thrsigbyfor} }}
\newcommand{\thrsigepsbyfor}{\ensuremath{ \frac{3\sigeps^2}{4} }}
\newcommand{\smallthrsigepsbyfor}{\ensuremath{ \texttt{\small\thrsigepsbyfor} }}
\newcommand{\trace}{\ensuremath{ \,\mbox{tr} }}
\newcommand{\vctr}{\ensuremath{ \,\mbox{vec}   }}

\newcommand{\inm}{\ensuremath{{Q}}}
\newcommand{\onm}{\ensuremath{{R}}}
\newcommand{\pdm}{\ensuremath{P^d }}
\newcommand{\pttm}{\ensuremath{ P^{\xi} }}
\newcommand{\rttm}{\ensuremath{ \mathcal{Y}^{\xi} }}
\newcommand{\onttm}{\ensuremath{ \onm^{\xi} }}
\newcommand{\kttm}{\ensuremath{ K^{\xi} }}


\newcommand{\av}{\ensuremath{ {\bf a}  }}
\newcommand{\ahat}{\ensuremath{ \widehat{\av} }}
\newcommand{\ahatv}{\ensuremath{ \widehat{\av} }}
\newcommand{\avcross}{\ensuremath{ \left[ \av \times \right]  }}

\newcommand{\alfav}{\ensuremath{ {\boldsymbol\alpha}  }}

\newcommand{\bv}{\ensuremath{ \,{\bf b}  }}
\newcommand{\bonev}{\ensuremath{ \bv_{_{1}} }}
\newcommand{\btwov}{\ensuremath{ \bv_{_{2}} }}
\newcommand{\biv}{\ensuremath{ \bv_{_{i}} }}
\newcommand{\bi}{\ensuremath{ \,{\bf b}_{_i}  }}
\newcommand{\bion}{\ensuremath{ \bi^{^1} }}
\newcommand{\bitw}{\ensuremath{ \bi^{^2} }}
\newcommand{\bko}{\ensuremath{ \,{\bf b}_{_{k+1}}  }}
\newcommand{\bkoi}{\ensuremath{ \bko^i }}
\newcommand{\bk}{\ensuremath{ \,{\bf b}_{_k}  }}
\newcommand{\bki}{\ensuremath{ \,{\bf b}_{_{k+i}}  }}
\newcommand{\bz}{\ensuremath{ {\,{\bf b}^o}  }}
\newcommand{\bkt}{\ensuremath{  \bz_{_k} }}
\newcommand{\bkot}{\ensuremath{  \bz_{_{k+1}} }}
\newcommand{\bt}{\ensuremath{ \bz }}
\newcommand{\bit}{\ensuremath{ \bt_{_i} }}
\newcommand{\bo}{\ensuremath{ {\bf b}_{_1}  }}
\newcommand{\btwo}{\ensuremath{ {\bf b}_{_2}  }}
\newcommand{\bn}{\ensuremath{ {\bf b}_{_n}  }}
\newcommand{\berv}{\ensuremath{\mathbf{\delta b}}}
\newcommand{\bers}{\ensuremath{{\sigma_b}}}
\newcommand{\bq}{\ensuremath{ \bv_{_q} }}
\newcommand{\be}{\ensuremath{ \widehat{\bv} }}
\newcommand{\bei}{\ensuremath{ \be_{_{i}} }}

\newcommand{\betav}{\ensuremath{ \,{\boldsymbol\beta}  }}
\newcommand{\betatv}{\ensuremath{ \betav_{_{\!t}} }}
\newcommand{\betativ}{\ensuremath{ \betav_{_{t_i}} }}
\newcommand{\betatov}{\ensuremath{ \betav_{_{\!\tau}} }}

\newcommand{\bital}{\ensuremath{ \emph{\bv} }}

\newcommand{\cv}{\ensuremath{{\bf c}}}
\newcommand{\ctv}{\ensuremath{{\cv_{_t}}}}
\newcommand{\cvto}{\ensuremath{ \cv^{\tau} }}
\newcommand{\cvf}{\ensuremath{ \cv^{f} }}

\newcommand{\dalfav}{\ensuremath{{\,\boldsymbol{\delta\!\alfa}}}}
\newcommand{\dalfavcross}{\ensuremath{ [\dalfav \times ] }}

\newcommand{\dbv}{\ensuremath{{\,\boldsymbol{\delta\!b}}}}
\newcommand{\dbi}{\ensuremath{ \dbv_{_{i}}  }}
\newcommand{\dbj}{\ensuremath{ \dbv_{_{j}}  }}
\newcommand{\dbk}{\ensuremath{ \dbv_{_{\!k}}  }}
\newcommand{\dbki}{\ensuremath{ \dbv_{_{k+i}}  }}
\newcommand{\dbko}{\ensuremath{ \dbv_{_{k+1}}  }}
\newcommand{\dbkoi}{\ensuremath{ {\dbko^i} }}
\newcommand{\dbkoj}{\ensuremath{ {\dbko^j} }}

\newcommand{\dv}{\ensuremath{{\bf d}}}
\newcommand{\dkv}{\ensuremath{ \dv_{_{k}} }}
\newcommand{\dk}{\ensuremath{ \dv_{_{k}} }}
\newcommand{\donev}{\ensuremath{ \dv_{_{1}} }}
\newcommand{\dtwov}{\ensuremath{ \dv_{_{2}} }}
\newcommand{\dko}{\ensuremath{ \dv_{_{k+1}} }}
\newcommand{\dumykk}{\ensuremath{  {\widehat{\left( \bullet \right)}}_{_{k/k}}    }}
\newcommand{\dumykok}{\ensuremath{  {\widehat{\left( \bullet \right)}}_{_{k+1/k}}    }}
\newcommand{\dumyerkk}{\ensuremath{  {\widetilde{\left( \bullet \right)}}_{_{k/k}}    }}
\newcommand{\dumyerkok}{\ensuremath{  {\widetilde{\left( \bullet \right)}}_{_{k+1/k}}    }}

\newcommand{\deltav}{\ensuremath{{\boldsymbol{\delta}}}}

\newcommand{\dev}{\ensuremath{{\,\boldsymbol{\delta e}}}}
\newcommand{\dqv}{\ensuremath{\boldsymbol{\delta q}  }}
\newcommand{\dqvs}{\ensuremath{\delta q}}
\newcommand{\dqkk}{\ensuremath{ \dqv_{_{k/k}} }}
\newcommand{\dqkok}{\ensuremath{ \dqv_{_{k+1/k}} }}

\newcommand{\dqtv}{\ensuremath{{\bf \dv\!\qtv}}}
\newcommand{\dqetv}{\ensuremath{{\bf \dv\!\qetv}}}
\newcommand{\dqatv}{\ensuremath{{\dv\!\qatv}}}
\newcommand{\dbetav}{\ensuremath{{\bf \dv\!\!\betav}}}
\newcommand{\dbetatv}{\ensuremath{ \dbetav_t }}
\newcommand{\dbetatov}{\ensuremath{{\bf \dv\!\!\betatov}}}
\newcommand{\dztv}{\ensuremath{{\bf \dv\ztv}}}
\newcommand{\dnutv}{\ensuremath{{\bf \dv\!\nutv}}}
\newcommand{\dxtv}{\ensuremath{{\bf \dv\!\xtv}}}
\newcommand{\dqtldtv}{\ensuremath{{\bf \dv\!\qtldtv}}}
\newcommand{\dntv}{\ensuremath{{\bf \dv\!\ntv}}}
\newcommand{\dytv}{\ensuremath{{\bf \dv\!\ytv}}}
\newcommand{\detav}{\ensuremath{{\bf \dv\!\etav}}}
\newcommand{\detatv}{\ensuremath{ \detav_t }}

\newcommand{\Dqv}{\ensuremath{\boldsymbol{\Delta\!q} }}
\newcommand{\Dqkk}{\ensuremath{\Dqv_{_{k/k}} }}
\newcommand{\Dqkok}{\ensuremath{ \Dqv_{_{k+1/k}} }}
\newcommand{\Dqz}{\ensuremath{ \Dqv^0 }}
\newcommand{\Dqzkok}{\ensuremath{ \Dqz_{_{k+1/k}} }}
\newcommand{\Dqi}{\ensuremath{ \Dqv^i }}
\newcommand{\Dqikok}{\ensuremath{ \Dqi_{_{k+1/k}} }}

\newcommand{\ev}{\ensuremath{ \,{\bf e}  }}
\newcommand{\etz}{\ensuremath{ \ev^0 }}
\newcommand{\ek}{\ensuremath{ \ev_{_k} }}
\newcommand{\eko}{\ensuremath{ \ev_{_{k+1}} }}
\newcommand{\ee}{\ensuremath{ \widehat{\ev}  }}
\newcommand{\eek}{\ensuremath{ \ee_{_{k}} }}
\newcommand{\eekk}{\ensuremath{ \ee_{_{k/k}} }}
\newcommand{\eeko}{\ensuremath{ \ee_{_{k+1}} }}
\newcommand{\ekok}{\ensuremath{ \ee_{_{k+1/k}}  }}
\newcommand{\eekok}{\ensuremath{ \ee_{_{k+1/k}}  }}
\newcommand{\ekoko}{\ensuremath{ \ee_{_{k+1/k+1}}  }}
\newcommand{\eekoko}{\ensuremath{ \ee_{_{k+1/k+1}}  }}
\newcommand{\ei}{\ensuremath{ \,{\bf e}_{_i}  }}
\newcommand{\eekolk}{\ensuremath{ \ee_{_{k+l-1/k}}  }}
\newcommand{\eeklk}{\ensuremath{ \ee_{_{k+l/k}}  }}

\newcommand{\eulv}{\ensuremath{ \,\boldsymbol{\theta} }}

\newcommand{\epsv}{\ensuremath{ \,\boldsymbol{\epsilon} }}
\newcommand{\epsk}{\ensuremath{ \epsv_{_k} }}
\newcommand{\epski}{\ensuremath{ \epsv_{_{k+i}} }}
\newcommand{\epsi}{\ensuremath{  \epsv_{_i} }}
\newcommand{\epsko}{\ensuremath{ \epsv_{_{k+1}} }}
\newcommand{\epsok}{\ensuremath{ \epsv_{_{k-1}} }}
\newcommand{\epsktt}{\ensuremath{  \epsv_{k,\,true} }}
\newcommand{\epsbar}{\ensuremath{ \overline{\epsv}  }}
\newcommand{\epsbark}{\ensuremath{ \epsbar_{_k}  }}
\newcommand{\epskbar}{\ensuremath{ \epsbar_{_k}  }}
\newcommand{\epse}{\ensuremath{ \widehat{\epsv} }}
\newcommand{\epseokk}{\ensuremath{ \epse_{_{k-1/k}} }}
\newcommand{\epsekk}{\ensuremath{ \epse_{_{k/k}} }}
\newcommand{\epstv}{\ensuremath{ \epsv{_{_t}} }}
\newcommand{\epstov}{\ensuremath{ \epsv{_{_\tau}} }}

\newcommand{\epsxtk}{\ensuremath{ \epsk^{16} }}

\newcommand{\etav}{\ensuremath{ {\boldsymbol\eta}  }}
\newcommand{\etatv}{\ensuremath{ \etav_{_t}  }}

\newcommand{\fv}{\ensuremath{ {\bf f}  }}
\newcommand{\Fv}{\ensuremath{ {\bf F}  }}

\newcommand{\fik}{\ensuremath{{ {\bf \varphi}_{_k}}}}

\newcommand{\gv}{\ensuremath{ {\bf g}  }}

\newcommand{\hv}{\ensuremath{ {\bf h}  }}
\newcommand{\hvf}{\ensuremath{ \hv^{f}  }}
\newcommand{\hvto}{\ensuremath{ \hv^{\tau}  }}

\newcommand{\inv}{\ensuremath{{{\bf w}}}}

\newcommand{\Iv}{\ensuremath{{\bf I}}}
\newcommand{\Jv}{\ensuremath{{\bf I}}}
\newcommand{\Kv}{\ensuremath{{\bf I}}}
\newcommand{\Lv}{\ensuremath{{\bf I}}}

\newcommand{\lv}{\ensuremath{ {\bf l}  }}
\newcommand{\lev}{\ensuremath{ \widehat{\lv} }}
\newcommand{\leiv}{\ensuremath{ \lev_i }}

\newcommand{\minv}{\ensuremath{\overline{\inv}}}
\newcommand{\monv}{\ensuremath{\overline{\onv}}}
\newcommand{\mberv}{\ensuremath{\mathbf{\mu_b}}}
\newcommand{\mr}{\ensuremath{{\bf{m}}_r}}
\newcommand{\mepsk}{\ensuremath{ \mv_{_{\eps_{k}}}    }}

\newcommand{\nv}{\ensuremath{ {\bf n}  }}
\newcommand{\no}{\ensuremath{  \nv_{_1} }}
\newcommand{\nt}{\ensuremath{   \nv_{_2}  }}
\newcommand{\nepsk}{\ensuremath{ \nv_{_{\eps_{k}}}    }}
\newcommand{\nwk}{\ensuremath{ \nv_{_{w_{k}}}    }}
\newcommand{\ntv}{\ensuremath{ \nv_{_{t}} }}
\newcommand{\ntov}{\ensuremath{ \nv_{_{\tau}} }}
\newcommand{\nev}{\ensuremath{ \widehat{\nv} }}

\newcommand{\nuv}{\ensuremath{ {\boldsymbol \nu}  }}
\newcommand{\nutv}{\ensuremath{ \nuv_{_{t}} }}

\newcommand{\Ov}{\ensuremath{ {\bf{0}}  }}
\newcommand{\Onev}{\ensuremath{ {\bf{1}}  }}

\newcommand{\omg}{\ensuremath{ \,\boldsymbol{\omega} }}
\newcommand{\omgv}{\ensuremath{ \,\boldsymbol{\omega} }}
\newcommand{\omk}{\ensuremath{  \omgv_{_k}  }}
\newcommand{\omkt}{\ensuremath{ {\omgv_{_k}^{o}}  }}
\newcommand{\omok}{\ensuremath{ \omgv^0_{_k} }}
\newcommand{\ome}{\ensuremath{  \widehat{\omgv} }}
\newcommand{\omekk}{\ensuremath{  \ome_{_{k/k}}  }}
\newcommand{\omgtv}{\ensuremath{  \omgv_{_t}  }}
\newcommand{\omgtzv}{\ensuremath{  \omgtv^{o}  }}
\newcommand{\omgtzvcross}{\ensuremath{ \,\left[ \omgtzv \times \right] }}

\newcommand{\onv}{\ensuremath{{{\bf v}}}}

\newcommand{\qv}{\ensuremath{{\bf q}}}
\newcommand{\qt}{\ensuremath{ \qv_{_t}}}
\newcommand{\qtv}{\ensuremath{ \qv_{_t} }}
\newcommand{\qonev}{\ensuremath{{\qv_{_{1}}}}}
\newcommand{\qtwov}{\ensuremath{{\qv_{_{2}}}}}
\newcommand{\qthrv}{\ensuremath{{\qv_{_{3}}}}}
\newcommand{\qforv}{\ensuremath{{\qv_{_{4}}}}}
\newcommand{\qonethrv}{\ensuremath{{\qv_{_{13}}}}}
\newcommand{\qoneforv}{\ensuremath{{\qv_{_{14}}}}}
\newcommand{\qtwothrv}{\ensuremath{{\qv_{_{23}}}}}
\newcommand{\qtwoforv}{\ensuremath{{\qv_{_{24}}}}}
\newcommand{\qiv}{\ensuremath{{\qv_{_{i}}}}}

\newcommand{\qtz}{\ensuremath{ {\bf q}^0  }}
\newcommand{\qto}{\ensuremath{ {\bf q}^1  }}
\newcommand{\qtw}{\ensuremath{ {\bf q}^2  }}
\newcommand{\qth}{\ensuremath{ {\bf q}^3  }}
\newcommand{\qk}{\ensuremath{{  {\bf q}_{_k}}}}
\newcommand{\qak}{\ensuremath{{  {\bf q}_{_k}}}}
\newcommand{\qko}{\ensuremath{ \qv_{_{k+1}} }}
\newcommand{\qok}{\ensuremath{ \qv_{_{k-1}} }}
\newcommand{\qz}{\ensuremath{{  {\bf q}_{_0}}}}
\newcommand{\qon}{\ensuremath{ \qv_{_{1}} }}
\newcommand{\qtwo}{\ensuremath{ \qv_{_{2}} }}
\newcommand{\qN}{\ensuremath{ \qv_{_N} }}
\newcommand{\qoN}{\ensuremath{ \qv_{_{N-1}} }}
\newcommand{\qNo}{\ensuremath{ \qv_{_{N+1}} }}

\newcommand{\qiko}{\ensuremath{ {\qko^i} }}
\newcommand{\qzko}{\ensuremath{ {\qko^0} }}
\newcommand{\qoko}{\ensuremath{ {\qko^1} }}
\newcommand{\qtwko}{\ensuremath{ {\qko^2} }}
\newcommand{\qthko}{\ensuremath{ {\qko^3} }}

\newcommand{\qkt}{\ensuremath{ \qk^{true} }}
\newcommand{\qkot}{\ensuremath{ \qko^{true} }}
\newcommand{\qzt}{\ensuremath{ \qv_0^{true} }}

\newcommand{\qev}{\ensuremath{ \widehat{\bf q} }}
\newcommand{\qe}{\ensuremath{ \widehat{\bf q} }}
\newcommand{\qek}{\ensuremath{ \qe_{_k} }}
\newcommand{\qeko}{\ensuremath{ \qe_{_{k+1}} }}
\newcommand{\qekk}{\ensuremath{   \qe_{_{k/k}} }}
\newcommand{\qeNN}{\ensuremath{   \qe_{_{N/N}} }}
\newcommand{\qeNoN}{\ensuremath{   \qe_{_{N+1/N}} }}
\newcommand{\qeKok}{\ensuremath{  \qe_{_{k/k-1}}  }}
\newcommand{\qeNon}{\ensuremath{  \qe_{_{N/N-1}}  }}
\newcommand{\qeoNoN}{\ensuremath{  \qe_{_{N-1/N-1}}  }}
\newcommand{\qeNoNo}{\ensuremath{   \qe_{_{N+1/N+1}} }}
\newcommand{\qekoko}{\ensuremath{ \qe_{_{k+1/k+1}} }}
\newcommand{\qeokok}{\ensuremath{ \qe_{_{k-1/k-1}} }}
\newcommand{\qekokoast}{\ensuremath{ \qekoko^{\ast} }}
\newcommand{\qekkast}{\ensuremath{ \qekk^{\ast} }}
\newcommand{\qekok}{\ensuremath{   \qe_{_{k+1/k}} }}
\newcommand{\qeklk}{\ensuremath{   \qe_{_{k+l/k}} }}
\newcommand{\qekolk}{\ensuremath{   \qe_{_{k+l-1/k}} }}
\newcommand{\qezz}{\ensuremath{    \qe_{_{0/0}} }}
\newcommand{\qeoz}{\ensuremath{    \qe_{_{1/0}} }}
\newcommand{\qeoo}{\ensuremath{     \qe_{_{1/1}} }}
\newcommand{\qzstar}{\ensuremath{{ {\star{\bf q}}_{_0} }}}
\newcommand{\qostar}{\ensuremath{{ {\star{\bf q}}_{_1} }}}
\newcommand{\qater}{\ensuremath{\tilde{\bf q}}}
\newcommand{\qaterkoko}{\ensuremath{ \qater_{_{k+1/k+1}}    }}
\newcommand{\qeki}{\ensuremath{{ \,\widehat{\bf q}_{_{k/i} }}}}

\newcommand{\qetv}{\ensuremath{  \qe_{_{t}} }}

\newcommand{\qezkok}{\ensuremath{ {\qekok^0} }}
\newcommand{\qeonekok}{\ensuremath{ {\qekok^1} }}
\newcommand{\qetwkok}{\ensuremath{ {\qekok^2} }}
\newcommand{\qethkok}{\ensuremath{ {\qekok^3} }}
\newcommand{\qeikok}{\ensuremath{ {\qekok^i} }}

\newcommand{\qb}{\ensuremath{ \overline{\bf q}  }}
\newcommand{\qbar}{\ensuremath{  \,\bar{\bf q}  }}
\newcommand{\qbarz}{\ensuremath{  \qbar_{_0} }}
\newcommand{\qbaro}{\ensuremath{  \qbar_{_1} }}
\newcommand{\qbarN}{\ensuremath{  \qbar_{_N} }}
\newcommand{\qbaroN}{\ensuremath{  \qbar_{_{N-1}} }}
\newcommand{\qbark}{\ensuremath{  \qbar_{_k} }}
\newcommand{\qbarok}{\ensuremath{  \qbar_{_{k-1}} }}
\newcommand{\qbari}{\ensuremath{  \qbar_{_i} }}
\newcommand{\qbarKok}{\ensuremath{  \qbar_{_{k/k-1}}   }}
\newcommand{\qbarbar}{\ensuremath{      \bar{\qbar}  }}
\newcommand{\qbarbarKok}{\ensuremath{  \qbarbar_{_{k/k-1}} }}

\newcommand{\qkbar}{\ensuremath{{  \bar{\bf q}_{_{k}} }}}
\newcommand{\qzbar}{\ensuremath{{  \bar{\bf q}_{_0} }}}
\newcommand{\qobar}{\ensuremath{{  \bar{\bf q}_{_1} }}}
\newcommand{\qkobar}{\ensuremath{{  \bar{\bf q}_{_{k+1}} }}}

\newcommand{\qkk}{\ensuremath{ {\bf q}_{_{k/k}}  }}
\newcommand{\qerkk}{\ensuremath{ {\boldsymbol{\delta}} \qkk }}
\newcommand{\qkok}{\ensuremath{ {\bf q}_{_{k+1/k}}  }}
\newcommand{\qerkok}{\ensuremath{ {\boldsymbol{\delta}} \qkok }}
\newcommand{\qkoko}{\ensuremath{ {\bf q}_{_{k+1/k+1}}  }}
\newcommand{\qerkoko}{\ensuremath{ {\boldsymbol{\delta}} \qkoko }}
\newcommand{\qerkokoast}{\ensuremath{ \qerkoko^{\ast} }}
\newcommand{\qierkok}{\ensuremath{ {\qerkok^i} }}
\newcommand{\qzerkok}{\ensuremath{ {\qerkok^0} }}

\newcommand{\qtldv}{\ensuremath{ \tilde{\bf q} }}
\newcommand{\qtldtv}{\ensuremath{ \qtldv_{_{t}} }}

\newcommand{\qav}{\ensuremath{ \qv^{\!a} }}
\newcommand{\qatv}{\ensuremath{ \qav_{_t} }}

\newcommand{\Qvo}{\ensuremath{  {\bf{e}}_{_1}  }}
\newcommand{\Qvtw}{\ensuremath{  {\bf{e}}_{_2}  }}
\newcommand{\Qvth}{\ensuremath{  {\bf{e}}_{_3}  }}
\newcommand{\Qv}{\ensuremath{  {\bf{e}}  }}
\newcommand{\Quatv}{\ensuremath{{\bf e}}}

\newcommand{\rv}{\ensuremath{ \,{\bf r}  }}
\newcommand{\ronev}{\ensuremath{ \rv_{_{1}} }}
\newcommand{\rtwov}{\ensuremath{ \rv_{_{2}} }}
\newcommand{\riv}{\ensuremath{ \rv_{_{i}} }}
\newcommand{\ri}{\ensuremath{ \,{\bf r}_{_i}  }}
\newcommand{\rj}{\ensuremath{ \,{\bf r}_{_j}  }}
\newcommand{\rko}{\ensuremath{ \rv_{_{k+1}} }}
\newcommand{\rkoi}{\ensuremath{ {\rv_{_{k+1}}^i} }}
\newcommand{\rz}{\ensuremath{ \,{\bf r}_{_0}  }}
\newcommand{\ro}{\ensuremath{ \,{\bf r}_{_1}  }}
\newcommand{\rt}{\ensuremath{ \,{\bf r}_{_2}  }}
\newcommand{\rtwo}{\ensuremath{ \,{\bf r}_{_2}  }}
\newcommand{\rn}{\ensuremath{ \,{\bf r}_{_n}  }}
\newcommand{\rttv}{\ensuremath{ \,\bf{\xi}_{r} }}
\newcommand{\res}{\ensuremath{ \,\widehat{\bf e}}}
\newcommand{\resid}{\ensuremath{ \,{\bf r}}}
\newcommand{\rkok}{\ensuremath{ \rv_{_{k+1/k}}  }}
\newcommand{\rk}{\ensuremath{ {\bf r}_{_k}  }}
\newcommand{\rvq}{\ensuremath{ \rv_{_q} }}
\newcommand{\rklk}{\ensuremath{ \rv_{_{k+l/k}}  }}
\newcommand{\rknk}{\ensuremath{ \rv_{_{k+n/k}}  }}
\newcommand{\rvo}{\ensuremath{ \rv_{_1} }}
\newcommand{\rvtw}{\ensuremath{ \rv_{_2} }}
\newcommand{\rvth}{\ensuremath{ \rv_{_3} }}
\newcommand{\rqv}{\ensuremath{ \rv^q }}
\newcommand{\rqko}{\ensuremath{ \rqv_{_{k+1}} }}

\newcommand{\rov}{\ensuremath{ {\boldsymbol\rho}  }}
\newcommand{\rovast}{\ensuremath{ \rov^{\!\ast} }}

\newcommand{\sv}{\ensuremath{{\bf s}}}
\newcommand{\sko}{\ensuremath{ \sv_{_{k+1}} }}
\newcommand{\sk}{\ensuremath{ \sv_{_{k}} }}
\newcommand{\sonev}{\ensuremath{ \sv_{_{1}} }}
\newcommand{\stwov}{\ensuremath{ \sv_{_{2}} }}
\newcommand{\siv}{\ensuremath{ \sv_{_{i}} }}

\newcommand{\tov}{\ensuremath{ {\boldsymbol \tau}  }}

\newcommand{\tetav}{\ensuremath{ \,\boldsymbol{\theta} }}
\newcommand{\tetavko}{\ensuremath{ \,\boldsymbol{\theta}_{_{k+1}} }}
\newcommand{\tetaev}{\ensuremath{ \,\widehat{\tetav} }}
\newcommand{\tetaeWLS}{\ensuremath{ \tetaev^{_{_{WLS}}} }}
\newcommand{\tetaeWLSk}{\ensuremath{ \tetaeWLS_{_k} }}
\newcommand{\tetaeWLSko}{\ensuremath{ \tetaeWLS_{_{k+1}} }}

\newcommand{\uv}{\ensuremath{\mathbf{u}}}
\newcommand{\ui}{\ensuremath{ \uv_{_i}   }}
\newcommand{\ujv}{\ensuremath{ \uv_{_j}   }}
\newcommand{\uonev}{\ensuremath{ \uv_{_1}   }}
\newcommand{\utwov}{\ensuremath{ \uv_{_2}   }}
\newcommand{\uthrv}{\ensuremath{ \uv_{_3}   }}
\newcommand{\uforv}{\ensuremath{ \uv_{_4}   }}
\newcommand{\uiv}{\ensuremath{ \uv_{_i}   }}

\newcommand{\uhat}{\ensuremath{ \widehat{\uv} }}

\newcommand{\vv}{\ensuremath{ \mathbf{v} }}
\newcommand{\vk}{\ensuremath{  \vv_{_{\!k}} }}
\newcommand{\vko}{\ensuremath{  \vv_{_{k+1}} }}
\newcommand{\vkon}{\ensuremath{  \vko^n }}
\newcommand{\vkoi}{\ensuremath{  \vko^i }}
\newcommand{\vi}{\ensuremath{  \vv_{_i} }}
\newcommand{\von}{\ensuremath{  \vv_{_1} }}
\newcommand{\vtw}{\ensuremath{  \vv_{_2} }}
\newcommand{\vth}{\ensuremath{  \vv_{_3} }}

\newcommand{\vvf}{\ensuremath{ \vv^f }}
\newcommand{\vvto}{\ensuremath{ \vv^{\tau} }}

\newcommand{\vbarv}{\ensuremath{ {\overline{\vv}} }}
\newcommand{\vbark}{\ensuremath{ \vbarv_{_{k}} }}

\newcommand{\wv}{\ensuremath{  \,{\bf w} }}
\newcommand{\wonev}{\ensuremath{  \wv_{_{1}} }}
\newcommand{\wtwov}{\ensuremath{  \wv_{_{2}} }}
\newcommand{\wk}{\ensuremath{  \,{\bf w}_{_k} }}
\newcommand{\wok}{\ensuremath{  \wv_{_{k-1}} }}
\newcommand{\wN}{\ensuremath{  {\bf w}_{_N} }}
\newcommand{\woN}{\ensuremath{  {\wv_{_{N-1}}}  }}
\newcommand{\wko}{\ensuremath{     \wv_{_{k+1}} }}
\newcommand{\wtwN}{\ensuremath{  \wv_{_{N-2}} }}
\newcommand{\wz}{\ensuremath{  {\bf w}_{_0}   }}
\newcommand{\wo}{\ensuremath{  {\bf w}_{_1}  }}
\newcommand{\wsxtk}{\ensuremath{  \wk^{16} }}
\newcommand{\wtenk}{\ensuremath{  \wk^{10} }}
\newcommand{\wnink}{\ensuremath{  \wk^{9} }}
\newcommand{\wi}{\ensuremath{  \wv_{_i} }}

\newcommand{\wbar}{\ensuremath{ \,\bar{\wv} }}
\newcommand{\wkbar}{\ensuremath{ \wbar_{_k} }}
\newcommand{\wbark}{\ensuremath{ \wbar_{_k} }}
\newcommand{\wbarok}{\ensuremath{ \wbar_{_{k-1}} }}
\newcommand{\wbarN}{\ensuremath{   \wbar_{_{N}} }}
\newcommand{\wbaroN}{\ensuremath{   \wbar_{_{N-1}} }}
\newcommand{\wkobar}{\ensuremath{     \wbar_{_{k+1}} }}
\newcommand{\wzbar}{\ensuremath{   \wbar_{_0} }}
\newcommand{\wbarz}{\ensuremath{   \wbar_{_0} }}
\newcommand{\wbartwk}{\ensuremath{  \wbar_{_{k-2}} }}
\newcommand{\wbartwN}{\ensuremath{  \wbar_{_{N-2}} }}
\newcommand{\wbarKok}{\ensuremath{  \wbar_{_{k/k-1}} }}
\newcommand{\wbarbar}{\ensuremath{  \bar{\wbar}  }}
\newcommand{\wbarbarokok}{\ensuremath{  \wbarbar_{_{k-1/k-1}}  }}

\newcommand{\wzstar}{\ensuremath{{  \star{\bf w}_{_0}}}}
\newcommand{\wostar}{\ensuremath{{  \star{\bf w}_{_1}}}}
\newcommand{\wstar}{\ensuremath{  \, {\wv}^{\star} }}
\newcommand{\wstarz}{\ensuremath{ \wstar_{_0}  }}
\newcommand{\wstaro}{\ensuremath{ \wstar_{_1}  }}
\newcommand{\wstarok}{\ensuremath{ \wstar_{_{k-1}}  }}
\newcommand{\wstarokqk}{\ensuremath{
 \wstarok \left( \qk \right) }}
\newcommand{\wstartwk}{\ensuremath{ \wstar_{_{k-2}}  }}
\newcommand{\wstarstar}{\ensuremath{ \wstar^{\star}  }}
\newcommand{\wstarstartwk}{\ensuremath{ \wstarstar_{_{k-2}}  }}

\newcommand{\wezz}{\ensuremath{{ \widehat{\bf w}_{_{0/0} }}}}
\newcommand{\weoz}{\ensuremath{{ \widehat{\bf w}_{_{1/0} }}}}
\newcommand{\weoo}{\ensuremath{{ \widehat{\bf w}_{_{1/1} }}}}
\newcommand{\wekk}{\ensuremath{{ \,\widehat{\bf w}_{_{k/k} }}}}
\newcommand{\wekok}{\ensuremath{{ \,\widehat{\bf w}_{_{k+1/k} }}}}
\newcommand{\weki}{\ensuremath{{ \,\widehat{\bf w}_{_{k/i} }}}}
\newcommand{\wekoK}{\ensuremath{{ \,\widehat{\bf w}_{_{k+1/k} }}}}
\newcommand{\weKok}{\ensuremath{{ \,\widehat{\bf w}_{_{k/k-1} }}}}
\newcommand{\weokk}{\ensuremath{{ \,\widehat{\bf w}_{_{k-1/k} }}}}

\newcommand{\wtv}{\ensuremath{{\bf \omega}^o}}
\newcommand{\wtm}{\ensuremath{{ \Omega}^o}}
\newcommand{\wm}{\ensuremath{{\Omega}}}
\newcommand{\werm}{\ensuremath{{\mathcal{E}}}}
\newcommand{\wers}{\ensuremath{{\sigma_{\varepsilon}}}}

\newcommand{\xettv}{\ensuremath{ \bf{\xi}_{\widehat{x}} }}
\newcommand{\xv}{\ensuremath{ \,{\bf x}  }}
\newcommand{\xedv}{\ensuremath{\widehat{\xv}^d }}
\newcommand{\xev}{\ensuremath{{\widehat{\xv}}}}
\newcommand{\xk}{\ensuremath{ \xv_{_k}  }}
\newcommand{\xko}{\ensuremath{ \xv_{_{k+1}}  }}
\newcommand{\xz}{\ensuremath{ \xv_{_0}  }}
\newcommand{\xo}{\ensuremath{ \xv_{_1}  }}
\newcommand{\xtw}{\ensuremath{ \xv_{_2}  }}
\newcommand{\xsxtk}{\ensuremath{ \xk^{16} }}
\newcommand{\xsxtko}{\ensuremath{ \xko^{16} }}
\newcommand{\xtenk}{\ensuremath{ \xk^{10} }}
\newcommand{\xtenko}{\ensuremath{ \xko^{10} }}
\newcommand{\xninko}{\ensuremath{ \xko^{9} }}
\newcommand{\xnink}{\ensuremath{ \xk^{9} }}
\newcommand{\xekk}{\ensuremath{  \xev_{_{k/k}} }}
\newcommand{\xekok}{\ensuremath{  \xev_{_{k+1/k}} }}
\newcommand{\xezz}{\ensuremath{  \xev_{_{0/0}} }}
\newcommand{\xekoko}{\ensuremath{  \xev_{_{k+1/k+1}} }}
\newcommand{\xzv}{\ensuremath{ \xv^0 }}
\newcommand{\xvast}{\ensuremath{ \xv^{\!\ast}  }}

\newcommand{\Xv}{\ensuremath{ {\bf X}  }}

\newcommand{\xtv}{\ensuremath{ \xv_{_t}  }}
\newcommand{\xav}{\ensuremath{ \xv_{_a}  }}

\newcommand{\yv}{\ensuremath{{\bf y}}}
\newcommand{\yzv}{\ensuremath{  \yv^0 }}
\newcommand{\yttv}{\ensuremath{ \bf{\xi}_y }}
\newcommand{\Yv}{\ensuremath{ {\bf y}  }}
\newcommand{\ydv}{\ensuremath{\yv^d }}
\newcommand{\yk}{\ensuremath{ \yv_{_k} }}
\newcommand{\yko}{\ensuremath{ \yv_{_{k+1}} }}
\newcommand{\yz}{\ensuremath{ \yv_{_0} }}
\newcommand{\ytv}{\ensuremath{ \yv_{_t}  }}

\newcommand{\yev}{\ensuremath{ \widehat{\yv} }}

\newcommand{\Zeps}{\ensuremath{ {\bf z}_{_\epsilon}  }}
\newcommand{\zeps}{\ensuremath{ {\bf z}_{_\epsilon}  }}
\newcommand{\Zb}{\ensuremath{ {\bf z}_{_b}  }}
\newcommand{\zv}{\ensuremath{ {\bf z}  }}
\newcommand{\zb}{\ensuremath{ {\bf z}_{_b}  }}
\newcommand{\zk}{\ensuremath{ \zv_{_{k}}  }}
\newcommand{\zko}{\ensuremath{ \zv_{_{k+1}}  }}
\newcommand{\dzko}{\ensuremath{ \delta \zv_{_{k+1}}  }}
\newcommand{\zkot}{\ensuremath{ \zv_{_{k+1}}^{true}  }}
\newcommand{\zer}{\ensuremath{ \widetilde{\zv} }}
\newcommand{\zerk}{\ensuremath{ \zer_{_k} }}
\newcommand{\zz}{\ensuremath{ \zv_{_{0}}  }}
\newcommand{\zo}{\ensuremath{ \zv_{_{1}}  }}
\newcommand{\zN}{\ensuremath{ \zv_{_{N}}  }}
\newcommand{\zbi}{\ensuremath{ {\zb^i} }}
\newcommand{\zbn}{\ensuremath{ {\zb^n} }}
\newcommand{\zev}{\ensuremath{ \widehat{\zv} }}
\newcommand{\zzero}{\ensuremath{ {\zv^{o}}  }}
\newcommand{\ztv}{\ensuremath{ \zv_{_{t}}  }}

\newcommand{\zvf}{\ensuremath{ \zv^{f}  }}
\newcommand{\zvto}{\ensuremath{ \zv^{\tau}  }}

\newcommand{\Zv}{\ensuremath{ {\bf Z}  }}


\newcommand{\ai}{\ensuremath{ \,a_{_i}  }}
\newcommand{\aj}{\ensuremath{ \,a_{_j}  }}
\newcommand{\ako}{\ensuremath{ \,a_{_{k+1}}  }}
\newcommand{\ak}{\ensuremath{ \,a_{_k}  }}
\newcommand{\ao}{\ensuremath{ \,a_{_1}  }}
\newcommand{\atw}{\ensuremath{ \,a_{_2}  }}

\newcommand{\alfa}{\ensuremath{ \alpha  }}
\newcommand{\alfaz}{\ensuremath{ \alpha_{_0}  }}
\newcommand{\alfao}{\ensuremath{ \alpha_{_1}  }}
\newcommand{\alfatw}{\ensuremath{ \alpha_{_2}  }}
\newcommand{\alfak}{\ensuremath{ \alpha_{_k}  }}
\newcommand{\alfaN}{\ensuremath{ \,\alpha_{_N}  }}
\newcommand{\alfako}{\ensuremath{ \alpha_{_{k+1}}  }}
\newcommand{\alfai}{\ensuremath{ \alpha_{_i}  }}
\newcommand{\alfaone}{\ensuremath{ \alpha_{_1}  }}
\newcommand{\alfatwo}{\ensuremath{ \alpha_{_2}  }}
\newcommand{\alfathr}{\ensuremath{ \alpha_{_3}  }}
\newcommand{\alfaij}{\ensuremath{ \alpha_{_{ij}}  }}
\newcommand{\alfaonefor}{\ensuremath{ \alpha_{_{14}}  }}
\newcommand{\alfacos}{\ensuremath{ c_{\alfa}  }}

\newcommand{\bos}{\ensuremath{ \,b_{_1}  }}
\newcommand{\btws}{\ensuremath{ \,b_{_2}  }}
\newcommand{\bths}{\ensuremath{ \,b_{_3}  }}

\newcommand{\betat}{\ensuremath{ \beta_{_t}  }}
\newcommand{\betazer}{\ensuremath{ \beta_{_0}  }}
\newcommand{\betaone}{\ensuremath{ \beta_{_1}  }}
\newcommand{\betatwo}{\ensuremath{ \beta_{_2}  }}
\newcommand{\betathr}{\ensuremath{ \beta_{_3}  }}
\newcommand{\betafor}{\ensuremath{ \beta_{_4}  }}
\newcommand{\betai}{\ensuremath{ \beta_{_{\!i}}  }}

\newcommand{\dsxone}{\ensuremath{ \delta x_1 }}
\newcommand{\dsxtwo}{\ensuremath{ \delta x_2 }}
\newcommand{\dsxthr}{\ensuremath{ \delta x_3 }}
\newcommand{\dsxfor}{\ensuremath{ \delta x_4 }}
\newcommand{\dsz}{\ensuremath{ \delta z }}
\newcommand{\dsvz}{\ensuremath{ \delta v_z }}
\newcommand{\dstheta}{\ensuremath{ \delta \theta }}
\newcommand{\dsomg}{\ensuremath{ \delta \omega }}
\newcommand{\dsd}{\ensuremath{ \delta d }}
\newcommand{\dsdd}{\ensuremath{ \delta \dot{d} }}
\newcommand{\dsyone}{\ensuremath{ \delta y_1 }}
\newcommand{\dsytwo}{\ensuremath{ \delta y_2 }}
\newcommand{\dsythr}{\ensuremath{ \delta y_3 }}
\newcommand{\dsyfor}{\ensuremath{ \delta y_4 }}

\newcommand{\dfi}{\ensuremath{ \;\delta \phi }}
\newcommand{\dfikk}{\ensuremath{ \dfi_{_{k/k}} }}
\newcommand{\dfie}{\ensuremath{ \widetilde{\delta\phi}  }}
\newcommand{\dfiekk}{\ensuremath{ \dfie_{_{k/k}} }}

\newcommand{\deltaok}{\ensuremath{ \,\delta_{_{k-1}}  }}
\newcommand{\deltatwk}{\ensuremath{ \,\delta_{_{k-2}}  }}
\newcommand{\deltaij}{\ensuremath{ \,\delta_{_{ij}}  }}
\newcommand{\dqes}{\ensuremath{ \,\delta q }}

\newcommand{\eos}{\ensuremath{ \,e_{_1}  }}
\newcommand{\etws}{\ensuremath{ \,e_{_2}  }}
\newcommand{\eths}{\ensuremath{ \,e_{_3}  }}

\newcommand{\etaz}{\ensuremath{ \eta_{_0}  }}
\newcommand{\etak}{\ensuremath{ \eta_{_k}  }}
\newcommand{\etaok}{\ensuremath{ \eta_{_{k-1}}  }}
\newcommand{\etai}{\ensuremath{ \eta_{_i}  }}
\newcommand{\etaoz}{\ensuremath{ \eta_{_{1/0}}  }}
\newcommand{\etaoo}{\ensuremath{ \,\eta_{_{1/1}}  }}
\newcommand{\etae}{\ensuremath{\widehat{\eta}}}
\newcommand{\etakast}{\ensuremath{ \etak^{\ast}  }}
\newcommand{\etaokast}{\ensuremath{ \etaok^{\ast}  }}
\newcommand{\etaoN}{\ensuremath{ \,\eta_{_{N-1}}  }}

\newcommand{\eps}{\ensuremath{ \epsilon  }}
\newcommand{\epso}{\ensuremath{ \epsilon_{_1}  }}
\newcommand{\epstw}{\ensuremath{ \epsilon_{_2}  }}
\newcommand{\epsth}{\ensuremath{ \epsilon_{_3}  }}

\newcommand{\fit}{\ensuremath{{ {\varphi}_{_t}}}}

\newcommand{\fie}{\ensuremath{ \widehat{\varphi} }}
\newcommand{\fiet}{\ensuremath{ \fie_{_{t}} }}

\newcommand{\gamazz}{\ensuremath{ \gamma_{_{0/0}}  }}
\newcommand{\gamaoz}{\ensuremath{ \gamma_{_{1/0}}  }}
\newcommand{\gamaoo}{\ensuremath{ \gamma_{_{1/1}}  }}
\newcommand{\gamaoostar}{\ensuremath{ \star{\gamaoo}  }}
\newcommand{\gamakk}{\ensuremath{ \,\gamma_{_{k/k}}  }}
\newcommand{\gamaokok}{\ensuremath{ \,\gamma_{_{ {k-1}/{k-1} }}  }}
\newcommand{\gamakoK}{\ensuremath{ \,\gamma_{_{ {k+1}/{k} }}  }}
\newcommand{\gamaKok}{\ensuremath{  \,\gamma_{_{    k /{k-1} }}  }}
\newcommand{\gamabar}{\ensuremath{  \,\bar{\gamma}  }}
\newcommand{\gamabarKok}{\ensuremath{  \,\gamabar_{_{    k /{k-1} }}  }}
\newcommand{\gamabarbar}{\ensuremath{  \,\bar{\gamabar}  }}
\newcommand{\gamabarbarKok}{\ensuremath{  \,\gamabarbar_{_{    k /{k-1} }}  }}

\newcommand{\gama}{\ensuremath{ \,\gamma  }}
\newcommand{\gamao}{\ensuremath{ \gama_{_{1}}  }}
\newcommand{\gamatw}{\ensuremath{ \gama_{_{2}}  }}
\newcommand{\gamath}{\ensuremath{ \gama_{_{3}}  }}

\newcommand{\gNN}{\ensuremath{  g_{_{N/N}} }}

\newcommand{\kes}{\ensuremath{ \widehat{k} }}

\newcommand{\Kkoo}{\ensuremath{  \Kk_{_{11}}  }}
\newcommand{\Kkotw}{\ensuremath{ \Kk_{_{12}}  }}
\newcommand{\Kkoth}{\ensuremath{ \Kk_{_{13}}  }}
\newcommand{\Kkof}{\ensuremath{  \Kk_{_{14}}  }}
\newcommand{\Kktwo}{\ensuremath{ \Kk_{_{21}}  }}
\newcommand{\Kktwtw}{\ensuremath{\Kk_{_{22}}  }}
\newcommand{\Kktwth}{\ensuremath{ \Kk_{_{23}}  }}
\newcommand{\Kktwf}{\ensuremath{ \Kk_{_{24}}  }}
\newcommand{\Kktho}{\ensuremath{  \Kk_{_{31}}  }}
\newcommand{\Kkthtw}{\ensuremath{ \Kk_{_{32}}  }}
\newcommand{\Kkthth}{\ensuremath{  \Kk_{_{33}}  }}
\newcommand{\Kkthf}{\ensuremath{ \Kk_{_{34}}  }}
\newcommand{\Kkfo}{\ensuremath{  \Kk_{_{41}}  }}
\newcommand{\Kkftw}{\ensuremath{ \Kk_{_{42}}  }}
\newcommand{\Kkfth}{\ensuremath{ \Kk_{_{43}}  }}
\newcommand{\Kkff}{\ensuremath{  \Kk_{_{44}}  }}
\newcommand{\Kkij}{\ensuremath{  {\Kk}_{_{ij}}  }}

\newcommand{\Kkooo}{\ensuremath{  \Kko_{_{11}}  }}
\newcommand{\Kkootw}{\ensuremath{ \Kko_{_{12}}  }}
\newcommand{\Kkooth}{\ensuremath{ \Kko_{_{13}}  }}
\newcommand{\Kkoof}{\ensuremath{  \Kko_{_{14}}  }}
\newcommand{\Kkotwo}{\ensuremath{ \Kko_{_{21}}  }}
\newcommand{\Kkotwtw}{\ensuremath{\Kko_{_{22}}  }}
\newcommand{\Kkotwth}{\ensuremath{ \Kko_{_{23}}  }}
\newcommand{\Kkotwf}{\ensuremath{ \Kko_{_{24}}  }}
\newcommand{\Kkotho}{\ensuremath{  \Kko_{_{31}}  }}
\newcommand{\Kkothtw}{\ensuremath{ \Kko_{_{32}}  }}
\newcommand{\Kkothth}{\ensuremath{  \Kko_{_{33}}  }}
\newcommand{\Kkothf}{\ensuremath{ \Kko_{_{34}}  }}
\newcommand{\Kkofo}{\ensuremath{  \Kko_{_{41}}  }}
\newcommand{\Kkoftw}{\ensuremath{ \Kko_{_{42}}  }}
\newcommand{\Kkofth}{\ensuremath{ \Kko_{_{43}}  }}
\newcommand{\Kkoff}{\ensuremath{  \Kko_{_{44}}  }}
\newcommand{\Kkoij}{\ensuremath{  {\Kko}_{_{ij}}  }}

\newcommand{\Kon}{\ensuremath{  K_{_{1}}  }}
\newcommand{\Ktw}{\ensuremath{ K_{_{2}}  }}
\newcommand{\Koo}{\ensuremath{  K_{_{11}}  }}
\newcommand{\Kotw}{\ensuremath{ K_{_{12}}  }}
\newcommand{\Koth}{\ensuremath{ K_{_{13}}  }}
\newcommand{\Kof}{\ensuremath{  K_{_{14}}  }}
\newcommand{\Ktwo}{\ensuremath{ K_{_{21}}  }}
\newcommand{\Ktwtw}{\ensuremath{  K_{_{22}}  }}
\newcommand{\Ktwth}{\ensuremath{ K_{_{23}}  }}
\newcommand{\Ktwf}{\ensuremath{ K_{_{24}}  }}
\newcommand{\Ktho}{\ensuremath{  K_{_{31}}  }}
\newcommand{\Kthtw}{\ensuremath{ K_{_{32}}  }}
\newcommand{\Kthth}{\ensuremath{  K_{_{33}}  }}
\newcommand{\Kthf}{\ensuremath{ K_{_{34}}  }}
\newcommand{\Kfo}{\ensuremath{  K_{_{41}}  }}
\newcommand{\Kftw}{\ensuremath{ K_{_{42}}  }}
\newcommand{\Kfth}{\ensuremath{ K_{_{43}}  }}
\newcommand{\Kff}{\ensuremath{  K_{_{44}}  }}
\newcommand{\Ksoneone}{\ensuremath{  K_{_{\!11}}  }}
\newcommand{\Ksonetwo}{\ensuremath{ K_{_{\!12}}  }}
\newcommand{\Ksonethr}{\ensuremath{ K_{_{\!13}}  }}
\newcommand{\Ksonefor}{\ensuremath{  K_{_{\!14}}  }}
\newcommand{\Kstwoone}{\ensuremath{ K_{_{\!21}}  }}
\newcommand{\Kstwotwo}{\ensuremath{  K_{_{\!22}}  }}
\newcommand{\Kstwothr}{\ensuremath{ K_{_{\!23}}  }}
\newcommand{\Kstwofor}{\ensuremath{ K_{_{\!24}}  }}
\newcommand{\Ksthrone}{\ensuremath{  K_{_{\!31}}  }}
\newcommand{\Ksthrtwo}{\ensuremath{ K_{_{\!32}}  }}
\newcommand{\Ksthrthr}{\ensuremath{  K_{_{\!33}}  }}
\newcommand{\Ksthrfor}{\ensuremath{ K_{_{\!34}}  }}
\newcommand{\Ksforone}{\ensuremath{  K_{_{\!41}}  }}
\newcommand{\Ksfortwo}{\ensuremath{ K_{_{\!42}}  }}
\newcommand{\Ksforthr}{\ensuremath{ K_{_{\!43}}  }}
\newcommand{\Ksforfor}{\ensuremath{  K_{_{\!44}}  }}

\newcommand{\lbd}{\ensuremath{{\,\lambda}}}
\newcommand{\lbdz}{\ensuremath{ \lambda^{^0}  }}
\newcommand{\lbdo}{\ensuremath{ \lambda_{_1}  }}
\newcommand{\lbdtw}{\ensuremath{ \lambda_{_2}  }}
\newcommand{\lbdth}{\ensuremath{ \lambda_{_3}  }}
\newcommand{\lbdfo}{\ensuremath{ \lambda_{_4}  }}
\newcommand{\lmx}{\ensuremath{\lambda_{max}}}
\newcommand{\lkok}{\ensuremath{ \,\lambda_{_{k+1/k}} }}
\newcommand{\likok}{\ensuremath{ {\lkok^i} }}
\newcommand{\lzkok}{\ensuremath{ {\lkok^0} }}
\newcommand{\lokok}{\ensuremath{ {\lkok^1} }}
\newcommand{\ltwkok}{\ensuremath{ {\lkok^2} }}
\newcommand{\lthkok}{\ensuremath{ {\lkok^3} }}
\newcommand{\lmxk}{\ensuremath{  {\lmx}_{_k} }}
\newcommand{\lbdi}{\ensuremath{ \lambda_{_i}  }}
\newcommand{\lko}{\ensuremath{ \lambda_{_{k+1}} }}
\newcommand{\liko}{\ensuremath{ \lko^i }}
\newcommand{\lzko}{\ensuremath{ \lko^0 }}
\newcommand{\loko}{\ensuremath{ \lko^1 }}
\newcommand{\ltwko}{\ensuremath{ \lko^2 }}
\newcommand{\lthko}{\ensuremath{ \lko^3 }}

\newcommand{\lbdtld}{\ensuremath{ \widetilde{\lambda}}}
\newcommand{\lbdtldone}{\ensuremath{ \lbdtld_{_1} }}
\newcommand{\lbdtldtwo}{\ensuremath{ \lbdtld_{_2} }}
\newcommand{\lbdtldthr}{\ensuremath{ \lbdtld_{_3} }}
\newcommand{\lbdtldfor}{\ensuremath{ \lbdtld_{_4} }}
\newcommand{\lbdtldi}{\ensuremath{ \lbdtld_{_i} }}

\newcommand{\mk}{\ensuremath{ m_{_k}  }}
\newcommand{\mko}{\ensuremath{ m_{_{k+1}}  }}
\newcommand{\dmko}{\ensuremath{ \delta \mko }}
\newcommand{\mz}{\ensuremath{ m_{_0}  }}
\newcommand{\dmz}{\ensuremath{ \delta \mz }}

\newcommand{\msij}{\ensuremath{ m_{_{ij}}  }}
\newcommand{\msonetwo}{\ensuremath{ m_{_{12}}  }}
\newcommand{\msonethr}{\ensuremath{ m_{_{13}}  }}
\newcommand{\msonefor}{\ensuremath{ m_{_{14}}  }}
\newcommand{\mstwothr}{\ensuremath{ m_{_{23}}  }}
\newcommand{\mstwofor}{\ensuremath{ m_{_{24}}  }}
\newcommand{\msthrfor}{\ensuremath{ m_{_{34}}  }}

\newcommand{\mast}{\ensuremath{ m^{\!\ast}  }}

\newcommand{\muko}{\ensuremath{ \mu_{_{k+1}}  }}
\newcommand{\mukoast}{\ensuremath{ {\muko^{\ast}}  }}
\newcommand{\muz}{\ensuremath{ \mu_{_0}  }}
\newcommand{\muezz}{\ensuremath{ \widehat{\mu}_{_{0/0}}  }}
\newcommand{\muk}{\ensuremath{ \,\mu_{_k}  }}
\newcommand{\mui}{\ensuremath{ \,\mu_{_i}  }}
\newcommand{\muo}{\ensuremath{ \,\mu_{_1}  }}
\newcommand{\mutw}{\ensuremath{ \,\mu_{_2}  }}
\newcommand{\muth}{\ensuremath{ \,\mu_{_3}  }}
\newcommand{\mufo}{\ensuremath{ \,\mu_{_4}  }}

\newcommand{\nk}{\ensuremath{ {n_{_k}} }}

\newcommand{\nuoz}{\ensuremath{ \nu_{_{1/0}}  }}
\newcommand{\nui}{\ensuremath{ \nu_{_i}  }}

\newcommand{\omo}{\ensuremath{ \,\omega_{_{1}}   }}
\newcommand{\omtw}{\ensuremath{ \,\omega_{_{2}}   }}
\newcommand{\omth}{\ensuremath{ \,\omega_{_{3}}   }}

\newcommand{\omot}{\ensuremath{ \,\omega_{_{1}}^{true}   }}
\newcommand{\omtwt}{\ensuremath{ \,\omega_{_{2}}^{true}   }}
\newcommand{\omtht}{\ensuremath{ \,\omega_{_{3}}^{true}   }}

\newcommand{\Pff}{\ensuremath{ P_{_{44}} }}
\newcommand{\Pfsv}{\ensuremath{ P_{_{47}} }}
\newcommand{\Pfni}{\ensuremath{ P_{_{49}} }}
\newcommand{\Psvsv}{\ensuremath{ P_{_{77}} }}
\newcommand{\Psvni}{\ensuremath{ P_{_{79}} }}
\newcommand{\Pnini}{\ensuremath{ P_{_{99}} }}

\newcommand{\pdf}{\ensuremath{ p_\bv \left( \bv,A \right)  }}

\newcommand{\qs}{\ensuremath{ \,\emph{q}  }}
\newcommand{\qsz}{\ensuremath{ \qs^0 }}
\newcommand{\qsk}{\ensuremath{ \qs_{_k} }}
\newcommand{\qsko}{\ensuremath{ \qs_{_{k+1}} }}
\newcommand{\qes}{\ensuremath{ \,\widehat{\qs}  }}
\newcommand{\qesk}{\ensuremath{ \qes_{_{k}} }}
\newcommand{\qesko}{\ensuremath{ \qes_{_{k+1}} }}
\newcommand{\qeskk}{\ensuremath{ \qes_{_{k/k}} }}
\newcommand{\qeskok}{\ensuremath{ \qes_{_{k+1/k}}  }}
\newcommand{\qeskoko}{\ensuremath{ \qes_{_{k+1/k+1}}  }}
\newcommand{\qeskolk}{\ensuremath{ \qes_{_{k+l-1/k}}  }}
\newcommand{\qesklk}{\ensuremath{ \qes_{_{k+l/k}}  }}
\newcommand{\quats}{\ensuremath{\,\emph{q}}}
\newcommand{\qeps}{\ensuremath{ \,q_{_\epsilon}  }}

\newcommand{\rl}{\ensuremath{ \,r^{\lambda}  }}
\newcommand{\rlko}{\ensuremath{ \rl_{_{k+1}}  }}

\newcommand{\ros}{\ensuremath{ \,r_{_1}  }}
\newcommand{\rtws}{\ensuremath{ \,r_{_2}  }}
\newcommand{\rths}{\ensuremath{ \,r_{_3}  }}

\newcommand{\rkos}{\ensuremath{ \, r_{_{k+1}} }}
\newcommand{\rkoso}{\ensuremath{ {\rkos}_{_1} }}
\newcommand{\rkostw}{\ensuremath{{\rkos}_{_2} }}
\newcommand{\rkosth}{\ensuremath{ {\rkos}_{_3} }}

\newcommand{\rooz}{\ensuremath{ \,\rho_{_{1/0}}  }}
\newcommand{\rozo}{\ensuremath{ \,\rho_{_{0/1}}  }}
\newcommand{\roko}{\ensuremath{ \,\rho_{_{k+1}}  }}
\newcommand{\rok}{\ensuremath{ \,\rho_{_{k}}  }}
\newcommand{\rokoast}{\ensuremath{ \roko^{\ast}  }}
\newcommand{\rokast}{\ensuremath{ \rok^{\ast}  }}
\newcommand{\rooo}{\ensuremath{  \,\rho_{_{1/1}}  }}
\newcommand{\roe}{\ensuremath{   \,\widehat{\rho}}}
\newcommand{\rookk}{\ensuremath{ \,\rho_{_{{k-1}/k}}  }}
\newcommand{\rokok}{\ensuremath{ \,\roe_{_{{k+1}/k}}  }}
\newcommand{\rotwkok}{\ensuremath{ \,\rho_{_{{k-2}/{k-1}}}  }}
\newcommand{\rookok}{\ensuremath{ \,\rho_{_{{k-1}/{k-1}}}  }}
\newcommand{\robar}{\ensuremath{  \,\bar{\rho}  }}
\newcommand{\robarokok}{\ensuremath{ \,\robar_{_{{k-1}/{k-1}}}  }}
\newcommand{\robarbar}{\ensuremath{  \,\bar{\robar}  }}
\newcommand{\robarbarokok}{\ensuremath{ \,\robarbar_{_{{k-1}/{k-1}}}  }}

\newcommand{\sigeps}{\ensuremath{ \sigma_{_{\!\!\!\epsilon}}   }}
\newcommand{\sigb}{\ensuremath{ \,\sigma_{_{\!\!b}}   }}
\newcommand{\kape}{\ensuremath{ \widehat{\kappa} }}

\newcommand{\sigk}{\ensuremath{ \,\sigma_{_{k}}   }}
\newcommand{\sigko}{\ensuremath{ \,\sigma_{_{k+1}}   }}
\newcommand{\dsigko}{\ensuremath{ \,\delta \sigma_{_{k+1}}   }}
\newcommand{\sigkot}{\ensuremath{ \sigko^{true} }}
\newcommand{\sigi}{\ensuremath{ \,\sigma_{_i}   }}
\newcommand{\sigtot}{\ensuremath{ \,\sigma_{_{tot}}   }}
\newcommand{\sigm}{\ensuremath{   \,\sigma_{_m}   }}
\newcommand{\sigbi}{\ensuremath{ {\sigb^i} }}
\newcommand{\sigbn}{\ensuremath{ {\sigb^n} }}
\newcommand{\sigz}{\ensuremath{ \,\sigma^{o} }}
\newcommand{\sig}{\ensuremath{ \sigma }}
\newcommand{\sige}{\ensuremath{ \widehat{\sigma} }}
\newcommand{\sigone}{\ensuremath{ \sig_{_{1}}   }}
\newcommand{\sigtwo}{\ensuremath{ \sig_{_{2}}   }}
\newcommand{\sigthr}{\ensuremath{ \sig_{_{3}}   }}

\newcommand{\tz}{\ensuremath{ \,t_{_0}   }}
\newcommand{\tone}{\ensuremath{ \,t_{_1}   }}
\newcommand{\tk}{\ensuremath{ \,t_{_k}   }}
\newcommand{\tko}{\ensuremath{ \,t_{_{k+1}}   }}
\newcommand{\tktw}{\ensuremath{ \,t_{_{k+2}}   }}
\newcommand{\tok}{\ensuremath{ \,t_{_{k-1}}   }}
\newcommand{\tN}{\ensuremath{ \,t_{_N}   }}
\newcommand{\toN}{\ensuremath{ \,t_{_{N-1}}   }}
\newcommand{\tkN}{\ensuremath{ \,t_{_{k+N}}   }}
\newcommand{\tNo}{\ensuremath{ \,t_{_{N+1}}   }}
\newcommand{\ti}{\ensuremath{ \,t_{_{i}}   }}
\newcommand{\tj}{\ensuremath{ \,t_{_{j}}   }}
\newcommand{\tkl}{\ensuremath{ \,t_{_{k+l}}   }}
\newcommand{\tnu}{\ensuremath{ \,t_{_\nu}   }}
\newcommand{\tro}{\ensuremath{ \,t_{_\rho}   }}

\newcommand{\tetaek}{\ensuremath{ \,\widehat{\Theta}_{_k} }}

\newcommand{\toes}{\ensuremath{ \widehat{\tau} }}

\newcommand{\usone}{\ensuremath{  u_{_{1}}  }}
\newcommand{\ustwo}{\ensuremath{  u_{_{2}}  }}
\newcommand{\usthr}{\ensuremath{  u_{_{3}}  }}
\newcommand{\usfor}{\ensuremath{  u_{_{4}}  }}

\newcommand{\vsone}{\ensuremath{  v_{_{1}}  }}
\newcommand{\vstwo}{\ensuremath{  v_{_{2}}  }}
\newcommand{\vsthr}{\ensuremath{  v_{_{3}}  }}
\newcommand{\vsfor}{\ensuremath{  v_{_{4}}  }}

\newcommand{\Vkoioo}{\ensuremath{  \Vkoi_{_{11}}  }}
\newcommand{\Vkoiotw}{\ensuremath{ \Vkoi_{_{12}}  }}
\newcommand{\Vkoioth}{\ensuremath{ \Vkoi_{_{13}}  }}
\newcommand{\Vkoiof}{\ensuremath{  \Vkoi_{_{14}}  }}
\newcommand{\Vkoitwo}{\ensuremath{ \Vkoi_{_{21}}  }}
\newcommand{\Vkoitwtw}{\ensuremath{\Vkoi_{_{22}}  }}
\newcommand{\Vkoitwth}{\ensuremath{ \Vkoi_{_{23}}  }}
\newcommand{\Vkoitwf}{\ensuremath{ \Vkoi_{_{24}}  }}
\newcommand{\Vkoitho}{\ensuremath{  \Vkoi_{_{31}}  }}
\newcommand{\Vkoithtw}{\ensuremath{ \Vkoi_{_{32}}  }}
\newcommand{\Vkoithth}{\ensuremath{  \Vkoi_{_{33}}  }}
\newcommand{\Vkoithf}{\ensuremath{ \Vkoi_{_{34}}  }}
\newcommand{\Vkoifo}{\ensuremath{  \Vkoi_{_{41}}  }}
\newcommand{\Vkoiftw}{\ensuremath{ \Vkoi_{_{42}}  }}
\newcommand{\Vkoifth}{\ensuremath{ \Vkoi_{_{43}}  }}
\newcommand{\Vkoiff}{\ensuremath{  \Vkoi_{_{44}}  }}
\newcommand{\Vkoiij}{\ensuremath{  {\Vkoi}_{_{ij}}  }}

\newcommand{\Wkoo}{\ensuremath{  \Wk_{_{11}}  }}
\newcommand{\Wkotw}{\ensuremath{ \Wk_{_{12}}  }}
\newcommand{\Wkoth}{\ensuremath{ \Wk_{_{13}}  }}
\newcommand{\Wkof}{\ensuremath{  \Wk_{_{14}}  }}
\newcommand{\Wktwo}{\ensuremath{ \Wk_{_{21}}  }}
\newcommand{\Wktwtw}{\ensuremath{\Wk_{_{22}}  }}
\newcommand{\Wktwth}{\ensuremath{ \Wk_{_{23}}  }}
\newcommand{\Wktwf}{\ensuremath{ \Wk_{_{24}}  }}
\newcommand{\Wktho}{\ensuremath{  \Wk_{_{31}}  }}
\newcommand{\Wkthtw}{\ensuremath{ \Wk_{_{32}}  }}
\newcommand{\Wkthth}{\ensuremath{  \Wk_{_{33}}  }}
\newcommand{\Wkthf}{\ensuremath{ \Wk_{_{34}}  }}
\newcommand{\Wkfo}{\ensuremath{  \Wk_{_{41}}  }}
\newcommand{\Wkftw}{\ensuremath{ \Wk_{_{42}}  }}
\newcommand{\Wkfth}{\ensuremath{ \Wk_{_{43}}  }}
\newcommand{\Wkff}{\ensuremath{  \Wk_{_{44}}  }}
\newcommand{\Wkij}{\ensuremath{  {\Wk}_{_{ij}}  }}

\newcommand{\xoo}{\ensuremath{ \,x_{_{11}}   }}
\newcommand{\xotw}{\ensuremath{ \,x_{_{12}}   }}
\newcommand{\xtwtw}{\ensuremath{ \,x_{_{22}}   }}
\newcommand{\xij}{\ensuremath{ \,x_{_{ij}}   }}
\newcommand{\xks}{\ensuremath{ \,x_{_k} }}
\newcommand{\xkso}{\ensuremath{ {\xks}_{_1} }}
\newcommand{\xkstw}{\ensuremath{ {\xks}_{_2} }}
\newcommand{\xksth}{\ensuremath{ {\xks}_{_3} }}
\newcommand{\xksf}{\ensuremath{ {\xks}_{_4} }}
\newcommand{\xksfv}{\ensuremath{ {\xks}_{_5} }}
\newcommand{\xkssx}{\ensuremath{ {\xks}_{_6} }}
\newcommand{\xkssv}{\ensuremath{ {\xks}_{_7} }}
\newcommand{\xksei}{\ensuremath{ {\xks}_{_8} }}
\newcommand{\xksni}{\ensuremath{ {\xks}_{_9} }}

\newcommand{\xst}{\ensuremath{ x_{_{t}}   }}

\newcommand{\xsone}{\ensuremath{ x_{_{1}}   }}
\newcommand{\xstwo}{\ensuremath{ x_{_{2}}   }}
\newcommand{\xsthr}{\ensuremath{ x_{_{3}}   }}
\newcommand{\xsfor}{\ensuremath{ x_{_{4}}   }}

\newcommand{\xes}{\ensuremath{ \,\widehat{x} }}
\newcommand{\xest}{\ensuremath{ \xes_{_{t}}   }}
\newcommand{\xeso}{\ensuremath{ {\xes}_{_1} }}
\newcommand{\xestw}{\ensuremath{ {\xes}_{_2} }}
\newcommand{\xesth}{\ensuremath{ {\xes}_{_3} }}
\newcommand{\xesf}{\ensuremath{ {\xes}_{_4} }}
\newcommand{\xesfv}{\ensuremath{ {\xes}_{_5} }}
\newcommand{\xessx}{\ensuremath{ {\xes}_{_6} }}
\newcommand{\xessv}{\ensuremath{ {\xes}_{_7} }}
\newcommand{\xesei}{\ensuremath{ {\xes}_{_8} }}
\newcommand{\xesni}{\ensuremath{ {\xes}_{_9} }}
\newcommand{\xesten}{\ensuremath{ {\xes}_{_{10}} }}

\newcommand{\xso}{\ensuremath{ {x}_{_1} }}
\newcommand{\xstw}{\ensuremath{ {x}_{_2} }}
\newcommand{\xsth}{\ensuremath{ {x}_{_3} }}
\newcommand{\xsf}{\ensuremath{ {x}_{_4} }}
\newcommand{\xsfv}{\ensuremath{ {x}_{_5} }}
\newcommand{\xssx}{\ensuremath{ {x}_{_6} }}
\newcommand{\xssv}{\ensuremath{ {x}_{_7} }}
\newcommand{\xsei}{\ensuremath{ {x}_{_8} }}
\newcommand{\xsni}{\ensuremath{ {x}_{_9} }}
\newcommand{\xsten}{\ensuremath{ {x}_{_{10}} }}

\newcommand{\xsoneone}{\ensuremath{ {x}_{_{11}} }}
\newcommand{\xstwotwo}{\ensuremath{ {x}_{_{22}} }}
\newcommand{\xsthrthr}{\ensuremath{ {x}_{_{33}} }}
\newcommand{\xsforfor}{\ensuremath{ {x}_{_{44}} }}
\newcommand{\xsonefor}{\ensuremath{ {x}_{_{14}} }}
\newcommand{\xstwofor}{\ensuremath{ {x}_{_{24}} }}
\newcommand{\xsthrfor}{\ensuremath{ {x}_{_{34}} }}
\newcommand{\xsonetwo}{\ensuremath{ {x}_{_{12}} }}
\newcommand{\xsonethr}{\ensuremath{ {x}_{_{13}} }}
\newcommand{\xstwothr}{\ensuremath{ {x}_{_{23}} }}

\newcommand{\xsii}{\ensuremath{ {x}_{_{ii}} }}
\newcommand{\xsjj}{\ensuremath{ {x}_{_{jj}} }}
\newcommand{\xsij}{\ensuremath{ {x}_{_{ij}} }}

\newcommand{\fx}{\ensuremath{ f \left( \xv \right)  }}


\newcommand{\dhat}{\ensuremath{ \,\widehat{D}  }}
\newcommand{\dhato}{\ensuremath{ \dhat_{_1}  }}

\newcommand{\fz}{\ensuremath{ \,F_{_0}  }}
\newcommand{\fN}{\ensuremath{ \,F_{_N}  }}
\newcommand{\foN}{\ensuremath{ \,F_{_{N-1}}  }}
\newcommand{\fk}{\ensuremath{ \,F_{_k}  }}
\newcommand{\foktwk}{\ensuremath{ \,F_{_{k-1,k-2}}  }}
\newcommand{\fKok}{\ensuremath{ \,F_{_{k,k-1}}  }}

\newcommand{\fNqN}{\ensuremath{ \fN \left( \qN \right) }}
\newcommand{\fkqk}{\ensuremath{ \fk \left( \qk \right) }}
\newcommand{\foNqoN}{\ensuremath{ \foN \left( \qoN \right) }}
\newcommand{\fzqz}{\ensuremath{ \fz \left( \qz\right) }}

\newcommand{\fstar}{\ensuremath{ \,{F}^{\star}  }}
\newcommand{\fstarz}{\ensuremath{ \fstar_{_0}  }}
\newcommand{\fstarzqz}{\ensuremath{ \fstarz \left( \qz \right)  }}
\newcommand{\fstarok}{\ensuremath{ \fstar_{_{k-1}}  }}
\newcommand{\fstarokqok}{\ensuremath{ \fstarok \left( \qok \right)  }}
\newcommand{\fstark}{\ensuremath{ \fstar_{_k}  }}
\newcommand{\fstarkqk}{\ensuremath{ \fstark \left( \qk \right)  }}
\newcommand{\fstaroktwk}{\ensuremath{ \fstar_{_{k-1,k-2}}  }}

\newcommand{\fhat}{\ensuremath{ \,\widehat{F}  }}
\newcommand{\fhatz}{\ensuremath{ \fhat_{_0}  }}
\newcommand{\fhatok}{\ensuremath{ \fhat_{_{k-1}}  }}
\newcommand{\fhatk}{\ensuremath{ \fhat_{_k}  }}
\newcommand{\fhatoktwk}{\ensuremath{ \fhat_{_{k-1,k-2}}  }}

\newcommand{\fq}{\ensuremath{ f(\qv) }}

\newcommand{\gz}{\ensuremath{ \,G_{_0}  }}
\newcommand{\gN}{\ensuremath{ \,G_{_N}  }}
\newcommand{\goN}{\ensuremath{ \,G_{_{N-1}}  }}
\newcommand{\gk}{\ensuremath{ \,G_{_k}  }}
\newcommand{\gkqk}{\ensuremath{ \gk \left( \qk \right)  }}
\newcommand{\goktwk}{\ensuremath{ \,G_{_{k-1,k-2}}  }}
\newcommand{\gokok}{\ensuremath{ \,G_{_{k-1,k-1}}  }}
\newcommand{\gNqN}{\ensuremath{ \gN \left( \qN \right) }}

\newcommand{\gstar}{\ensuremath{ \,{G}^{\star}  }}
\newcommand{\gstark}{\ensuremath{ \gstar_{_k}  }}
\newcommand{\gstarkqk}{\ensuremath{ \gstark \left( \qk \right)  }}

\newcommand{\ghat}{\ensuremath{ \,\widehat{G}  }}
\newcommand{\ghatk}{\ensuremath{ \ghat_{_k}  }}

\newcommand{\hN}{\ensuremath{ \,H_{_N}  }}
\newcommand{\hk}{\ensuremath{ \,H_{_k}  }}
\newcommand{\hokok}{\ensuremath{ \,H_{_{k-1,k-1}}  }}
\newcommand{\hKok}{\ensuremath{ \,H_{_{k,k-1}}  }}

\newcommand{\hstar}{\ensuremath{ \,\star{H}  }}
\newcommand{\hstarokok}{\ensuremath{ \hstar_{_{k-1,k-1}}  }}
\newcommand{\hstarKok}{\ensuremath{ \hstar_{_{k,k-1}}  }}

\newcommand{\hhat}{\ensuremath{ \,\widehat{H}  }}
\newcommand{\hhatz}{\ensuremath{ \hhat_{_0}  }}
\newcommand{\hhato}{\ensuremath{ \hhat_{_1}  }}
\newcommand{\hhatokok}{\ensuremath{ \hhat_{_{k-1,k-1}}  }}
\newcommand{\hhatKok}{\ensuremath{ \hhat_{_{k,k-1}}  }}

\newcommand{\iN}{\ensuremath{ \,I_{_N}  }}
\newcommand{\iKok}{\ensuremath{ \,I_{_{k,k-1}}  }}

\newcommand{\istar}{\ensuremath{ \,\star{I}  }}
\newcommand{\istarKok}{\ensuremath{ \istar_{_{k,k-1}}  }}

\newcommand{\ihat}{\ensuremath{ \,\widehat{I}  }}
\newcommand{\ihatKok}{\ensuremath{ \ihat_{_{k,k-1}}  }}

\newcommand{\Jko}{\ensuremath{ \,J_{_{k+1}}  }}
\newcommand{\Jzz}{\ensuremath{ \,J_{_{0/0}}  }}
\newcommand{\Joz}{\ensuremath{ \,J_{_{1/0}}  }}
\newcommand{\Joo}{\ensuremath{ \,J_{_{1/1}}  }}
\newcommand{\Je}{\ensuremath{  \,\widehat{J}}}
\newcommand{\Jezz}{\ensuremath{ \,\Je_{_{0/0}}  }}
\newcommand{\Jeoz}{\ensuremath{ \,\Je_{_{1/0}}  }}
\newcommand{\Jeoo}{\ensuremath{ \,\Je_{_{1/1}}  }}
\newcommand{\Jekk}{\ensuremath{ \,\Je_{_{k/k}}  }}
\newcommand{\Jeokok}{\ensuremath{ \,\Je_{_{k-1/k-1}}  }}
\newcommand{\Jstar}{\ensuremath{  \,\star{J}}}
\newcommand{\Jzzstar}{\ensuremath{ \,\Jstar_{_{0/0}}   }}
\newcommand{\Jozstar}{\ensuremath{ \,\Jstar_{_{1/0}}   }}
\newcommand{\Joostar}{\ensuremath{ \,\Jstar_{_{1/1}}   }}
\newcommand{\Jbar}{\ensuremath{  {  \,\bar{J}} }}
\newcommand{\Joobar}{\ensuremath{  \,\Jbar_{{1/1}} }}
\newcommand{\Joobarstar}{\ensuremath{ \,\Joobar^\star    }}
\newcommand{\Jeoobar}{\ensuremath{  \,\widehat{\Joobar} }}
\newcommand{\JNN}{\ensuremath{ \,J_{_{N/N}}  }}
\newcommand{\JN}{\ensuremath{ \,J_{_{N}}  }}
\newcommand{\JNoN}{\ensuremath{ J_{_{N+1/N}}  }}
\newcommand{\JNoNo}{\ensuremath{ J_{_{N+1/N+1}}  }}


\newcommand{\Aml}{\ensuremath{{ \,\widehat{A}{_{_{ML}}} }}}
\newcommand{\At}{\ensuremath{{ A_{_t}} }}
\newcommand{\Ae}{\ensuremath{  \widehat{A} }}
\newcommand{\Aet}{\ensuremath{  \Ae \left( t \right) }}
\newcommand{\Aetp}{\ensuremath{   \Ae \left( t' \right) }}
\newcommand{\Ako}{\ensuremath{ \,A_{_{k+1}}  }}
\newcommand{\Atp}{\ensuremath{{ \,A \left( t' \right)  }}}
\newcommand{\Atrue}{\ensuremath{ \,A^{\mbox{true}}  }}

\newcommand{\AAm}{\ensuremath{{ \mathcal{A}} }}

\newcommand{\Be}{\ensuremath{ \widehat{B} }}
\newcommand{\Beps}{\ensuremath{{ \,B_{_\epsilon} }}}
\newcommand{\Bb}{\ensuremath{{ \,B_{_b} }}}
\newcommand{\Bt}{\ensuremath{{ \,B \left( t \right)  }}}
\newcommand{\Btp}{\ensuremath{{ \,B \left( t' \right)  }}}
\newcommand{\Btx}{\ensuremath{{ \,B \left( t , \xv\right)  }}}
\newcommand{\Bk}{\ensuremath{{ B_{_{k}}  }}}
\newcommand{\Bko}{\ensuremath{{ \,B_{_{k+1}}  }}}
\newcommand{\dBko}{\ensuremath{{ \,\delta B_{_{k+1}}  }}}
\newcommand{\Bkk}{\ensuremath{{ \,B_{_{k/k}}  }}}
\newcommand{\Bkok}{\ensuremath{{ \,B_{_{k+1/k}}  }}}
\newcommand{\Bkoko}{\ensuremath{{ \,B_{_{k+1/k+1}}  }}}
\newcommand{\bicross}{\ensuremath{ \left[ {\bf b}_{_i} \times \right]  }}
\newcommand{\bcross}{\ensuremath{ \left[ {\bf b} \times \right]  }}
\newcommand{\bkcross}{\ensuremath{ \left[ {\bk} \times \right]  }}

\newcommand{\Bbi}{\ensuremath{ \Bb^i }}
\newcommand{\Bbn}{\ensuremath{ \Bb^n }}
\newcommand{\Bz}{\ensuremath{{ \,B^o }}}

\newcommand{\Btrue}{\ensuremath{ \,B^o }}
\newcommand{\Bkkt}{\ensuremath{ \Btrue_{_{k/k}} }}
\newcommand{\Bkot}{\ensuremath{ \Btrue_{_{k+1}} }}

\newcommand{\Coo}{\ensuremath{{ \,C{_{_{11}}} }}}
\newcommand{\Cotw}{\ensuremath{{ \,C{_{_{12}}} }}}
\newcommand{\Ctwo}{\ensuremath{{ \,C{_{_{21}}} }}}
\newcommand{\Ctwtw}{\ensuremath{{ \,C{_{_{22}}} }}}

\newcommand{\Cbar}{\ensuremath{\overline{C} }}
\newcommand{\Cbark}{\ensuremath{ \Cbar_{_{k}} }}

\newcommand{\Cl}{\ensuremath{  \,\mbox{Cl} }}
\newcommand{\Clo}{\ensuremath{  \Cl_{_1} }}
\newcommand{\Cltw}{\ensuremath{  \Cl_{_2} }}
\newcommand{\Clth}{\ensuremath{  \Cl_{_3} }}
\newcommand{\Clf}{\ensuremath{  \Cl_{_4} }}
\newcommand{\Clfv}{\ensuremath{  \Cl_{_5} }}
\newcommand{\Clsx}{\ensuremath{  \Cl_{_6} }}
\newcommand{\Clsv}{\ensuremath{  \Cl_{_7} }}
\newcommand{\Clei}{\ensuremath{  \Cl_{_8} }}
\newcommand{\Clni}{\ensuremath{  \Cl_{_9} }}
\newcommand{\Clten}{\ensuremath{  \Cl_{_{10}} }}
\newcommand{\Clel}{\ensuremath{  \Cl_{_{11}} }}
\newcommand{\Cltwl}{\ensuremath{  \Cl_{_{12}} }}
\newcommand{\Clthr}{\ensuremath{  \Cl_{_{13}} }}
\newcommand{\Clfrt}{\ensuremath{  \Cl_{_{14}} }}
\newcommand{\Clfft}{\ensuremath{  \Cl_{_{15}} }}
\newcommand{\Clsxt}{\ensuremath{  \Cl_{_{16}} }}

\newcommand{\Dtw}{\ensuremath{  \,D_{_2} }}

\newcommand{\Deltay}{\ensuremath{  \Delta_{_{\!Y}} }}

\newcommand{\Di}{\ensuremath{ D_{_{i}} }}
\newcommand{\Dt}{\ensuremath{{ D_{_t}} }}

\newcommand{\dbicross}{\ensuremath{ \left[ \delta {\bf b}_{_i} \times \right]  }}
\newcommand{\dbjcross}{\ensuremath{ \left[ \delta {\bf b}_{_j} \times \right]  }}
\newcommand{\dbkocross}{\ensuremath{ \left[ \dbko \times \right]  }}
\newcommand{\donevcross}{\ensuremath{ \left[ \donev \times \right]  }}

\newcommand{\Dalfast}{\ensuremath{ D_{\!\alfav^{\!\ast}} }}
\newcommand{\Dalfav}{\ensuremath{ D_{\!\alfav} }}

\newcommand{\Eij}{\ensuremath{{ \, \Xer{_{_{ij}}} }}}
\newcommand{\Eji}{\ensuremath{{ \,\Xer{_{_{ji}}} }}}
\newcommand{\Eik}{\ensuremath{{ \,\Xer{_{_{ik}}} }}}
\newcommand{\Ejk}{\ensuremath{{ \,\Xer{_{_{jk}}} }}}
\newcommand{\Ekk}{\ensuremath{{ \,\Xer{_{_{k/k}}} }}}
\newcommand{\Ekok}{\ensuremath{{ \,\Xer{_{_{k+1/k}}} }}}
\newcommand{\Ekoko}{\ensuremath{{ \,\Xer{_{_{k+1/k+1}}} }}}
\newcommand{\ekcross}{\ensuremath{ \,\left[ \ek \times \right] }}
\newcommand{\ekocross}{\ensuremath{ \,\left[ \eko \times \right] }}
\newcommand{\ecross}{\ensuremath{ \,\left[ \ev \times \right] }}
\newcommand{\evcross}{\ensuremath{ \,\left[ \ev \times \right] }}
\newcommand{\eekcross}{\ensuremath{ \,\left[ \eek \times \right] }}
\newcommand{\eekkcross}{\ensuremath{ \,\left[ \eekk \times \right] }}
\newcommand{\eekocross}{\ensuremath{ \,\left[ \eeko \times \right] }}
\newcommand{\eekokcross}{\ensuremath{ \,\left[ \eekok \times \right] }}
\newcommand{\eekokocross}{\ensuremath{ \,\left[ \eekoko \times \right] }}
\newcommand{\eekolkcross}{\ensuremath{ \,\left[ \eekolk \times \right] }}
\newcommand{\eeklkcross}{\ensuremath{ \,\left[ \eeklk \times \right] }}

\newcommand{\Epsk}{\ensuremath{{ \,\mathcal{E}_{_k}}}}
\newcommand{\Epsko}{\ensuremath{{ \,\mathcal{E}_{_{k+1}}}}}
\newcommand{\Eps}{\ensuremath{ \,\mathcal{E} }}
\newcommand{\epscross}{\ensuremath{ \,\left[ \epsv \times \right] }}
\newcommand{\epskocross}{\ensuremath{ \,\left[ \epsko \times \right] }}
\newcommand{\epskcross}{\ensuremath{ \,\left[ \epsk \times \right] }}

\newcommand{\Fko}{\ensuremath{ \,F_{_{k+1}} }}

\newcommand{\Fe}{\ensuremath{ \widehat{F} }}
\newcommand{\Fet}{\ensuremath{ \Fe_{_{t}} }}

\newcommand{\Ftt}{\ensuremath{ \,F_{_{\theta \theta}}  }}
\newcommand{\Fk}{\ensuremath{ \,\mathcal{F}_{_k}  }}

\newcommand{\Fik}{\ensuremath{   \,\Phi_{_k} }}
\newcommand{\FiN}{\ensuremath{   \,\Phi_{_N} }}
\newcommand{\Fiko}{\ensuremath{   \,\Phi_{_{k+1}} }}
\newcommand{\Fikt}{\ensuremath{   {\Phi_{_k}^{o}} }}
\newcommand{\Fiqk}{\ensuremath{   \,\Phi_{4_k} }}
\newcommand{\Fiz}{\ensuremath{   \,\Phi_{_0} }}
\newcommand{\dFik}{\ensuremath{  \,\Delta \Phi_{_k} }}
\newcommand{\Fitt}{\ensuremath{  \, \Phi \left( t' , t \right)   }}
\newcommand{\fikcross}{\ensuremath{ \,\left[ \fik \times \right] }}
\newcommand{\Fikol}{\ensuremath{ \, \Phi_{_{k+l-1}} }}
\newcommand{\Fiok}{\ensuremath{{   \,\Phi_{_{k-1}}  }}}
\newcommand{\FioN}{\ensuremath{{   \,\Phi_{_{N-1}}  }}}
\newcommand{\Fiktrue}{\ensuremath{{  \,\Phi^0_{_k}}}}
\newcommand{\Fie}{\ensuremath{   \,\widehat{\Phi} }}
\newcommand{\Fiekk}{\ensuremath{  \Fie_{_{k/k}}  }}
\newcommand{\Fieokok}{\ensuremath{  \Fie_{_{k-1/k-1}}  }}
\newcommand{\Fiezz}{\ensuremath{  \Fie_{_{0/0}}  }}

\newcommand{\Fito}{\ensuremath{ F_{_{\!\!I}} }}
\newcommand{\Flgv}{\ensuremath{ F_{_{\!\!L}} }}
\newcommand{\Fy}{\ensuremath{ F_{_{\!\!Y}} }}

\newcommand{\Fisxtk}{\ensuremath{   \Fik^{16} }}
\newcommand{\Fitenk}{\ensuremath{   \Fik^{10} }}
\newcommand{\Finink}{\ensuremath{   \Fik^{9} }}

\newcommand{\fitm}{\ensuremath{{ \Phi}^o}}
\newcommand{\fim}{\ensuremath{{ \Phi}}}
\newcommand{\fierm}{\ensuremath{{\Delta \Phi}}}
\newcommand{\fikrm}{\ensuremath{\fim_{[2]_k} }}

\newcommand{\Gam}{\ensuremath{ \,\Gamma }}
\newcommand{\Gamk}{\ensuremath{ \Gam_{_{k}} }}
\newcommand{\Gamae}{\ensuremath{ \widehat{\Gam} }}
\newcommand{\Gamekk}{\ensuremath{ \Gamae_{_{k/k}} }}

\newcommand{\Gamsxtk}{\ensuremath{ \Gamk^{16} }}
\newcommand{\Gamtenk}{\ensuremath{ \Gamk^{10} }}
\newcommand{\Gamnink}{\ensuremath{ \Gamk^{9} }}

\newcommand{\Gko}{\ensuremath{ \,G_{_{k+1}} }}
\newcommand{\GWLS}{\ensuremath{ \,G^{_{_{WLS}}} }}
\newcommand{\GWLSko}{\ensuremath{ \GWLS_{_{k+1}} }}

\newcommand{\Hk}{\ensuremath{ \,H_{_{k}} }}
\newcommand{\Hko}{\ensuremath{ \,H_{_{k+1}} }}
\newcommand{\Hz}{\ensuremath{ \,H_{_{0}} }}
\newcommand{\Hzz}{\ensuremath{ \,H^{0}_{_{0}} }}
\newcommand{\Hzo}{\ensuremath{ \,H^{1}_{_{0}} }}
\newcommand{\Ho}{\ensuremath{ \,H_{_{1}} }}
\newcommand{\HN}{\ensuremath{ \,H_{_{N}} }}
\newcommand{\HNo}{\ensuremath{ \,H_{_{N+1}} }}
\newcommand{\Hktw}{\ensuremath{ \,H_{_{k+2}} }}
\newcommand{\Hkl}{\ensuremath{ \,H_{_{k+l}} }}
\newcommand{\Hkot}{\ensuremath{ \Hko^{true} }}
\newcommand{\Herko}{\ensuremath{ \,\Delta H_{_{k+1}} }}
\newcommand{\Hglobal}{\ensuremath{ \,\mathcal{H} }}
\newcommand{\Hglobaloo}{\ensuremath{ \Hglobal_{_{1/1}} }}
\newcommand{\HglobalNN}{\ensuremath{ \Hglobal_{_{N/N}} }}
\newcommand{\HglobalNoN}{\ensuremath{ \Hglobal_{_{N+1/N}} }}
\newcommand{\HglobalNoNo}{\ensuremath{ \Hglobal_{_{N+1/N+1}} }}
\newcommand{\Hglobalkk}{\ensuremath{ \Hglobal_{_{k/k}} }}
\newcommand{\Hglobalkok}{\ensuremath{ \Hglobal_{_{k+1/k}} }}
\newcommand{\Hglobalkoko}{\ensuremath{ \Hglobal_{_{k+1/k+1}} }}
\newcommand{\Hglobalzz}{\ensuremath{ \Hglobal_{_{0/0}} }}

\newcommand{\Hbar}{\ensuremath{\overline{H} }}
\newcommand{\Hbark}{\ensuremath{ \Hbar_{_{k}} }}
\newcommand{\Hbarko}{\ensuremath{ \Hbar_{_{k+1}} }}

\newcommand{\Itwo}{\ensuremath{ \,I_{_2}  }}
\newcommand{\Ithree}{\ensuremath{ \,I_{_3}  }}
\newcommand{\Ifour}{\ensuremath{  \,I_{_4}  }}
\newcommand{\Inine}{\ensuremath{  \,I_{_9}  }}
\newcommand{\Idm}{\ensuremath{ \,I_{_m}  }}
\newcommand{\Isixteen}{\ensuremath{  \,I_{_{16}}  }}

\newcommand{\Kk}{\ensuremath{{ K_{_k} }}}
\newcommand{\Kko}{\ensuremath{ K_{_{k+1}} }}
\newcommand{\Ky}{\ensuremath{{ K_{_{\!Y}} }}}
\newcommand{\Kepsk}{\ensuremath{{ K_{_{\epsk}} }}}
\newcommand{\Kbko}{\ensuremath{{ K_{_{b_{k+1}}} }}}
\newcommand{\Kbk}{\ensuremath{{ K_{_{b_{k}}} }}}
\newcommand{\Kkk}{\ensuremath{{ \,K{_{_{\!\!k/k}}}  }}}
\newcommand{\Kij}{\ensuremath{{ \,K{_{_{\!\!i/j}}}  }}}
\newcommand{\Kijt}{\ensuremath{{ \,K_{_{\!\!i/j}}^{o}  }}}
\newcommand{\KNN}{\ensuremath{{ \,K{_{_{\!\!N/N}}}  }}}
\newcommand{\Kkok}{\ensuremath{{ \,K_{_{\!\!k+1/k}}  }}}
\newcommand{\Kkoko}{\ensuremath{{ \,K_{_{\!\!k+1/k+1}}  }}}
\newcommand{\Kzz}{\ensuremath{{ \,K{_{_{\!\!0/0}}}  }}}
\newcommand{\dK}{\ensuremath{ \,\delta\!K }}
\newcommand{\dKko}{\ensuremath{ \,\delta\!K_{_{\!k+1}} }}
\newcommand{\dKk}{\ensuremath{ \,\delta\!K_{_{\!k}} }}
\newcommand{\dKi}{\ensuremath{ \,\delta\!K_{_{\!i}} }}
\newcommand{\dKit}{\ensuremath{ \delta\!K_{_{i}}^{o}  }}
\newcommand{\dKz}{\ensuremath{ \,\delta\!K_{_{0}} }}
\newcommand{\dKon}{\ensuremath{ \,\delta\!K_{_{1}} }}
\newcommand{\dKtw}{\ensuremath{ \,\delta\!K_{_{2}} }}
\newcommand{\Kwerm}{\ensuremath{{{{\Delta\!K}^\varepsilon} }}}
\newcommand{\DKkokeps}{\ensuremath{ \Kwerm_{_{k+1/k}} }}
\newcommand{\Kweronem}{\ensuremath{ \Kwerm^{(1)}}}
\newcommand{\Kberm}{\ensuremath{{{{\Delta K}^b} }}}
\newcommand{\Ker}{\ensuremath{  \Delta\!K  }}
\newcommand{\Kzer}{\ensuremath{  {\Ker^0} }}
\newcommand{\Kerij}{\ensuremath{  \Ker_{_{i/j}}  }}
\newcommand{\Kerkk}{\ensuremath{  \Ker_{_{k/k}}  }}
\newcommand{\Kerkok}{\ensuremath{  \Ker_{_{k+1/k}}  }}
\newcommand{\Kzerkok}{\ensuremath{  {\Kerkok^0} }}
\newcommand{\Kzerkoko}{\ensuremath{  {\Kerkoko^0} }}
\newcommand{\Kzerkk}{\ensuremath{  {\Kerkk^0} }}
\newcommand{\Kierkok}{\ensuremath{  {\Kerkok^i} }}
\newcommand{\Kerkoko}{\ensuremath{  \Ker_{_{k+1/k+1}}  }}
\newcommand{\Ke}{\ensuremath{ \widehat{K} }}
\newcommand{\Kekk}{\ensuremath{ \Ke_{_{k/k}} }}
\newcommand{\Ket}{\ensuremath{ \Ke_{_t} }}

\newcommand{\Kt}{\ensuremath{ \,K^o  }}
\newcommand{\Ktij}{\ensuremath{ \Kt_{_{i/j}}  }}
\newcommand{\dKt}{\ensuremath{  \dK^o }}
\newcommand{\dKti}{\ensuremath{ \dKt_{_i}  }}
\newcommand{\dKkot}{\ensuremath{ \dKt_{_{k+1}} }}
\newcommand{\Kzzt}{\ensuremath{ \Kt_{_{0/0}} }}
\newcommand{\Kkokt}{\ensuremath{ \Kt_{_{k+1/k}} }}
\newcommand{\Kkkt}{\ensuremath{ \Kt_{_{k/k}} }}
\newcommand{\Kkokot}{\ensuremath{ \Kt_{_{k+1/k+1}} }}
\newcommand{\Kkokap}{\ensuremath{ \Kkokt }}
\newcommand{\Kkkap}{\ensuremath{ \Kkkt }}
\newcommand{\Kkokoap}{\ensuremath{ \Kkokot }}
\newcommand{\Kzzap}{\ensuremath{ \Kzzt }}

\newcommand{\Kgaint}{\ensuremath{ K_{_{t}}  }}

\newcommand{\KW}{\ensuremath{ \,K^{_{_W} } }}
\newcommand{\KWko}{\ensuremath{ \KW_{_{k+1}} }}

\newcommand{\Lko}{\ensuremath{ \,L_{_{k+1}} }}
\newcommand{\Ltw}{\ensuremath{ \,L_{_{2}} }}
\newcommand{\LN}{\ensuremath{ \,L_{_{N}} }}
\newcommand{\Lt}{\ensuremath{ L_{_{t}} }}
\newcommand{\Lonefor}{\ensuremath{ L_{_{14}} }}
\newcommand{\Ltwofor}{\ensuremath{ L_{_{24}} }}
\newcommand{\Lthrfor}{\ensuremath{ L_{_{34}} }}
\newcommand{\Ljk}{\ensuremath{ L_{_{jk}} }}
\newcommand{\Lij}{\ensuremath{ L_{_{ij}} }}

\newcommand{\Ll}{\ensuremath{  \,\mbox{Li} }}
\newcommand{\Llo}{\ensuremath{  \Ll_{_1} }}
\newcommand{\Lltw}{\ensuremath{  \Ll_{_2} }}
\newcommand{\Llth}{\ensuremath{  \Ll_{_3} }}
\newcommand{\Llf}{\ensuremath{  \Ll_{_4} }}
\newcommand{\Llfv}{\ensuremath{  \Ll_{_5} }}
\newcommand{\Llsx}{\ensuremath{  \Ll_{_6} }}
\newcommand{\Llsv}{\ensuremath{  \Ll_{_7} }}
\newcommand{\Llei}{\ensuremath{  \Ll_{_8} }}
\newcommand{\Llni}{\ensuremath{  \Ll_{_9} }}
\newcommand{\Llten}{\ensuremath{  \Ll_{_{10}} }}
\newcommand{\Llel}{\ensuremath{  \Ll_{_{11}} }}
\newcommand{\Lltwl}{\ensuremath{  \Ll_{_{12}} }}
\newcommand{\Llthr}{\ensuremath{  \Ll_{_{13}} }}
\newcommand{\Llfrt}{\ensuremath{  \Ll_{_{14}} }}
\newcommand{\Llfft}{\ensuremath{  \Ll_{_{15}} }}
\newcommand{\Llsxt}{\ensuremath{  \Ll_{_{16}} }}

\newcommand{\Lam}{\ensuremath{ \,\Lambda }}
\newcommand{\Lamko}{\ensuremath{ \Lam_{_{k+1}} }}
\newcommand{\Lami}{\ensuremath{ \Lam^i }}
\newcommand{\Lamiko}{\ensuremath{ {\Lami_{_{k+1}}} }}
\newcommand{\Lamj}{\ensuremath{ \Lam^j }}
\newcommand{\Lamjko}{\ensuremath{ {\Lamj_{_{k+1}}} }}

\newcommand{\Mk}{\ensuremath{ \,M_{_k} }}
\newcommand{\Mtw}{\ensuremath{ \,M_{_2} }}
\newcommand{\MN}{\ensuremath{ \,M_{_N} }}

\newcommand{\Me}{\ensuremath{ \widehat{M} }}
\newcommand{\Met}{\ensuremath{ \Me_{_{t}} }}

\newcommand{\onem}{\ensuremath{\,\widehat{R}}}
\newcommand{\Omg}{\ensuremath{ \,\Omega }}
\newcommand{\Omk}{\ensuremath{ \,\Omega_{_k} }}
\newcommand{\Omkt}{\ensuremath{ \,\Omega_{_k}^{true} }}
\newcommand{\omgcross}{\ensuremath{ \,\left[ \omg \times \right] }}
\newcommand{\omkcross}{\ensuremath{ \,\left[ \omk \times \right] }}
\newcommand{\omktcross}{\ensuremath{ \,\left[ \omkt \times \right] }}
\newcommand{\omekkcross}{\ensuremath{ \,\left[ \omekk \times \right] }}
\newcommand{\Ome}{\ensuremath{ \widehat{\Omg} }}
\newcommand{\Omekk}{\ensuremath{ \Ome_{_{k/k}}  }}

\newcommand{\Omghat}{\ensuremath{ \widehat{\Omega} }}

\newcommand{\Ofour}{\ensuremath{ \,O_{_4} }}
\newcommand{\Othree}{\ensuremath{ \,O_{_3} }}

\newcommand{\Pij}{\ensuremath{{ \,P{_{_{i/j}}} }}}
\newcommand{\Pkk}{\ensuremath{ \,P_{_{k/k}}  }}
\newcommand{\Pkok}{\ensuremath{ \,P_{_{k+1/k}} }}
\newcommand{\Pklk}{\ensuremath{{ \,P_{_{k+l/k}} }}}
\newcommand{\Pkolk}{\ensuremath{{ \,P_{_{k+l-1/k}} }}}
\newcommand{\Pkoko}{\ensuremath{ \,P_{_{k+1/k+1}} }}
\newcommand{\Past}{\ensuremath{ \,P^{\ast} }}
\newcommand{\Pkokoast}{\ensuremath{ \Past_{_{k+1/k+1}} }}
\newcommand{\Pkkast}{\ensuremath{ \Past_{_{k/k}} }}
\newcommand{\Pzz}{\ensuremath{{ \,P_{_{0/0}} }}}
\newcommand{\Ptt}{\ensuremath{ \,P_{_{\theta \theta}}  }}
\newcommand{\Pxx}{\ensuremath{{ \,P{_{_{XX}}} }}}
\newcommand{\Pxxoo}{\ensuremath{ \Pxx_{_{11}}       }}
\newcommand{\Prr}{\ensuremath{{ \,P{_{_{\!\!\!rr}}} }}}
\newcommand{\Prrkok}{\ensuremath{ \Prr_{_{k+1/k}} }}
\newcommand{\Prrklk}{\ensuremath{ \Prr_{_{k+l/k}} }}
\newcommand{\Pmm}{\ensuremath{{ \,P{_{_{mm}}} }}}
\newcommand{\Pko}{\ensuremath{ \,P_{_{k+1}} }}
\newcommand{\Pk}{\ensuremath{ \,P_{_{k}} }}

\newcommand{\Pt}{\ensuremath{ P_{_{t}} }}

\newcommand{\Pq}{\ensuremath{ \,P^{q} }}
\newcommand{\Pqkok}{\ensuremath{ \Pq_{_{\!\!k+1/k}} }}
\newcommand{\Pqkoko}{\ensuremath{ \Pq_{_{\!\!k+1/k+1}} }}
\newcommand{\Pqkk}{\ensuremath{ \Pq_{_{\!\!k/k}} }}

\newcommand{\Pee}{\ensuremath{ P_{_{ee}} }}
\newcommand{\Pzzz}{\ensuremath{ P_{_{zz}} }}

\newcommand{\Pv}{\ensuremath{ P^{v} }}
\newcommand{\Pvk}{\ensuremath{ \Pv_{_{\!\!k}} }}
\newcommand{\Pvij}{\ensuremath{ P^{v_{ij}} }}
\newcommand{\Pvonefor}{\ensuremath{ P^{v_{14}} }}

\newcommand{\Pvbar}{\ensuremath{ P^{\bar{v}} }}
\newcommand{\Pvbark}{\ensuremath{ \Pvbar_{_{\!\!k}} }}
\newcommand{\Pvbarko}{\ensuremath{ \Pvbar_{_{\!\!k+1}} }}

\newcommand{\Pe}{\ensuremath{ \widehat{P} }}
\newcommand{\Pet}{\ensuremath{ \Pe_{_{t}} }}

\newcommand{\Ptld}{\ensuremath{ \widetilde{P} }}
\newcommand{\Ptldt}{\ensuremath{ \Ptld_{_{t}} }}

\newcommand{\Psik}{\ensuremath{ \,\Psi_{_k} }}

\newcommand{\Qk}{\ensuremath{ \,\mathcal{Q}{_{_k}} }}
\newcommand{\Qoo}{\ensuremath{{ \,\mathcal{Q}{_{_{11}}} }}}
\newcommand{\Qot}{\ensuremath{{ \,\mathcal{Q}{_{_{12}}} }}}
\newcommand{\Qto}{\ensuremath{{ \,\mathcal{Q}{_{_{21}}} }}}
\newcommand{\Qtt}{\ensuremath{{ \,\mathcal{Q}{_{_{22}}} }}}
\newcommand{\Qwk}{\ensuremath{  \,Q_{_{w_k}}   }}
\newcommand{\Qeps}{\ensuremath{ Q^{\eps} }}
\newcommand{\Qkeps}{\ensuremath{ \Qeps_{_k} }}
\newcommand{\Qepsk}{\ensuremath{ \Qeps_{_k} }}
\newcommand{\Qkleps}{\ensuremath{ \Qeps_{_{k+l}} }}
\newcommand{\Qq}{\ensuremath{ Q^{q} }}
\newcommand{\Qqk}{\ensuremath{ \Qq_{_{k}} }}
\newcommand{\Qthrfor}{\ensuremath{ Q_{_{\!34}} }}

\newcommand{\Qt}{\ensuremath{ Q{_{_t}} }}

\newcommand{\Rk}{\ensuremath{{ {R}{_{_k}} }}}
\newcommand{\Rko}{\ensuremath{{ {R}{_{_{k+1}}} }}}

\newcommand{\Rz}{\ensuremath{{ \,\mathcal{R}{_{_0}} }}}
\newcommand{\Rb}{\ensuremath{ \,\mathcal{R}^{b}  }}
\newcommand{\Rbi}{\ensuremath{ {\Rb^i} }}
\newcommand{\Rkob}{\ensuremath{ \Rb_{_{k+1}} }}
\newcommand{\Rkobi}{\ensuremath{ {{\Rkob}^i} }}
\newcommand{\Rktwb}{\ensuremath{ \Rb_{_{k+2}} }}
\newcommand{\Rklb}{\ensuremath{ \Rb_{_{k+l}} }}
\newcommand{\Rq}{\ensuremath{ R^{q} }}
\newcommand{\Rqko}{\ensuremath{ \Rq_{_{k+1}} }}
\newcommand{\Rkoq}{\ensuremath{ \Rqko }}
\newcommand{\Roo}{\ensuremath{{ \,\mathcal{R}{_{_{11}}} }}}
\newcommand{\Rot}{\ensuremath{{ \,\mathcal{R}{_{_{12}}} }}}
\newcommand{\Rto}{\ensuremath{{ \,\mathcal{R}{_{_{21}}} }}}
\newcommand{\Rtt}{\ensuremath{{ \,\mathcal{R}{_{_{22}}} }}}
\newcommand{\rkocross}
{\ensuremath{ \,\left[ \rko \times \right] }}

\newcommand{\Rt}{\ensuremath{ R_{_{t}} }}

\newcommand{\Seps}{\ensuremath{{ \,S_{_\epsilon} }}}
\newcommand{\Sk}{\ensuremath{ \,S_{_{k}} }}
\newcommand{\Sko}{\ensuremath{ \,S_{_{k+1}} }}
\newcommand{\dSko}{\ensuremath{ \,\delta S_{_{k+1}} }}
\newcommand{\Skot}{\ensuremath{ \,S_{_{k+1}}^{true} }}
\newcommand{\Sbb}{\ensuremath{{ \,\mathcal{S}{_{_b}} }}}
\newcommand{\scross}{\ensuremath{ \,\left[ \sv \times \right] }}
\newcommand{\skcross}{\ensuremath{ \,\left[ \sk \times \right] }}
\newcommand{\skocross}{\ensuremath{ \,\left[ \sko \times \right] }}
\newcommand{\Sbbi}{\ensuremath{ {\Sbb^i} }}
\newcommand{\Sbbn}{\ensuremath{ {\Sbb^n} }}
\newcommand{\Sz}{\ensuremath{ \,S^{o} }}

\newcommand{\Tetako}{\ensuremath{ \,\Theta_{_{k+1}}    }}
\newcommand{\Tetakok}{\ensuremath{ \,\Theta_{_{k+1/k}} }}
\newcommand{\Tetae}{\ensuremath{ \widehat{\Theta} }}
\newcommand{\Tetaeko}{\ensuremath{ \Tetae_{_{k+1}} }}

\newcommand{\Ukok}{\ensuremath{{ \,U_{_{k+1/k}} }}}
\newcommand{\Ucovkok}{\ensuremath{ \,\mathcal{U}_{_{k+1/k}}  }}
\newcommand{\ucross}{\ensuremath{ \,\left[ \uv \times \right] }}
\newcommand{\uivcross}{\ensuremath{ \,\left[ \uiv \times \right] }}

\newcommand{\Vk}{\ensuremath{{ \,V_{_k} }}}
\newcommand{\Vi}{\ensuremath{{ \,V_{_i} }}}
\newcommand{\Vko}{\ensuremath{{ \,V_{_{\!\!k+1}} }}}
\newcommand{\Voj}{\ensuremath{{ \,V_{_{1j}} }}}
\newcommand{\Voi}{\ensuremath{{ \,V_{_{1i}} }}}
\newcommand{\vcross}{\ensuremath{ \,\left[ \vv \times \right] }}
\newcommand{\Vkoi}{\ensuremath{ {\,V_{_{k+1}}^i} }}
\newcommand{\Vkon}{\ensuremath{ \,V_{_{k+1}}^n }}
\newcommand{\Vki}{\ensuremath{{ \,V_{_{k+i}} }}}

\newcommand{\Vz}{\ensuremath{{ \,V^0 }}}
\newcommand{\Vzko}{\ensuremath{{ \,\Vz_{_{\hspace{-1.5ex}k+1}} }}}

\newcommand{\Vbar}{\ensuremath{ \overline{V} }}
\newcommand{\Vbark}{\ensuremath{ \Vbar_{_{k}} }}

\newcommand{\Wk}{\ensuremath{{ \,W_{_k} }}}
\newcommand{\Wz}{\ensuremath{{ \,W_{_0} }}}
\newcommand{\Wzz}{\ensuremath{ \,W^{0}_{_0} }}
\newcommand{\Wzo}{\ensuremath{ \,W^{1}_{_0} }}
\newcommand{\Wo}{\ensuremath{{ \,W_{_1} }}}
\newcommand{\Wko}{\ensuremath{{ \,W_{_{k+1}} }}}
\newcommand{\Wki}{\ensuremath{{ \,W_{_{k+i}} }}}
\newcommand{\Woj}{\ensuremath{{ \,W_{_{1j}} }}}
\newcommand{\Woi}{\ensuremath{{ \,W_{_{1i}} }}}
\newcommand{\Wnk}{\ensuremath{{ \,Wn_{_k} }}}
\newcommand{\Wi}{\ensuremath{{ \,W_{_i} }}}
\newcommand{\WN}{\ensuremath{{ \,W_{_N} }}}
\newcommand{\WNo}{\ensuremath{ \,W_{_{N+1}} }}
\newcommand{\Wkt}{\ensuremath{ W_{_k}^o }}

\newcommand{\Xk}{\ensuremath{{ \,X_{_k} }}}
\newcommand{\Xko}{\ensuremath{{ \,X_{_{k+1}} }}}
\newcommand{\Xz}{\ensuremath{{ \,X_{_0} }}}
\newcommand{\Xinf}{\ensuremath{{ \,X_{_\infty} }}}

\newcommand{\Xkk}{\ensuremath{{ \,{\widehat{X}}_{_{k/k}} }}}
\newcommand{\Xkok}{\ensuremath{{ \,{\widehat{X}}_{_{k+1/k}} }}}
\newcommand{\Xkoko}{\ensuremath{{ \,{\widehat{X}}_{_{k+1/k}} }}}
\newcommand{\Xzz}{\ensuremath{{ \,{\widehat{X}}_{_{0/0}} }}}
\newcommand{\Xer}{\ensuremath{{ \,{\widetilde{X}} }}}

\newcommand{\Xt}{\ensuremath{{ \,X_{_t} }}}

\newcommand{\Xik}{\ensuremath{ \Xi_{_k} }}
\newcommand{\Xiko}{\ensuremath{ \Xi_{_{k+1}} }}
\newcommand{\Xiz}{\ensuremath{{ \Xi(\qz) }}}
\newcommand{\Xio}{\ensuremath{{ \Xi(\qo) }}}
\newcommand{\Xiezz}{\ensuremath{{ \Xi(\qezz) }}}
\newcommand{\Xieoo}{\ensuremath{{ \Xi(\qeoo) }}}
\newcommand{\Xie}{\ensuremath{ \widehat{\Xi} }}
\newcommand{\Xiek}{\ensuremath{ \Xie_{_k} }}
\newcommand{\Xiet}{\ensuremath{ \Xie_{_t} }}
\newcommand{\Xieko}{\ensuremath{ \Xie_{_{k+1}} }}
\newcommand{\Xiekok}{\ensuremath{ \Xie_{_{k+1/k}} }}
\newcommand{\Xiektwk}{\ensuremath{ \Xie_{_{k+2/k}} }}
\newcommand{\Xiekoko}{\ensuremath{ \Xie_{_{k+1/k+1}} }}
\newcommand{\Xiekolk}{\ensuremath{ \Xie_{_{k+l-1/k}} }}
\newcommand{\Xieklk}{\ensuremath{ \Xie_{_{k+l/k}} }}
\newcommand{\Xiekk}{\ensuremath{ \Xie_{_{k/k}} }}

\newcommand{\Xitld}{\ensuremath{ \widetilde{\Xi} }}
\newcommand{\Xitldk}{\ensuremath{ \Xitld_{_k} }}
\newcommand{\Xitldt}{\ensuremath{ \Xitld_{_t} }}

\newcommand{\Yt}{\ensuremath{{ Y_{_t} }}}

\newcommand{\Ytld}{\ensuremath{ \widetilde{Y} }}
\newcommand{\Ytldt}{\ensuremath{ \Ytld_{_{t}} }}

\newcommand{\Yko}{\ensuremath{{ \,Y_{_{k+1}} }}}

\newcommand{\zvcross}{\ensuremath{ \,\left[ \zv \times \right] }}
\newcommand{\zbcross}{\ensuremath{ \,\left[ \zb \times \right] }}
\newcommand{\zbicross}{\ensuremath{ \,\left[ \zbi \times \right] }}
\newcommand{\zbncross}{\ensuremath{ \,\left[ \zbn \times \right] }}
\newcommand{\zkocross}{\ensuremath{ \,\left[ \zko \times \right] }}
\newcommand{\zepscross}{\ensuremath{ \,\left[ \zeps \times \right] }}

\newcommand{\Zt}{\ensuremath{{ Z_{_t} }}}


\newcommand{\AD}{\ensuremath{{ \bf{AD}\,  }}}
\newcommand{\EMF}{\ensuremath{{ \bf{EMF}\,  }}}
\newcommand{\EKF}{\ensuremath{{ \bf{EKF}\,  }}}
\newcommand{\RLS}{\ensuremath{{ \bf{RLS}\,  }}}
\newcommand{\KVKF}{\ensuremath{{ \bf{KVKF}\,  }}}
\newcommand{\quatKF}{\ensuremath{{ \bf{quatKF}\,  }}}
\newcommand{\REQUEST}{ \mbox{REQUEST} }

\newcommand{\xnorm}{\ensuremath{ \, \| \xv \|  }}
\newcommand{\xonorm}{\ensuremath{ \, \| \xo \|  }}
\newcommand{\xtwnorm}{\ensuremath{ \, \| \xtw \|  }}
\newcommand{\xounit}{\ensuremath{ \,\frac{\xo}{\| \xo \|}  }}
\newcommand{\axsum}{\ensuremath{  \,\left( \ao\,\xo\;+\;\atw\,\xtw \right)  }}
\newcommand{\axsumnorm}{\ensuremath{  \|\ao\,\xo\;+\;\atw\,\xtw \| }}
\newcommand{\aoatwsum}{\ensuremath{ \,\left( \ao + \atw \right)  }}

\newcommand{\ipo}{\ensuremath{ \,\mbox{ip}_{_1} }}
\newcommand{\iptw}{\ensuremath{ \,\mbox{ip}_{_2} }}
\newcommand{\ipmoo}{\ensuremath{ \,\mbox{ipm}_{_{11}} }}
\newcommand{\ipmtt}{\ensuremath{ \,\mbox{ipm}_{_{22}} }}

\newcommand{\pcmoo}{\ensuremath{ \,\mbox{pcm}_{_{11}} }}
\newcommand{\pcmtt}{\ensuremath{ \,\mbox{pcm}_{_{22}} }}
\newcommand{\gainoo}{\ensuremath{ \,\mbox{gain}_{_{11}} }}
\newcommand{\gaintt}{\ensuremath{ \,\mbox{gain}_{_{22}} }}

\newcommand{\Bfr}{\textit{B}}
\newcommand{\Gfr}{\textit{G}}
\newcommand{\Hfr}{\textit{H}}
\newcommand{\Ifr}{\textit{I}}
\newcommand{\Nfr}{\textit{N}}
\newcommand{\Rfr}{\textit{R}}
\newcommand{\Sfr}{\textit{S}}
\newcommand{\Szfr}{\textit{S}_0}


\section{Introduction}

 \par Rendezvous and docking (RvD) is a key operational technology involving more than one spacecraft required for different missions such as exchange of crew in orbital stations, repair of spacecraft in orbit, and space debris removal~\cite{Fehse}.  Spacecraft RvD enables novel space capabilities like on-orbit servicing of satellites. Many research projects were conducted on on-orbit servicing technologies~\cite{Coleshill2009869, Friend, NASA, Stoll, Yoshida}. Yet a few organizations worldwide accomplished such missions~\cite{NASA}. The German Orbital Servicing Mission (DEOS) is an example of an ongoing mission by the German Aerospace Center (DLR) to capture a non-cooperative tumbling satellite for manipulation and/or deorbiting purposes~\cite{DEOS}. Another project is the Orbit Life Extension Vehicle (OLEV) designed for service and life extension of geostationary communication satellites suffering from propellant depletion~\cite{OLEV}. Critical steps in a satellite on-orbit service mission are the rendezvous, the docking, and the capture of the target satellite. Autonomously performing these tasks increases the technical challenge and thus the risk. Technologies of servicing spacecraft must be thoroughly tested before launch in simulated micro-gravity environments.

\subsection*{RvD Simulation Technologies}
 \par Several technologies are available for testing and validation in a simulated micro-gravity environment. Air-bearing tables~\cite{Nakanishi, Menon} are limited to planar motions. Spherical air-bearing simulators are limited to small angular displacements and experience approximate equilibrium set-ups. Free-fall methods~\cite{Menon, Sawa} enable micro-gravity in a three dimensional environment, but for 20-30s only and in a typically very limited cargo space. Neutral Buoyancy methods~\cite{Menon} have been extensively used for astronauts training, but are not suitable for hardware testing - in particular because of the water-induced drag that alters the dynamics characteristics of the tested system. Suspended systems methods~\cite{Fehse, Sato} are effectively used to simulate micro-gravity in three dimensions, but exhibit difficulty in compensating for the kinetic friction within the tension control system. On the other hand, robotics-based hardware-in-the-loop simulators implement effective active gravity compensation, can accommodate complex systems for the RvD simulation, and enable full translation and rotational motions.
 \par There are several examples of hardware-in-the-loop (HIL) simulators for space systems RvD simulation. The DLR developed the European Proximity Operation Simulator (EPOS) a decade ago~\cite{Rupp}. The EPOS facility hosted test campaigns for rendezvous sensors of the autonomous spacecraft ATV and HTV. The NASA/MSFC developed a HIL docking simulator using a 3D Stewart platform for simulating the Space Shuttle berthing to the International Space Station (ISS)~\cite{Friend,Alder}. The Canadian Space Agency built an SPDM (Special Purpose Dexterous Manipulator) Task Verification Facility (STVF) using a giant 3D hydraulic robot to simulate the manipulator performance of ISS maintenance tasks~\cite{CSA, Ouma}. The US Naval Research Laboratory used two 3D robotic arms to simulate satellite rendezvous for HIL testing rendezvous sensors~\cite{bell}.

  \par The DLR has been upgrading the EPOS facility as shown in Fig. 1. The unique features of this new facility~\cite{Boge}, in comparison with the previously described simulators, are the two heavy-duty industrial robots. These robots can handle payloads up to 250 kg. In addition, the facility allows relative motion between the robots with a range up to 25m. The new EPOS facility is aimed at providing test and verification capabilities for complete RvD procedures of on-orbit servicing missions.

\begin{figure}[h]
	\centering
		\includegraphics[width=0.90\textwidth]{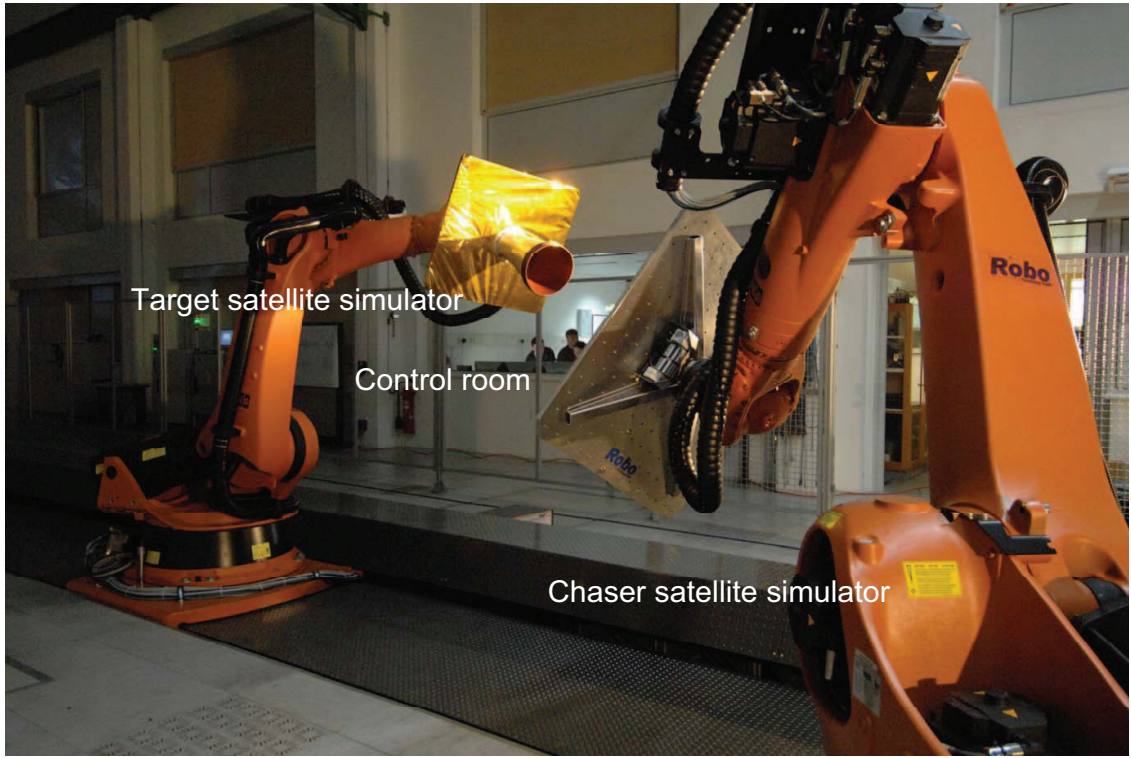}
	\caption{The EPOS facility: two robots (in orange) holding a satellite mock-up and docking interfaces, the target simulator robot mounted on a linear rail}
	\label{fig:eposmainparts}
\end{figure}
\subsection*{Challenges of Robotics-based Docking Simulator }
Using industrial robots as the key robotic components of such an important HIL simulation facility is a highly challenging approach because these robots are designed as accurate positioning machines. As such, they are typically very stiff and do not naturally comply with the particular contact dynamics that satellite boundaries experience during contact. In addition, initially designed for typical industrial applications, like automotive assembly, their speed of response may be too slow. For example, due to communication channels delays, the EPOS robots control system shows an average delay of 16ms between the positioning command signal and the actual position signal, and an average delay of 16ms between the actual position signal and the measured position signal. Thus, a closed-loop controller using positioning command and measurement of the robots experience an average delay of 32ms. This value is relatively high compared to the smaller characteristic times of contact dynamics for high stiffness contact case such as the stiffness of the given robots end-effectors. Industrial robots have been used for rendezvous simulation purpose but rarely for docking simulation. Existing technologies provide incomplete and unreliable solutions to the challenges of good compliance and quick response. In order to simulate docking (contact dynamics) in a HIL loop, the robots must have a good control of its compliance and should quickly respond the docking action. An ideal control approach would be to apply an impedance control strategy such as the one described in~\cite{hogan}. However, impedance control typically requires torque control capabilities at joint level. This is not possible for the industrial robots which controllers are designed for end-effector position control. Furthermore, their low-level control software is inaccessible. Similarly, many other advanced and proven  robot control strategies, such as the computed torque control~\cite{Middleton} cannot be implemented on industrial robots. The only option is to use admittance control as an outer loop on top of the built-in inner-loop position control system of the industrial robot using the measured contact force as its feedback input~\cite{Ma_melak}. Yet, because of the before-mentioned  high stiffness of the robots, the contact duration is shorter than the time delay of the robot controllers, leading to inconsistencies in the HIL docking simulation results. Instability might occur yielding damages to the robots. Some researchers have already proposed the use of passive compliance~\cite{wang98}, for impact or contact control to soften the stiffness of the contacting objects. Soft force sensors themselves were used as compliance devices in~\cite{roberts1984compliance, Xu}. In all these cases, the contact frequency was decreased, allowing adding an additional active control component, with the aim of improving the transient behavior of the system during transition from the non-contact to the contact phases. Adding a passive compliance device solves the problem of high stiffness. This requires however physically changing the device for different test scenarios. In addition, significant work is added in order to identify the contact parameters and to perform gravity compensation each time a new compliance device is installed.
\subsection*{Solution Strategies}
This paper presents an approach to mitigate the combined effect of the robots high stiffness and controller time delay. Furthermore, it analyzes under some assumptions the stability of the delayed HIL closed-loop system. The envisioned outcome is to provide to the robotic facility operator an operations ``envelope'' where the facility can be used safely. It is proposed to add passive compliance between the robot end-effector and the docking interface, combined with a virtual contact model. Following this novel approach, referred to as the ``hybrid contact dynamics emulation'' method, the virtual contact model parameters can be tuned to arrive at desired contact force characteristics. The effect of the passive compliance is to lengthen the duration of the impact, thus avoiding the undesired consequences of the aforementioned time delay, but at the loss of some accuracy in position. A new passive compliance device integrated with the probe is designed. This device is designed to have very small gravitational force readings on the force/torque (F/T) sensor.

\par This work continues earlier efforts~\cite{Zebenay,Melak,Melak2013,Melak2013may} and presents models, a stability analysis, and test results for an extension of the proof-of-concept of the hybrid docking simulator to more than 1D. The main contributions are: 1) the 3D nonlinear and 2D linearized design models, 2) the stability analysis for the 2D case, 3) the extension of the hybrid docking simulator concept for 2D and 3D, and 4) the experimental results for 1D and 3D docking cases. The nonlinear models are developed assuming a stationary target satellite and a contact with sliding without friction. The first assumption can be easily relaxed, and does not impair the generality of the findings, while the second assumption is very common.

\par Section 2 presents the  hybrid docking simulation concept of operations. Section 3 presents the mathematical modeling of the concept including the nonlinear and linearized state-space models. Section 4 is concerned with the stability analysis.  The analysis of the compliance device is pesented in section 5. Section 6 presents the test results in 1D and 3D. Finally, section 7 presents a summary, conclusions, and future work.

 \section {Hybrid Docking Simulator Concept of Operations}
   \begin{figure}[h]
	\centering
		\includegraphics[width=0.700\textwidth]{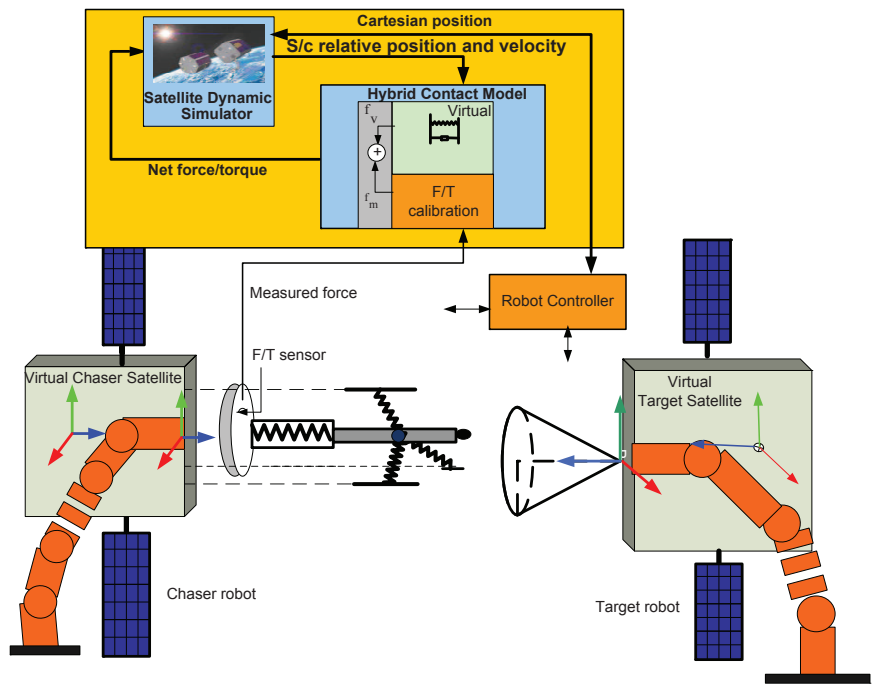}
	\caption{Concept of operations of the hybrid docking simulator}
	\label{fig:HILcon}
\end{figure}

Figure~\ref{fig:HILcon} shows the schematics of the hybrid docking simulation concept. The concept consists of three elementary subsystems. The first subsystem is a real-time computer simulator used to compute the dynamic response of the chaser and the target satellites, based on free-floating multi-body dynamics. This software element can implement any space environment effect at will. Yet, the combination of short contact time and high force/torque amplitudes typically renders the contact force/torque very dominant compared to other effects. The second subsystem consists of two real-time controlled industrial robots which are commanded in position and track, after a significant settling time, the 3D trajectories generated by the computer simulator. The third subsystem is a hardware mock-up of the docking mechanism mounted on the robots end-effector, which will make physical contact during docking and capturing. A force/torque (F/T) sensor is mounted together with the docking interface in order to measure the real contact force/torque. The sensors readings are provided via a feedback loop as inputs to the computer simulator. They are the physical F/T feedback. The computer simulator generates a virtual F/T, according to a conventional contact dynamics model. The combination of the software F/T signal and the hardware F/T signal as inputs to the computer simulator explains in essence the concept of hybrid docking simulation.  This method is called the \emph{hybrid contact dynamics emulation method} as reported in~\cite{Zebenay,Melak2013}.
It consists of combining a real(hardware) passive compliance between the robot end-effector and the docking interface with a virtual (software) contact dynamics model in the satellite numerical simulator. The advantage of this method is that the real passive compliance can remain unchanged while the virtual contact model can be tuned to arrive at the desired stiffness characteristics. The effect of the passive compliance is to lengthen the duration of the impact, thus avoiding the undesired consequences of the aforementioned time delay, but at the  cost of a loss of some accuracy in position. 
\section{Mathematical Modeling}
\subsection{Physical Model}
The end-effector of the chaser robot, servicer satellite simulator, is equipped with a tool frame onto which a force-torque (F/T) sensor is rigidly attached. A compliance device is attached to the tool frame, in series with the F/T sensor, and consists of a rigid shaft and a set of several springs that soften the contact dynamics between the two robots. A detailed treatment of this device is deferred to a subsequent section. The contact force is applied at the probe tip to both the chaser and target robots. The end-effector of the target robot holds a tool flange or fixtures where a conic shape device (the nozzle) is rigidly attached. Contact happens when the probe tip hits the interior of the nozzle.
\begin{figure}[htb]
      \centering
      \includegraphics[width=0.850\textwidth]{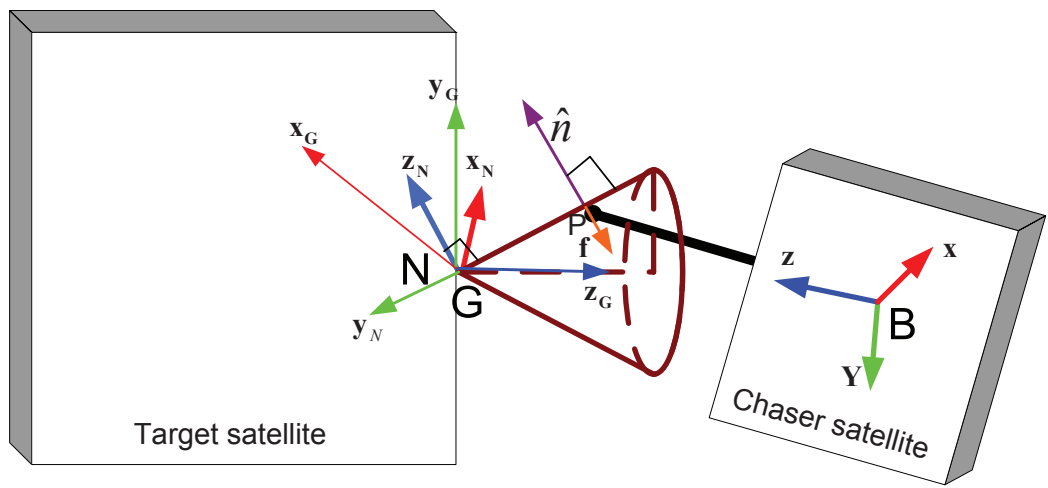}
      \caption{Block-diagram of two satellites in contact.}
       \label{fig:6dideal}
\end{figure}
 As a result of the applied force, the robots' controllers move the probe and the nozzle in three dimensions, both in translation and rotation. The robotic system is assumed to perfectly follow the desired steady-state after a time-invariant delay. The contact is assumed to be pointwise (point $P$ in Fig.~\ref{fig:6dideal}), without constraint in rotation and without friction parallel to the local tangential plane at $P$. Similar assumption has been made by other researchers in order to simplify the modeling task~\cite{UyamaIROS,Xu1}. The force direction due to contact is thus assumed to be normal to that plane~\cite{Gilardi}, and the local motion of the probe tip is assumed to take place along the local normal. The magnitude of the contact force is assumed to be a linear function of the penetration depth and penetration rate of the contacting point to the contacting surface~\cite{Gilardi}. The contact duration is very small, and the mass and inertia of the target are assumed to be significantly higher than those of the chaser (typical of a scenario of a LEO spacecraft chasing a GEO satellite): as a consequence the target is considered inertially fixed during contact. The contact force is assumed to be dominant and is the only force considered. For simplicity, the contact point, the probe axis, and the chaser center of mass are assumed aligned.

\subsection{Nonlinear Mathematical Model}

\paragraph{Three-Dimensional Generalized State-Space Model}
The relevant Cartesian frames are described as follows. The Global frame $\Gfr$ is an inertial frame, attached to the laboratory room, with center $G$, and its three axes as shown in Fig.~\ref{fig:6dideal}. The Nozzle frame $\Nfr$ is inertial, with center $G$, the $z$-axis coincides with the normal vector at the contact point, and the other two axes lie in an arbitrary orientation in the local tangent plane. The chaser Body frame $\Bfr$ is a rigid body frame centered at the chaser center of mass, $B$. For simplicity, but without loss of generality, the center $G$ is chosen at the origin of the nozzle (point $N$ in Fig.~\ref{fig:6dideal}). In the sequel the following notation is adopted: $\uv_{_\Gfr}$ denotes the $3 \times 1$ projection of a physical vector $\uv$ onto frame $\Gfr$, $\textbf{D}^{\Gfr}_{\Bfr}$ denotes the rotation matrix from $\Gfr$ to $\Bfr$, $\omgv_{\Bfr\Gfr}$ denotes the angular velocity vector of $\Bfr$ with respect to $\Gfr$, and $\dot{\textbf{r}}_{\Gfr}$ denotes the time-differentiation of the vector $\textbf{r}$ with respect to frame $\Gfr$. Applying the physical assumptions described in the previous subsection, it is straightforward to derive the generalized mathematical model, which is summarized as follows:
\begin{align}
\label{s2eq00}
& m \ddot{\textbf{r}}_{\Gfr} = \fv^{\Gfr} \\
\label{s2eq05}
& \textbf{J}\dot{\omgv}_{\Bfr\Gfr} =  \tov^{\Gfr} \
\end{align}
where $\rv$ denotes the vector $GB$, $\fv$ denotes the force applied to the chaser at $P$, $\tov$ denotes the torque applied to the chaser about its center of mass $B$ due to $\fv$, $m$ and $\textbf{J}$ denote the mass and inertia tensor of the chaser, and
\begin{align}
\label{s2eq01a}
& \tov =  \av \times \fv \\
\label{s2eq01}
& \fv = f \nev \\
\label{s2eq02}
& f = - k d- b \dot{d}\\
\label{s2eq04}
& d = \rov^T \nev \\
\label{s2eq03}
& \rov(t) = [\rv + \av](t-h)\
\end{align}
where $\av$ denotes the body-fixed vector $BP$, $\nev$ denotes the unit-norm normal vector (outward) at the local tangent plane, $f$ is the algebraic intensity of the force (positive outward) adopting a one-dimensional spring-dashpot model~\cite{Gilardi}, the time-invariant stiffness and damping coefficients,  $k$ and $b$, are given positive scalars, $d$ and $\dot{d}$ denote the penetration depth and depth rate, respectively, and $\rov$ denotes the inertial position vector of the contact point, $GP$. Notice that the current value of $\rov(t)$ at time $t$ is a delayed function  due to the robotics system delay, $h$.

\paragraph{State-Space Representation in Frames $\Gfr$ and $\Bfr$}
As is classically done for the sake of simplicity, Eq.~\eqref{s2eq00} is projected in an inertial frame while Eq.~\eqref{s2eq05} is projected in the body-fixed frame, where $J$ is time-invariant. Projection of the contact force in $\Gfr$ yields
\begin{align}
 \label{s2eq06}
  \fv  = f \nev \
\end{align}
where $\nev$ is known and time-invariant in $\Gfr$. From Eq.~\eqref{s2eq04}, the penetration depth $d$ is expressed using vector projections in $\Gfr$ and $\Bfr$ as follows:
\begin{align}
\nonumber
d(t)  & = \rov^T \nev \\
\nonumber
& = (\rv_{_\Gfr} + \av_{_\Gfr})^T|_{_{(t-h)}} \nev \\
\nonumber
& = (\rv_{_\Gfr}|_{_{(t-h)}} + \textbf{D}^{\Bfr}_{\Gfr}(t-h) \av_{_\Bfr})^T \,\nev  \\
\label{s2eq07}
& = \rv_{_\Gfr}^T|_{_{(t-h)}} \nev  +  \av_{_\Bfr}^T  \textbf{D}^{\Gfr}_{\Bfr}(t-h)\,  \nev \
\end{align}
where $\av_{_\Bfr}$ is known and time-invariant. The time derivative for $d$ is therefore derived as follows:
\begin{align}
\nonumber
\dot{d}(t) & = \dot{\rv}_{_\Gfr}^T|_{_{(t-h)}}\, \nev + \av_{_\Bfr}^T \dot{\textbf{D}}^{\Gfr}_{\Bfr}|_{_{(t-h)}}\, \nev \\
 \label{s2eq08}
 & = \dot{\rv}_{_\Gfr}^T|_{_{(t-h)}}\,  \nev + \av_{_\Bfr}^T \left\{[-\omgv_{_\Bfr} \times] \textbf{D}^{\Gfr}_{\Bfr}\right\}_{_{(t-h)}} \nev .\
\end{align}
The second line in Eq.~\eqref{s2eq08} stems from the rigid body kinematics equation in terms of the rotation matrix, i.e.:
\begin{align}
\label{s2eq09}
& \dot{\textbf{D}}^{\Gfr}_{\Bfr} = [-\omgv_{_\Bfr} \times] \textbf{D}^{\Gfr}_{\Bfr}. \
\end{align}
Hence, the variables $d$ and $\dot{d}$ are expressed as functions of the variables $\rv_{_\Gfr}$, $\dot{\rv}_{_\Gfr}$, and $\textbf{D}^{\Gfr}_{\Bfr}$. The dynamics equation~\eqref{s2eq05} projected in $\Bfr$ yields:
\begin{align}
 \nonumber
 \textbf{J} \dot{\omgv}_{_\Bfr}
 & = (\textbf{J} \omgv_{_\Bfr}) \times \omgv_{_\Bfr} + \av_{_\Bfr} \times \fv_{_\Bfr} \\
 \label{s2eq10}
 & = (\textbf{J} \omgv_{_\Bfr}) \times \omgv_{_\Bfr} + \av_{_\Bfr} \times \textbf{D}^{\Gfr}_{\Bfr} (f \nev). \
\end{align}
To conclude, the following state-space equations describe the motion of the chaser in rotation and translation due to the contact force, in terms of the state variables $\{\rv_{_\Gfr},\vv_{_\Gfr}, \textbf{D}, \omgv\}$:
\begin{align}
\label{s2eq11}
& \dot{\rv}_{_\Gfr} = \vv_{_\Gfr} \\
\label{s2eq12}
& \dot{\vv}_{_\Gfr} = \frac{f}{m} \nev \\
\label{s2eq13}
& \dot{\textbf{D}} = [-\omgv \times] \textbf{D} \\
\label{s2eq14}
& \dot{\omgv} = J^{-1}
\left\{  [- \omgv\times] J \omgv + \tov_{_\Bfr}  \right\}\
\end{align}
where
\begin{align}
\label{s2eq15ab}
& \tov_{_\Bfr}(t) = f(t) [\av\times] \textbf{D}|_{_{(t-h)}} \nev \\
\label{s2eq15a}
& f(t) = - k d(t) - b \dot{d}(t) \\
\label{s2eq15}
 & d(t) = \rv^T_{_\Gfr}|_{_{(t-h)}} \nev  +  \av^T  \textbf{D}|_{_{(t-h)}}\,  \nev \\
 \label{s2eq16}
 & \dot{d}(t) = \vv_{_\Gfr}^T|_{_{(t-h)}}\,  \nev + \av^T \left\{[-\omgv \times] \textbf{D}\right\}_{_{(t-h)}} \nev. \
\end{align}
In Eqs.~\eqref{s2eq11}-\eqref{s2eq16},subscripts and superscripts were dropped for notational simplicity.

\paragraph{State-Space Representation in Frames $\Nfr$ and $\Bfr$}
This subsection is concerned with the derivation of a simpler expression for the state-space model equations. It appears from the physical modeling assumptions that the vector $\nev$ defines a privileged direction along which the contact force is developing and the probe tip is moving. This is emphasized by realizing that the force, as defined in Eqs.~\eqref{s2eq15a}-\eqref{s2eq16}, is the orthogonal projection along $\nev$ of the physical vector $(- k\rov - b\dot{\rov}_{\Gfr})$, as shown next:
\begin{align}
 \nonumber
 \fv & = f \nev \\
 \nonumber
   & = \nev\nev^T \left[ - k (\rv_{_\Gfr} + \textbf{D}^T \av_{_\Bfr}) - b (\vv_{_\Gfr} + \textbf{D}^T [\omgv\times] \av_{_\Bfr}) \right] \\
 \label{s2eq17}
   & = \nev\nev^T (-k \rov_{_\Gfr} - b \dot{\rov}_{_\Gfr}). \
\end{align}
Furthermore, since $\nev$ coincides with the z-axis in the frame $\Nfr$ the quantity $\textbf{D}\nev$ represents the third column of $\textbf{D}^{\Nfr}_{\Bfr}$, $\dv_{_{\!c3}}$. Since $\textbf{D}\nev$ rather than the whole matrix $\textbf{D}$ is needed in the state-space equations, a reduced order model is obtained by projecting the generalized model on the inertial frame $\Nfr$ rather than on $\Gfr$. Notice that, since $\Nfr$ is an inertial frame, the angular velocity vectors $\omgv^{\Bfr\Nfr}$ and $\omgv^{\Bfr\Gfr}$ are identical.  Let $\{\rv, \vv, \dv_{_{\!c3}}, \omgv\}$ denote the state variables, i.e., the inertial position of $\Bfr$ along $\Nfr$, the inertial velocity of $B$ along $\Nfr$, the third column of the rotation matrix $D^{\Nfr}_{\Bfr}$, and the angular velocity vector of $\Bfr$ with respect to $\Nfr$ along $\Bfr$. The associated state-space equations are expressed as follows:
\begin{align}
\label{s2eq18}
& \dot{\rv} = \vv \\
\label{s2eq19}
& \dot{\vv} = \frac{f}{m} \nev \\
\label{s2eq20}
& \dot{\dv}_{_{\!c3}} = [-\omgv \times] \dv_{_{\!c3}} \\
\label{s2eq21}
& \dot{\omgv} = J^{-1}
\left\{  [- \omgv\times] J \omgv + \tov_{_\Bfr}   \right\}\
\end{align}
where
\begin{align}
\label{s2eq22a}
& \tov_{_\Bfr}(t) = f(t) [\av\times] \dv_{_{c3}}|_{_{(t-h)}} \\
\label{s2eq22}
& f(t) = - k d(t) - b \dot{d}(t) \\
\label{s2eq23}
 & d(t) = \rv^T|_{_{(t-h)}} \nev  +  \av^T  \dv_{_{\!c3}}|_{_{(t-h)}} \\
 \label{s2eq23a}
 & \dot{d}(t) = \vv^T|_{_{(t-h)}}\,  \nev + \av^T \left\{[-\omgv \times] \dv_{_{\!c3}}\right\}_{_{(t-h)}}. \
\end{align}

The order of the model is thus reduced from 18 to 12 states.

\begin{figure}[htb]
      \centering
      \includegraphics[width=0.850\textwidth]{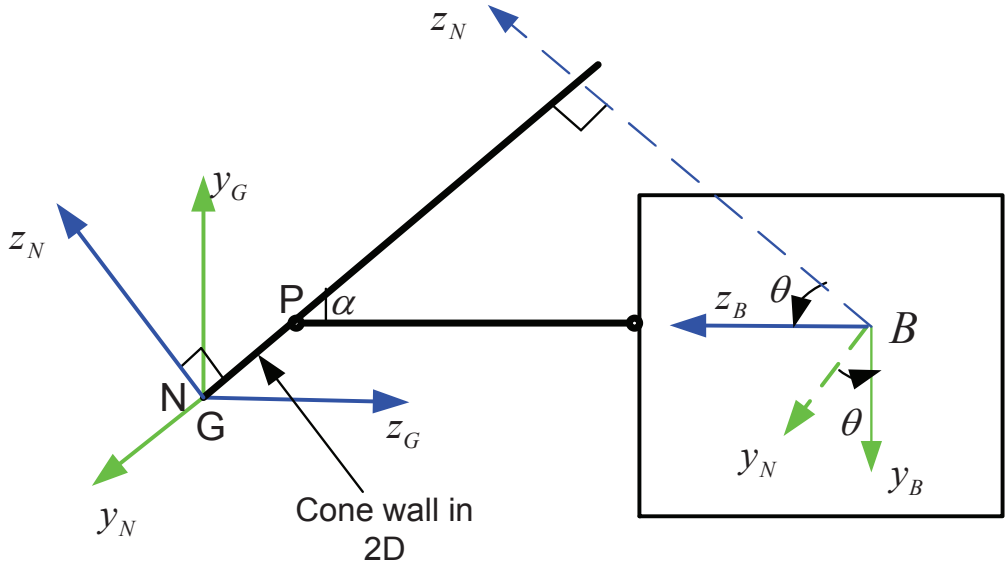}
      \caption{Free-body diagram of two bodies in contact in 2D.}
       \label{fig:3dideal}
\end{figure}

\paragraph{Two-Dimensional State-Space Model}
This subsection is concerned with the development of a particular case of the previous state-space model which is used for stability analysis of the docking simulator concept in 2D. The motion of the chaser center of mass $B$ is restricted to the (yz)-plane (see Fig.~\ref{fig:3dideal}), the rotation around $B$ is restricted to the x-axis, i.e.
\begin{align}
\label{s2eq24}
& \omgv = \{\omega, 0, 0\}\
\end{align}
where
\begin{align}
\label{s2eq25a}
& \omega = \dot{\theta} \
\end{align}
and $\theta$ denote the angle of the rotation around the x-axis bringing frame $\Nfr$ onto frame $\Bfr$, i.e.
\begin{align}
 \label{s2eq25}
 & \dv_{_{c3}} = \{0, \sin \theta, \cos \theta\}.\
\end{align}
By definition, the unit vector $\nev$ in $\Nfr$ and the body-fixed vector $\av$ in $\Bfr$ are simply expressed as follows:
\begin{align}
 \label{s2eq26}
 & \nev = \{0,0,1\} \\
 \label{s2eq27}
 & \av= \{ 0,0,a\}. \
\end{align}
Let $\rv=\{x,y, z\}$  and $\vv= \{v_x,v_y, v_z\}$ denote the inertial position and the inertial velocity of $\Bfr$ along $\Nfr$ respectively. Applying the above assumptions and using Eqs.~\eqref{s2eq24}-\eqref{s2eq27} in Eqs.~\eqref{s2eq18}-\eqref{s2eq23a} yields a state-space model for the variables $\{y,z,v_y, v_z, \theta, \omega\}$, as follows. Using Eq.~\eqref{s2eq26} in Eqs.~\eqref{s2eq18},\eqref{s2eq19} yields:
\begin{align}
 \label{s2eq28}
 & \dot{y}= v_y \\
 \label{s2eq29}
 & \dot{v}_y = 0. \
\end{align}
The expression for $d(t)$, the penetration depth, as given in Eq.~\eqref{s2eq23}, becomes
\begin{align}
 \nonumber
 d(t) & = \rv^T|_{_{(t-h)}} \nev  +  \av^T  \dv_{_{\!c3}}|_{_{(t-h)}} \\
 \label{s2eq30}
 & = ( z +a \cos \theta )_{_{(t-h)}}. \
\end{align}
The expression for $\dot{d}(t)$, the penetration depth rate, as given in Eq.~\eqref{s2eq23a}, becomes
\begin{align}
 \nonumber
 d(t) & = \vv^T|_{_{(t-h)}}\,  \nev + \av^T \left\{[-\omgv \times] \dv_{_{\!c3}}\right\}_{_{(t-h)}} \\
 \label{s2eq31}
 & = ( v_z - a \omega \sin \theta )_{_{(t-h)}}. \
\end{align}
The torque $\tov_{_\Bfr}(t)$, as given in Eq.~\eqref{s2eq22a}, is expressed by the following scalar quantity:
\begin{align}
 \label{s2eq32}
 & \tau_{_\Bfr}(t) = - a f(t) \sin \theta{{(t-h)}}. \
\end{align}
Notice that from physical considerations the angle $\theta$ is always acute and that $\sin \theta$ is thus always positive. Therefore, according to the adopted convention, a positive force $f(t)$ (directed upward along $\nev$) creates a negative torque (about the anti x-axis).
\subparagraph{Summary:} The two-dimensional state-space model for $\{y, v_y, z, v_z, \theta, \omega\}$ is summarized as follows:
\begin{align}
 \label{s2eq33}
& \dot{y}=v_y \\
\label{s2eq34}
& \dot{v}_y= 0 \\
\label{s2eq35}
& \dot{z} = v_z \\
\label{s2eq36}
& \dot{v}_z= \frac{f(t)}{m} \\
\label{s2eq37}
& \dot{\theta}= \omega \\
\label{s2eq38}
& \dot{\omega} = \frac{\tau(t)}{J_x} \
\end{align}
where
\begin{align}
\label{s2eq39a}
& \tau(t) = - a f(t) \sin \theta_{{(t-h)}} \\
& f(t) = - k d(t) - b \dot{d}(t)\\
& d(t) = ( z +a \cos \theta )_{{(t-h)}}\\
\label{s2eq39}
& \dot{d}(t) = ( v_z - a \omega \sin \theta )_{{(t-h)}} \
\end{align}
and $a$, $k$, $d$, $m$, $J_x$ are constant and subscripts were dropped for notational simplicity. Notice that the dynamics of the states $\{y, v_y \}$ are decoupled from the other states dynamics: these states represent the unstable and uncontrollable motion parallel to the nozzle wall. The reduced representation that includes the four states $\{z, v_z, \theta, \omega\}$ is relevant for the investigation of the delay $h$ on the system stability. The proposed decoupled and reduced representation greatly simplifies the stability analysis of this nonlinear delay system, as shown next.
\subsection{Linearized Two-Dimensional State-Space Model}

\paragraph{Original Two-Dimensional Model}
This subsection is concerned with the development of a linearized dynamical model for small perturbations, $\{\dsz, \dsvz, \dstheta, \dsomg\}$, about nominal values of the nonlinear delay system~\eqref{s2eq33}-\eqref{s2eq39}. In the proposed linear framework, it will be shown that the dynamics of the penetration depth $d(t)$ and rate $d(t)$ is enough for the system analysis . This allows for the definition of a different state representation, with $d$ and $\dot{d}$ as state variables. The resulting intuitive state-space model has a block-triangular dynamics matrix, which enables a simple stability analysis of the linear delay system using results from the 1D analysis presented in \cite{Melak2013}.
The relevant states are the position and velocity normal to the nozzle wall, $z$ and $v_z$, and the angle and angular rate of $\Bfr$ with respect to $\Nfr$, that is $\theta$ and $\omega$. The geometry is provided in Fig.~\ref{fig:3dideal}. Under nominal conditions, it is assumed that the chaser probe approaches the target nozzle with an orientation parallel to the nozzle axis of symmetry, with small translational and rotational velocities, and that the penetration depth and depth rate stay small during contact. It stems from the above assumptions that the nominal values for $\{z, v_z, \theta, \omega\}$ are given as follows:
\begin{align}
\label{s2eq40}
& z^{\ast}= - a \sin \alfa\\
& v_z^{\ast}= 0\\
& \theta^{\ast} = \frac{\pi}{2} - \alpha \\
\label{s2eq41}
& \omega^{\ast}= 0.\
\end{align}
Using Eqs.~\eqref{s2eq40}-\eqref{s2eq41} in Eqs.~\eqref{s2eq39}-\eqref{s2eq39a} it is straightforward to show that the nominal values for the penetration depth and depth rate are zero, and thus that the nominal force and torque are zero, too, i.e.
\begin{align}
\label{s2eq41a}
& d^{\ast}=0\\
& \dot{d}^{\ast}= 0\\
& f^{\ast}= 0\\
& \tau^{\ast} = 0.\
\end{align}
Let $\xv = \{\xsone,\xstwo, \xsthr, \xsfor\} $  denote the four state variables $\{z, v_z, \theta, \omega\}$, and let $\xvast$ denote the set of nominal values $\{z^{\ast}, v_z^{\ast}, \theta^{\ast}, \omega^{\ast}\}$ as given in Eqs.~\eqref{s2eq40}-\eqref{s2eq41}. Let $\phi_i$, $i=1,2,3,4$ denote the four nonlinear functions of $\xv$, as given in the right-hand sides of Eqs.~\eqref{s2eq35}-\eqref{s2eq38}, i.e.
\begin{align}
\label{s2eq42}
 & \phi_1(\xv) = x_2 \\
 \label{s2eq43}
 & \phi_2(\xv) = \frac{f(t)}{m} \\
 \label{s2eq44}
 & \phi_3(\xv) = x_4 \\
 \label{s2eq45}
 & \phi_4(\xv) = \frac{\tau(t)}{J_x} \
\end{align}
where
\begin{align}
\label{s2eq46}
& \tau(t) = - a f(t) (\sin x_3)_{_{(t-h)}} \\
\label{s2eq47}
& f(t) = - k d(t) - b \dot{d}(t)\\
\label{s2eq48}
& d(t) = ( x_1 + a \cos x_3 )_{_{(t-h)}}\\
\label{s2eq49}
& \dot{d}(t) = ( x_2 - a x_4 \sin x_3 )_{_{(t-h)}}. \
\end{align}
The dependence upon the time delay $h$ is essential for the upcoming stability analysis. It will be dropped however for notational simplicity since it formally does not impact the derivation of the partial derivatives of the functions $\phi_i$. Given Eqs.~\eqref{s2eq42}, \eqref{s2eq44}, the partial derivatives of $\phi_2(\xv)$ and $\phi_4(\xv)$ with respect to $\xv$ are written as follows:
\begin{align}
 &\frac{\partial \phi_1}{\partial \xv^T} =
 \begin{bmatrix}
  0 & 1 & 0 & 0 \\
 \end{bmatrix} \\
  &\frac{\partial \phi_3}{\partial \xv^T} =
 \begin{bmatrix}
  0 & 0 & 0 & 1 \\
 \end{bmatrix}. \
 \end{align}
 Using Eqs.~\eqref{s2eq48},\eqref{s2eq49} in Eq.~\eqref{s2eq47}, the partial derivatives of the force $f(t)$ with respect to $x_i$ are expressed as follows:
 \begin{align}
 \label{s2eq50}
  & \frac{\partial f}{\partial x_i} =
  \left\{
 \begin{array}{ll}
   - k & \hspace{1cm} i=1 \\
   - b & \hspace{1cm} i=2 \\
   k a \sin x_3 + b a x_4 \cos x_3 & \hspace{1cm} i=3 \\
   b a \sin x_3 & \hspace{1cm} i=4. \\
 \end{array}
  \right.
 \end{align}
Using Eq.~\eqref{s2eq50} in the expression for $\phi_2(\xv)$ as given in Eq.~\eqref{s2eq43} yields the partial derivative of $\phi_2(\xv)$ with respect to the state vector:
\begin{align}
\nonumber
\frac{\partial \phi_2}{\partial \xv^T}
& =   \frac{1}{m}
 \begin{bmatrix}
 -k  & -b & k a \sin x_3 + b a x_4 \cos x_3 & b a \sin x_3 \\
 \end{bmatrix} \\
\label{s2eq51}
& \stackrel{(\xv=\xvast)}{=} \frac{1}{m}
 \begin{bmatrix}
 -k  & -b & k a \alfacos & b a \alfacos \\
 \end{bmatrix} \
\end{align}
where the second line was obtained by inserting the nominal state $\xvast$, as given in Eqs.~\eqref{s2eq40}-\eqref{s2eq41}, and $\alfacos$ denotes $\cos \alpha$. Using Eqs.~\eqref{s2eq46},\eqref{s2eq50}, the partial derivatives of the torque $\tau(t)$ with respect to $x_i$ are expressed as follows:
 \begin{align}
 \label{s2eq52}
  & \frac{\partial \tau}{\partial x_i} =
  (-a)
  \left\{
 \begin{array}{ll}
   - k \sin x_3 & \hspace{1cm} i=1 \\
   - b \sin x_3 & \hspace{1cm} i=2 \\
   (\cos x_3) f + ( k a \sin x_3 + b a x_4 \cos x_3) \sin x_3 & \hspace{1cm} i=3 \\
   b a \sin^2 x_3 & \hspace{1cm} i=4. \\
 \end{array}
  \right.
 \end{align}
Using Eq.~\eqref{s2eq52} in the expression for $\phi_4(\xv)$ as given in Eq.~\eqref{s2eq45} yields the partial derivative of $\phi_4(\xv)$ with respect to the state vector:
\begin{align}
\nonumber
\frac{\partial \phi_4}{\partial \xv^T}
& =   \frac{a}{J_x}
 \begin{bmatrix}
 k \sin x_3  & b \sin x_3 & - (\cos x_3) f - ( k a \sin x_3 + b a x_4 \cos x_3) \sin x_3 & - b a \sin^2 x_3 \\
 \end{bmatrix} \\
\label{s2eq53}
& \stackrel{(\xv=\xvast)}{=} \frac{a}{J_x}
 \begin{bmatrix}
 k \alfacos  & b\alfacos & - k a \alfacos^2 & - b a \alfacos^2 \\
 \end{bmatrix}. \
\end{align}
where the second line was obtained by inserting the nominal state $\xvast$, as given in Eqs.~\eqref{s2eq40}-\eqref{s2eq41}, and $\alfacos^2$ denotes $\cos^2 \alpha$. To summarize, the gradient matrix of the functions $\phi_i(\xv)$ $i=1,2,3,4$ with respect to $\xv$, evaluated at $\xvast$, is expressed as follows:
\begin{align}
\label{s2eq54a}
& F_{\!\!x}^{\ast} =
\begin{bmatrix}
 0 & 1 & 0 & 0 \\
 -\frac{k}{m}  & -\frac{b}{m} & \frac{k a \alfacos}{m} & \frac{b a \alfacos}{m} \\
   0 & 0 & 0 & 1 \\
 \frac{k a \alfacos}{J_x}  & \frac{b a \alfacos}{J_x} & - \frac{k a^2 \alfacos^2}{J_x} & - \frac{b a^2 \alfacos^2}{J_x} \\\end{bmatrix} \
\end{align}
Taking into account the delay $h$, the dynamics for the perturbations $\delta x_i$ $ i=1,2,3,4$ are governed by the following linear time-invariant differential-delay equations:
\begin{align}
\label{s2eq54ab}
 & \delta \dot{x_1}(t) = \delta x_2(t) \\
 & \dot{\delta x_2}(t) = \left[ -\frac{k}{m} \delta x_1 - \frac{b}{m} \delta x_2 + \frac{ka\alfacos}{m} \delta x_3 + \frac{b a \alfacos}{m} \delta x_4 \right]_{(t-h)} \\
 & \dot{\delta x_3}(t) = \delta x_4(t) \\
 \label{s2eq54ac}
 & \dot{\delta x_4}(t) = \left[ \frac{ka\alfacos}{J_x} \delta x_1 + \frac{b a \alfacos}{m} \delta x_2 - \frac{k a^2 \alfacos^2}{J_x} \delta x_3 -
 \frac{ba^2 \alfacos^2}{J_x} \delta x_4 \right]_{(t-h)} \
\end{align}
with initial conditions $\delta x_i(0)$ $i=1,2,3,4$. Notice that the above equations are approximations to first-order in $\delta x_i$. Also notice that in the absence of delay, Eqs.~\eqref{s2eq54ab}-\eqref{s2eq54ac}, reduced to the classical state vector equation:
\begin{align}
\label{s2eq54}
 & \dot{\deltav\!\!\xv} =
 F_{\!\!x}^{\ast}\,\deltav\!\!\xv\
\end{align}
where $\deltav\!\!\xv$ denotes the vector of the small perturbations about $\xvast$. Based on the linear time-invariant state-space model for the delay system, Eqs.~\eqref{s2eq54ab}-\eqref{s2eq54ac}, standard tools may be applied from the realm of multivariable linear delay systems theory in order to analyze the stability as a function of the system's characteristics: the delay $h$, the mass $m$, the inertia $J$, the length $a$, the angle $\alpha$, the stiffness and damping coefficients, $k$ and $b$.

 \paragraph{Dynamics of the penetration rate $d(t)$}
As understood from the physical assumptions of the contact model, the direction normal to the nozzle wall has a particular role since a point contact force model is computed along this direction~\cite{Gilardi}. In the following the differential equation governing the penetration depth in the normal direction, i.e. $d(t)$, is developed. It will be shown that the equation is an autonomous second-order differential equation with characteristics $k$, $b$, and a reduced mass $m_{a}$. By definition, the depth $d$ is expressed as follows:
\begin{align}
\label{s2eq55}
& d= x_1 + a \cos x_3 \
\end{align}
where the delay dependence was dropped for simplicity. Let $\delta\!d$ denote the perturbation about $d^{\ast}$, i.e.
\begin{align}
 & \delta\!d = d - d^{\ast}.
\end{align}
A direct differentiation of Eq.~\eqref{s2eq55} yields
\begin{align}
\label{s2eq55a}
& \delta d = \delta x_1 -a s_\alpha \delta x_3 \
\end{align}
Using Eqs.~\eqref{s2eq54}, \eqref{s2eq54a}, the expression for $\dot{\delta\!d}$ is developed as follows:
\begin{align}
\nonumber
\delta \dot{d} & = \dot{\delta x_1} - a \alfacos \dot{\delta x_3} \\
\label{s2eq56}
& = \delta x_2 - a \alfacos \,\delta x_4. \
\end{align}
Taking the time-differential on both sides of Eq.~\eqref{s2eq56} yields the following expression for $\ddot{\delta\!d}$:
\begin{align}
 \nonumber
 \delta \ddot{d} & = \dot{\delta x_2} - a \alfacos \dot{\delta x_4} \\
 \nonumber
 & = - k \left( \frac{1}{m} +\frac{a^2 \alfacos^2}{J_x} \right) \underbrace{(\delta x_1 - a \alfacos \delta x_3)}_{\delta\!d}
 - b  \left( \frac{1}{m} +\frac{a^2 \alfacos^2}{J_x} \right) \underbrace{(\delta x_2 - a \alfacos \delta x_4)}_{\dot{\delta\!d}} \\
 \nonumber
 & = \left( \frac{1}{m} +\frac{a^2 \alfacos^2}{J_x} \right) (- k \,\delta\!d - b \,\dot{\delta\!d} )\\
 \label{s2eq57}
 & = \frac{1}{m_{\!a}} (- k \,\delta\!d - b \,\dot{\delta\!d} )\
\end{align}
where Eqs.~\eqref{s2eq55a}, \eqref{s2eq56} were used in the third line, and the reduced mass $m_{\!a}$ is defined from the last line. Recalling that the nominal penetration depth $d^{\ast}$ is zero (Eq.~\ref{s2eq41a}), the perturbation $\delta\!d$ and the depth variable $d(t)$ are identical. To conclude, the dynamics of $d(t-h)$ is governed by the following second-order differential-delay equation:
\begin{align}
\label{s2eq58}
 & m_{\!a} \ddot{d}_t + b \dot{d}{(t-h)}+k d{(t-h)} = 0 \
\end{align}
with initial conditions $d(0), \dot{d}(0)$, where
\begin{align}
\label{s2eq59}
& m_{\!a} =  \frac{m}{1+ \frac{m (a\alfacos)^2}{J_x} }.\
\end{align}
It appears from Eq.~\eqref{s2eq58} that the dynamics of $d(t-h)$ is governed by a homogeneous equation that is decoupled from the dynamics of the other state variables. It is function of the stiffness and damping coefficients, $k$ and $b$, and of the mass $m_{\!a}$, which includes mass $m$, the inertia about the x-axis $J_x$, and the arm length $a\alfacos$. It is instructive to consider  specific cases for $m_{\!a}$.

{\bf I.} If the contact is frontal and the probe is aligned with the direction normal to the wall, then the nominal angle $\theta^{\ast}$ is zero, that is, $\cos \alfa = 0$, and $m_{\!a}= m$. There is no rotation, and the system degenerates to a single-dimensional system acting in translation only, with states $z, v_z$.

{\bf II.} The opposite limiting case corresponds to an approach where the probe is parallel to the nozzle wall. The angle $\theta^{\ast}$ is thus 90 deg, thus $\cos \alfa = 1$ and $m_{\!a}=  \frac{m}{1+ ma^2/J_x}$. This is the minimal value that the mass $m_a$ can reach.

{\bf III.} When the inertia $J$ is very high with respect to the Steiner term $m (a\alfacos)^2$ then $m_{\!a}\simeq m$. Here, the dynamics in $d$ will almost exclusively result from the translation of the chaser center $B$, and (almost) no rotation will take place.

{\bf IV.} When the inertia $J$ is negligible compared to the Steiner term then $m_{\!a} \simeq J_x/(a\alfacos)^2$, and the dynamics of $d$ will mainly result from the rotation of the probe tip about the center $B$. The main conclusion from the above results, as given in Eqs.~\eqref{s2eq58}, \eqref{s2eq59}, is that the two-dimensional case study involves a single-dimensional dynamical delay system with $d$ and $\dot{d}$ as states. Hence, the same tools can be applied as in the single-dimensional study presented in \cite{Melak2013}.

\paragraph{Transformed Two-Dimensional Model}
Based on the previous results, this subsection presents a transformed state representation for the four-states model and develops the dynamics equation of the transformed system. The resulting dynamics matrix simplifies the stability analysis of this multivariable linear delay system. Let $\delta y_i$ $i=1,2,3,4$ denote the following four state variables as linear combinations of $\delta x_i$ $i=1,2,3,4$, as follows:
\begin{align}
\label{s2eq60}
 & \delta y_1 = \delta x_1 \\
 & \delta y_2 = \delta x_2 \\
 & \delta y_3 = \delta x_1 -a \alfacos \delta x_3 \\
 \label{s2eq61}
 & \delta y_4 = \delta x_2- a \alfacos \delta x_4. \
\end{align}
Notice that the first two variables are identical to the previous states variables and that the last two are, by definition, the penetration depth $\delta d$ and depth rate $\dot{\delta d}$. In vector-matrix form, Eqs.~\eqref{s2eq60}-\eqref{s2eq61} are re-written as follows:
\begin{align}
 \label{s2eq61a}
 \deltav\!\yv = T \deltav\!\xv \
\end{align}
where
\begin{align}
 & T =
 \begin{bmatrix}
 1 & 0 & 0 & 0 \\
 0 & 1 & 0 & 0 \\
 1 & 0 & - a \alfacos & 0 \\
 0 & 1 & 0 & -a \alfacos \\
 \end{bmatrix}. \
\end{align}
Assuming, for the time-being, that there is no delay, it is a well known result from linear systems theory that the dynamics matrix of the transformed system is expressed by the following similarity transformation:
\begin{align}
& F_{\!\!y} = T F_{\!\!x} T^{-1}
\end{align}
where $F_{\!\!y}$ denotes the dynamics matrix for the state vector $\deltav\!\!\yv$. The derivation of $F_{\!\!y}$, which is straightforward and is omitted here for the sake of brevity, yields the following expression:
\begin{align}
\label{s2eq61aa}
 & F_{\!\!y} = \begin{bmatrix}
 0 & 1 & 0 & 0 \\
 0 & 0 & \frac{k}{m} & \frac{b}{m} \\
 0 & 0 & 0 & 1 \\
 0 & 0 & -\frac{k}{m_{\!a}} & -\frac{b}{m_{\!a}}\\
 \end{bmatrix}\
\end{align}
The dynamics matrix $F_{\!\!y}$ in Eq.~\eqref{s2eq61aa} reveals  the uncoupled dynamics of the states $\{\delta y_3, \delta y_4\}$ ( i.e. $\{d, \dot{d}\}$).  The dynamics of $\{d, \dot{d}\}$ should be stable during the contact process as nature does when two bodies are in contact. Thus, this dynamics shall be used to analyze the stability of the contact process for two dimensional contact case. Recalling the impact of the delay $h$, the delay system dynamics for the transformed states is expressed as follows:
\begin{align}
 \label{s2eq62}
 & \dot{\delta y_1}(t) = \delta y_2(t) \\
 & \dot{\delta y_2}(t) = \left[ \frac{k}{m} \delta y_3 + \frac{b}{m} \delta y_4 \right]_{(t-h)} \\
  & \dot{\delta y_3}(t) = \delta y_4(t) \\
 \label{s2eq63}
 & \dot{\delta y_4}(t) = \left[ - \frac{k}{m_{\!a}} \delta y_3 - \frac{b}{m_{\!a}} \delta y_4 \right]_{(t-h)} \
\end{align}
where $m_a$ is given in Eq.~\eqref{s2eq59}.

\paragraph{Concluding Remarks}
The linearized time-invariant delay $4^{th}$-order system, with dynamics governed by Eqs.~\eqref{s2eq61a}-\eqref{s2eq61aa}, lends itself to stability analysis results for second-order systems, as developed in \cite{Melak2013}. The state transformation introduced in the linear analysis is intuitive and exploits the assumption of single-dimensional motion of the probe tip during contact. The mode related to the motion parallel to the wall is not relevant to the stability analysis since the contact force model is computed perpendicular to the contacting surface~\cite{Gilardi}. In addition, when friction is considered in the parallel direction motion, it  has a stabilizing effect as it dissipates energy. 
\section{Stability Analysis}

This section is concerned with a linear stability analysis of the linear time-invariant dynamical delay system described in Eqs.~\eqref{s2eq62}-\eqref{s2eq63}. The characteristic polynomial is easily derived as the product of two second-order polynomials. This decomposition enables the straightforward application of the pole location method for stability analysis. Exact expressions for the critical values of the delay and for the associated crossing frequencies are developed as functions of the mass, the inertia, the angle $\alpha$, the stiffness and damping coefficients. A numerical example is provided for illustration using realistic values.

\subsection{Characteristic Polynomial of 4$^{th}$-order}
Consider the set of four first-order linear differential-delay equations for the states $\delta y_i$ with  $i=1,2,3,4$, as given in Eqs.~\eqref{s2eq62}-\eqref{s2eq63}. Rewriting these equations as two second-order equations in $\delta y_1$ and $\delta y_3$, and bringing all terms to the left-hand side, yields:
\begin{align}
\label{s3eq64}
 & m \,\delta \ddot{y_1}(t) + \left[ 2 b \,\delta \dot{y_1} + 2 k  \, \delta y_1 \right]_{(t-h)}
 - \left[ b \,\delta \dot{y_3} + k \, \delta y_3 \right]_{(t-h)} = 0  \\
 \label{s3eq65}
 & m_{\!a} \, \delta \ddot{y_3}(t) + \left[ b \,\delta \dot{y_3} + k  \, \delta y_3 \right]_{(t-h)}  = 0. \
\end{align}
Applying the Laplace transform on both sides of Eqs.~\eqref{s3eq64}, \eqref{s3eq65}, and denoting by $\delta Y_1(s)$ and $\delta Y_3(s)$ the Laplace transforms of $\delta y_1(t)$ and $\delta y_3(t)$, respectively, yields:
\begin{align}
\label{s3eq66}
& \begin{bmatrix}
 m s^2  & - ( bs + k ) \\
 0 & m_{\!a} s^2 + e^{-sh} ( b s + k) \\
 \end{bmatrix}
 \,
 \begin{pmatrix}
 \delta Y_1(s) \\
  \delta Y_3(s) \\
 \end{pmatrix} =
 \begin{pmatrix}
 0 \\ 0
 \end{pmatrix}. \
\end{align}
By inspection of Eq.~\eqref{s3eq66}, it is straightforward to express the characteristic polynomial of the 4$^{th}$-order system:
\begin{align}
\label{s3eq67}
&    \chi_h(s)= \left[ m s^2 \right] \left[ m_{\!a} s^2+ e^{-sh} ( b s + k ) \right] \
\end{align}
where $m_{\!a}$ is given in Eq.~\eqref{s2eq59}. The roots of the first polynomial in $\chi_h(s)$ are related to the dynamics of the states $\delta y_1$ and $\delta y_2$, i.e. the displacements of the center $B$. The roots of the second polynomial are the poles of the mode related to the penetration depth and rate, $\delta y_3$, $\delta y_4$.

\subsection{Stability Analysis using the Pole Location Method}

\subsubsection{Application to a Standard Second-Order System}
The following analysis is based on~\cite{Melak2013} and is presented here for the sake of completeness. Let the characteristic equation of a standard second-order loop delay system be expressed as follows:
 \begin{align}
 \label{eq:cheq}
&    \chi_h(s)=  \mu s^2+ e^{-sh} (\beta s + \kappa) \
\end{align}
where $\mu$, $\beta$, and $\kappa$, denote mass, damping, and stiffness coefficients, respectively.
Following the pole location method~\cite{Marshall}, the stability of Eq.~\eqref{eq:cheq} is analyzed  by studying the behavior of the system roots as $h$ increases from zero. The condition for stability is that all the roots of $\chi_h(s)$ lie in the open left half-plane (OLHP) of the complex plane. The pole location method provides an analytical mean to determine the value(s) of the delay $h$, as a function of the system's parameters $\mu$, $\kappa$, and $\beta$, such that some roots of $\chi(s)$ lie on the imaginary axis.  The first step consists in examining the delay-free characteristic polynomial, i.e.
\begin{equation}
	\chi(s)= \mu s^2+ \beta s + \kappa
	\label{eq:transhzero}
\end{equation}
Necessary and sufficient conditions of stability are that all coefficients are positive. The delay-free loop system is thus stable as long as there is some stiffness and damping in the feedback force. The second step consists in analyzing the roots as $h$ increases from zero. Some of the (infinite number of) roots will cross the imaginary axis for a critical value of $h$. Let $D(s)$ and $N(s)$ be defined as follows:
\begin{align}
\label{s3eq10}
& D(s) = \mu s^2 \\
\label{s3eq11}
& N(s) = \beta s + \kappa\
\end{align}
The general condition for $\chi(s)$ to have roots on $(j\omega)$ is expressed as follows:
\begin{align}
\label{s3eq12}
& \chi(j\omega) = 0
\;\;\Leftrightarrow \;\;\left\{
\begin{tabular}{l}
$| \frac{N(j \omega)}{D(j \omega)}| = 1$ \\
$\arg[\frac{N(j \omega)}{D(j \omega)}] = -\omega h \pm 2 \pi n$ \\
\end{tabular}
\right.
\
\end{align}
where $n = 0,1,2,\ldots$. Using Eqs.~\eqref{s3eq10} and \eqref{s3eq11} in Eq.~\eqref{s3eq12} yields
\begin{align}
\label{s3eq13a}
&  \mu^2 \omega^4 - \beta^2 \omega^2 - \kappa^2 = 0 \\
\label{s3eq13b}
& \omega h  = \arctan(\frac{\omega \beta}{\kappa}) \pm 2 \pi n \
\end{align}
where $n = 0,1,2,\ldots$ Selecting the positive root for $\omega$ yields the following final expressions:
\begin{align}
\label{s3eq14a}
&\omega = \sqrt{\frac{\beta^2}{2 \mu^2} + \sqrt{ \frac{\beta^4}{4 \mu^4} + \frac{\kappa^2}{\mu^2}}} \\
\label{s3eq14b}
& h_n  = \frac{1}{\omega}\arctan(\frac{\omega \beta}{\kappa}) \pm \frac{2 \pi}{\omega} n \
\end{align}
where $n = 0,1,2,\ldots$ Equations~\eqref{s3eq14a}, \eqref{s3eq14b} show that, as the delay increases, poles are crossing the imaginary axis each time $h$ reaches one of the values $h_n$ of the described set. The first value, denoted by $h_c$, is computed at $n=0$, thus
\begin{align}
\label{s3eq15}
& h_c = \frac{1}{\omega_c}\arctan(\frac{\omega_c \beta}{\kappa}) \
\end{align}
where the natural frequency at the $j\omega$-crossing, $\omega_c$, is expressed from Eq.~\eqref{s3eq14a}. Notice that the value of $\omega_c$ is independent of the delay.\footnote{This general result stems from the fact that a pure delay is a unitary operator that does not change the loop gain.} The criteria that determines whether poles are crossing on their way out of the OLHP (switch) or on their way into the OLHP (reversal) is the sign of the following quantity, $\sigma(\omega)$:
\begin{align}
\label{eq:gainqu1}
& \sigma (\omega_c) \eqdef \left[\frac{d}{d\omega}(\mid D(j\omega)\mid^2-\mid N(j\omega)\mid^2)\right]_{\omega= \omega_c}
=  \sqrt{\frac{\beta^4}{4 \mu^4}+\frac{\kappa ^2}{\mu^2}} \
\end{align}
where the second equality in Eq.~\eqref{eq:gainqu1} results from using Eqs.~\eqref{s3eq10}, \eqref{s3eq11}, and \eqref{s3eq14a}. A switch occurs if $\sigma(\omega_c) > 0$, a reversal occurs if $\sigma(\omega_c)  < 0$, and no crossing occurs if $\sigma(\omega_c) = 0$. Obviously, in the present case, only switches occur, i.e., poles successively leave the OLHP as $h$ takes on the values $h_n$. A particular case consists of the absence of damping, i.e. $\beta=0$. It is straightforward to check that the critical delay is simply $0$, and the associated crossing frequency is $\omega_c = \sqrt{\frac{\kappa}{\mu}}$, which are the expected values. Notice that for relative large values of the stiffness $\kappa$, the frequency $\omega_c$ is of order $\Order(\sqrt{\kappa})$ (see Eq.~\ref{s3eq14a}), which yields an order $\Order(\frac{1}{\kappa})$ for the critical delay $h_c$ (Eq.~\ref{s2eq15}). This illustrates the known phenomenon that higher values of a proportional feedback gain - here $\kappa$ - are adverse to stability in presence of delays. Further, since typical values of the stiffness $\kappa$ yield low ratios $\frac{\omega_c \beta}{\kappa}$, and using the equivalence $\arctan(x) \sim x$ for small $x$, it appears from Eq.~\eqref{s3eq15} that the critical delay $h_c$ is equivalent to $\frac{\beta}{\kappa}$, independently from the frequency $\omega_c$.  In other words, if $\omega_c \beta << \kappa$, one can use the following approximation formula in order to compute the critical delay:
\begin{align}
\label{s3eq16}
& h_c = \frac{\beta}{\kappa} \
\end{align}
As a conclusion, Eqs.~\eqref{s3eq15} and \eqref{s3eq14a} provide analytical expressions for the critical delay that will destabilize the closed-loop system, and for the natural frequency at which this happens, as functions of the system's parameters, $\mu$, $\kappa$, and $\beta$. \\

\subsubsection{Numerical Example}
\begin{figure}[h]
\centering     
\subfigure[$b$ vs $h$, m=60 kg, k=1000 N/m]{\label{fig:a}\includegraphics[width=60mm, height=60mm]{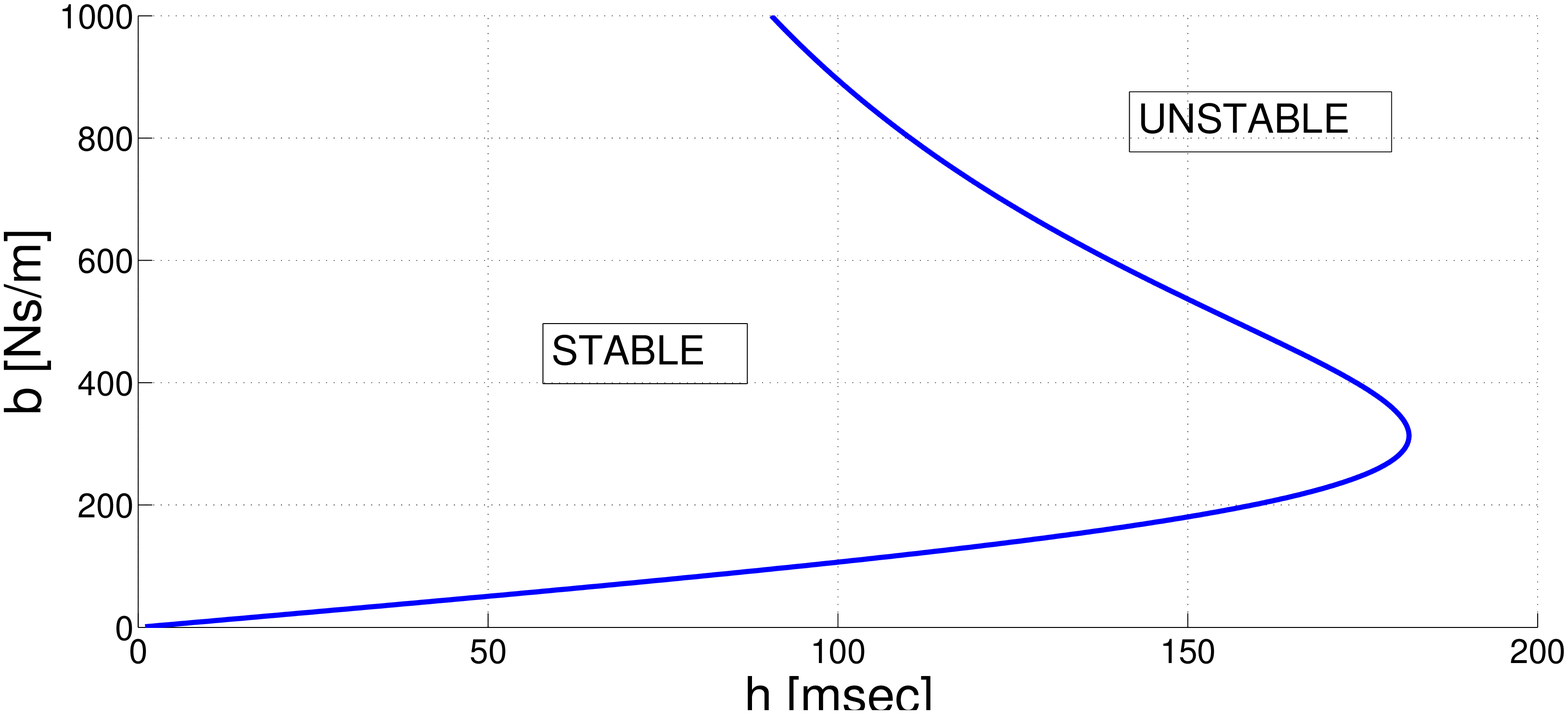}}
\subfigure[$k$ vs $h$, m=60 kg, b=50 Ns/m]{\label{fig:b}\includegraphics[width=60mm, height=60mm]{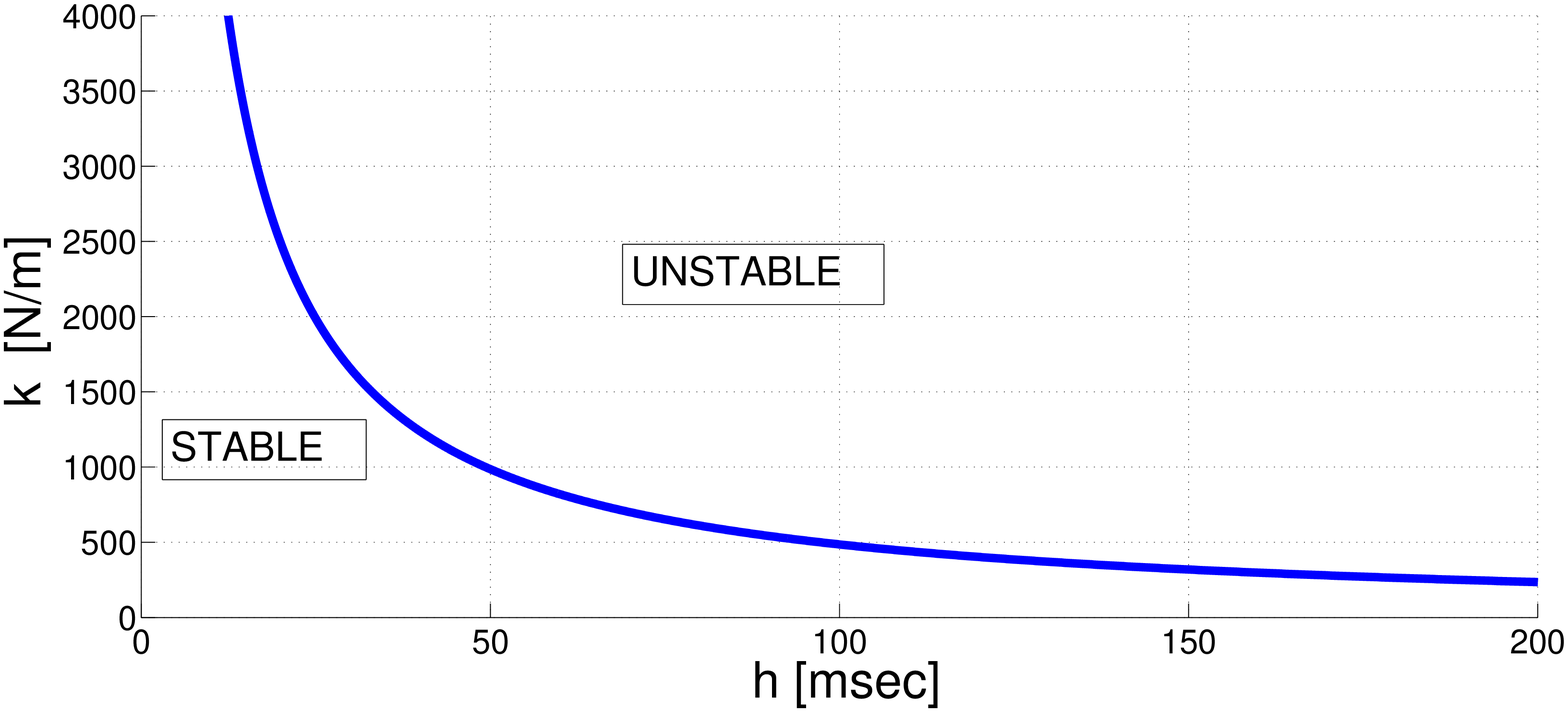}}
\subfigure[$m$ vs $h$, b=50 Ns/m, k=1000 N/m]{\label{fig:c}\includegraphics[width=60mm, height=60mm]{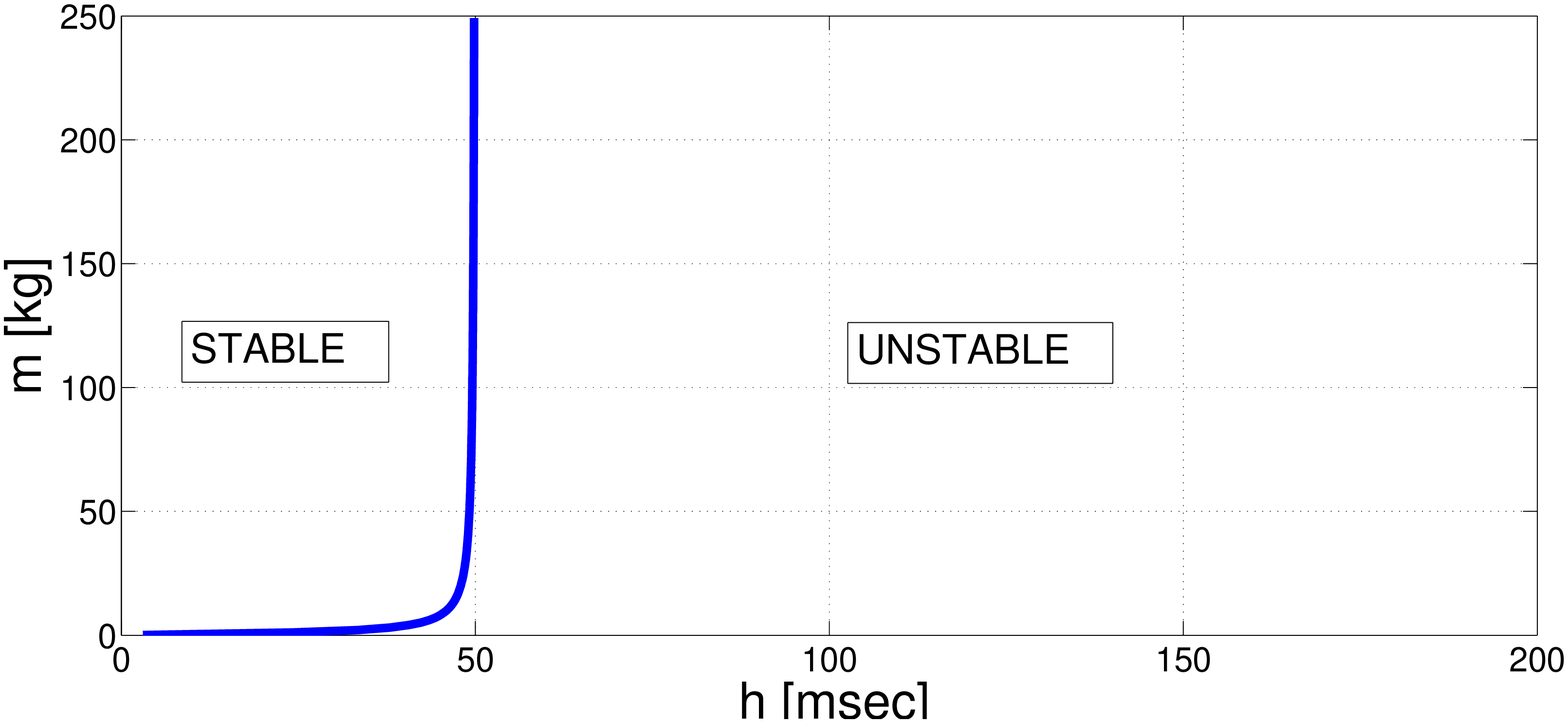}}
\caption{Stability domains for a typical operational point m=60kg, k=1000 N/m, b=50 Ns/m.}
   \label{f2}
\end{figure}

As an example, Figure~\ref{f2} depicts the stability regions for typical values of the delay $h$, the mass $m$, the stiffness $\kappa$, and the damping coefficient $\beta$. The plot in Fig.~\ref{f2}a illustrates the existence of a minimum required damping which ensures stability for a given value of the delay. It also shows that there exists an upper limit for the delay beyond which damping cannot ensure stability. Figure~\ref{f2}b provides the values of the stiffness $\kappa$ beyond which instability occurs for a given delay. Figure~\ref{f2}c illustrates the existence of a region of $\mu$ in which the critical delay becomes independent of $\mu$.

\begin{figure}[h!]
\centering     
\subfigure[$\beta$ vs $h$, $\mu=$60 kg, varying stiffness]{\label{fig:a}\includegraphics[width=60mm, height=60mm]{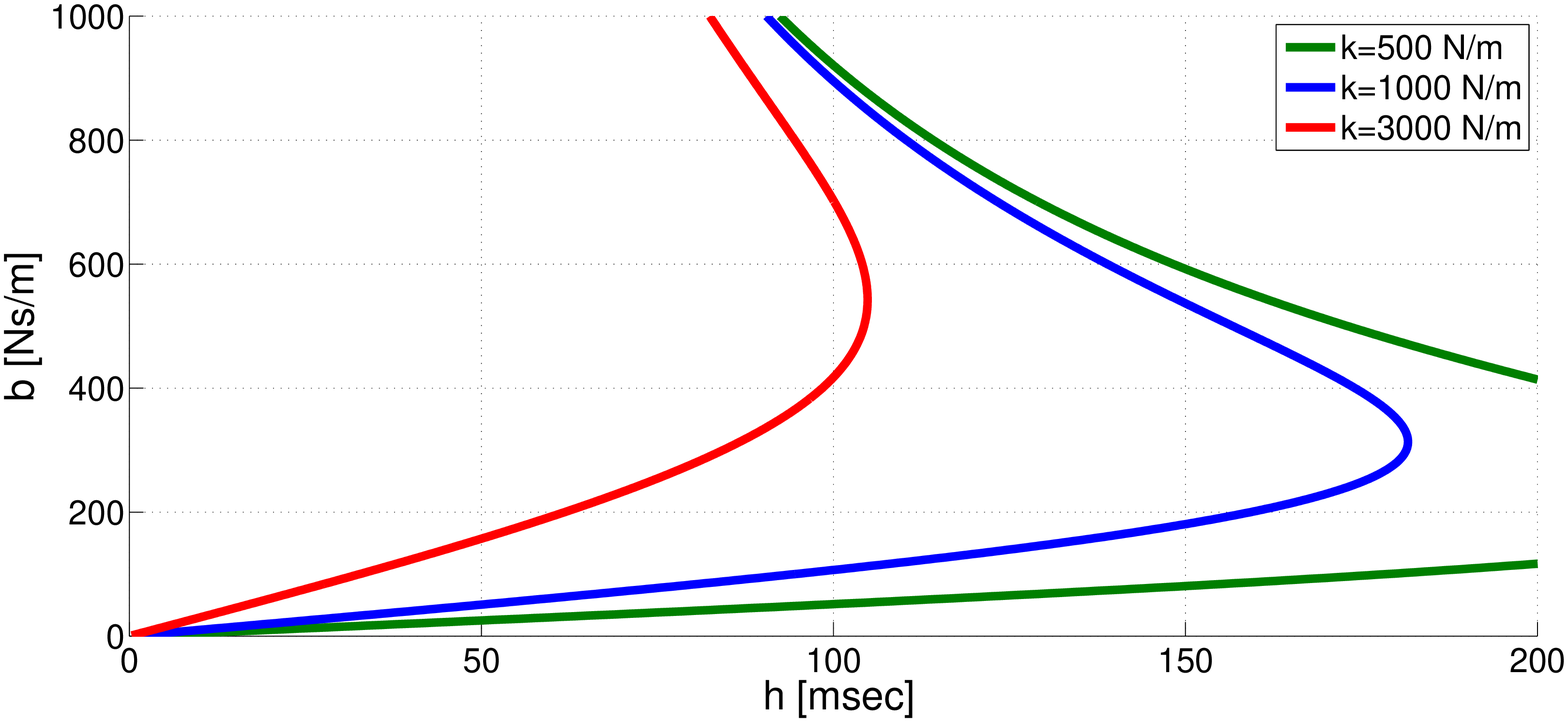}}
 \subfigure[$\beta$ vs $h$, $\kappa=$1000 N/m, varying mass]{\label{fig:b}\includegraphics[width=60mm, height=60mm]{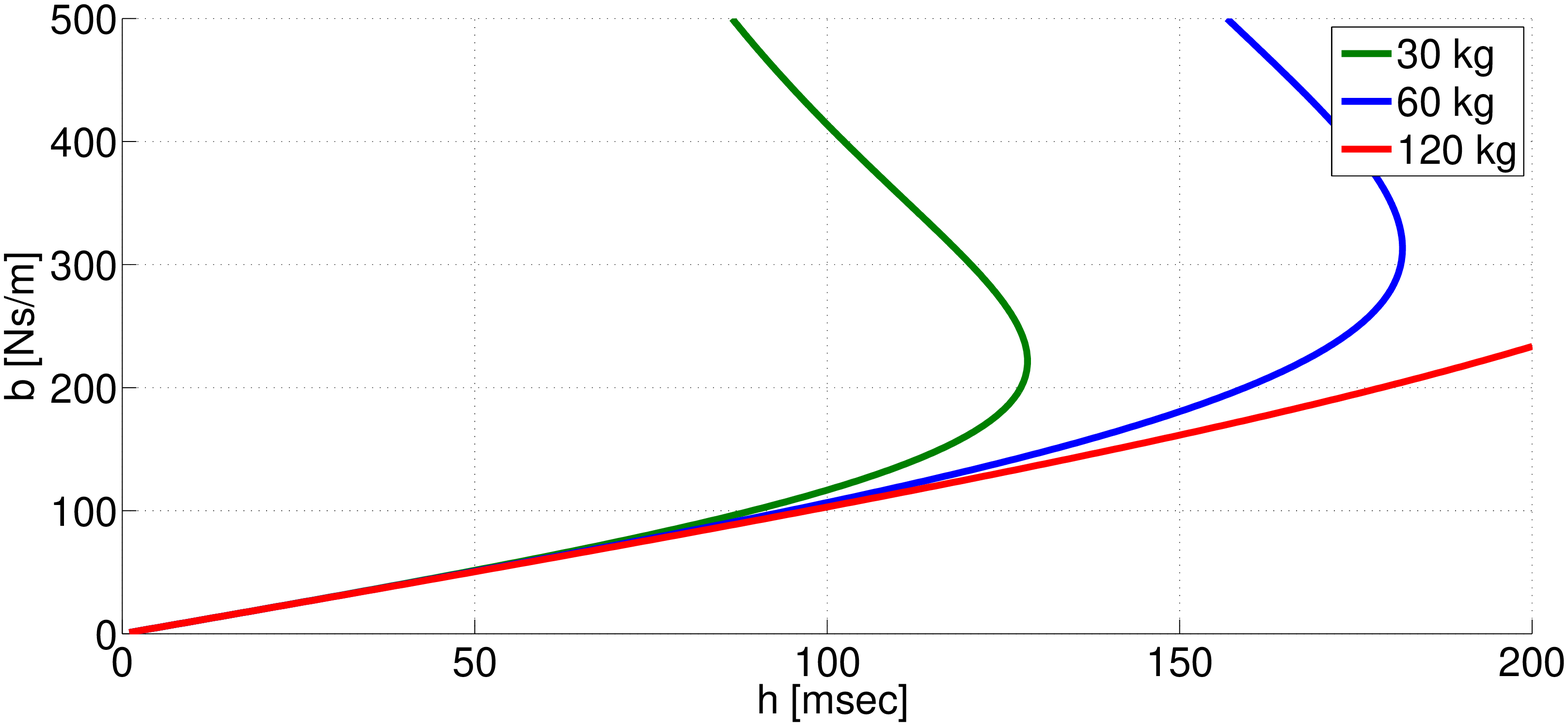}}
\subfigure[$\kappa$ vs $h$, $\beta=$50 Ns/m, varying mass]{\label{fig:c}\includegraphics[width=60mm, height=60mm]{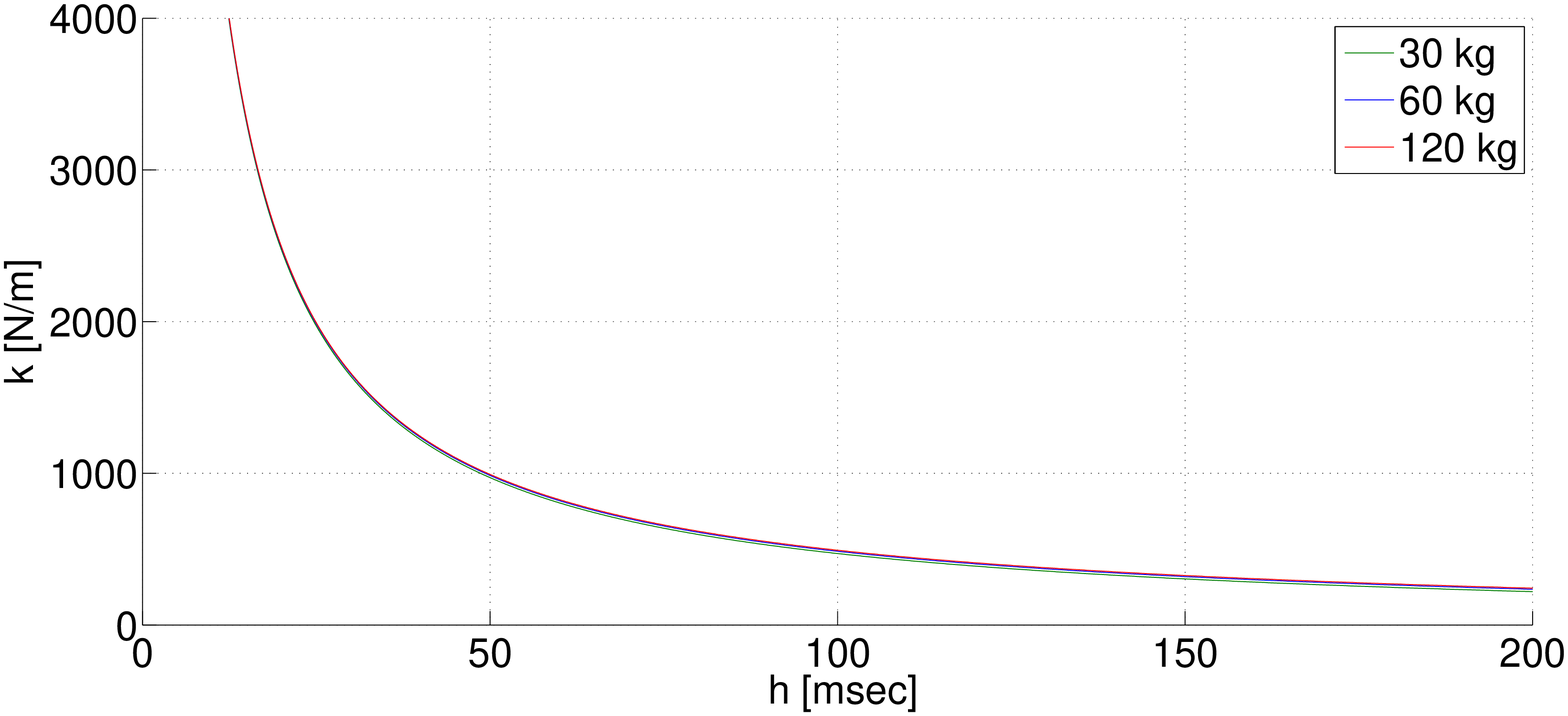}}
 \subfigure[$\kappa$ vs $h$, $\mu$=60 kg, varying damping]{\label{fig:d}\includegraphics[width=60mm, height=60mm]{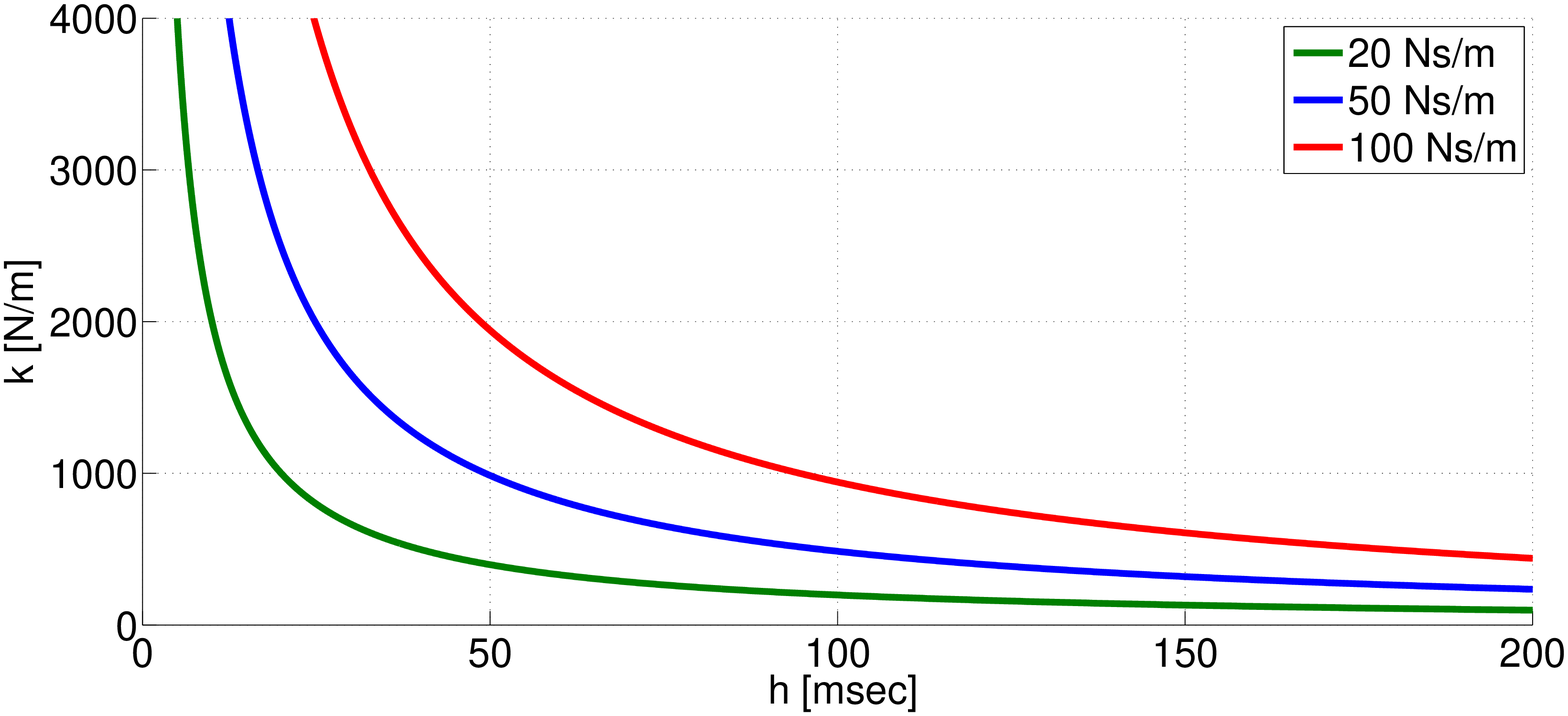}}
 \subfigure[$\mu$ vs $h$, $\kappa=$1000 N/m, varying damping]{\label{fig:e}\includegraphics[width=60mm, height=60mm]{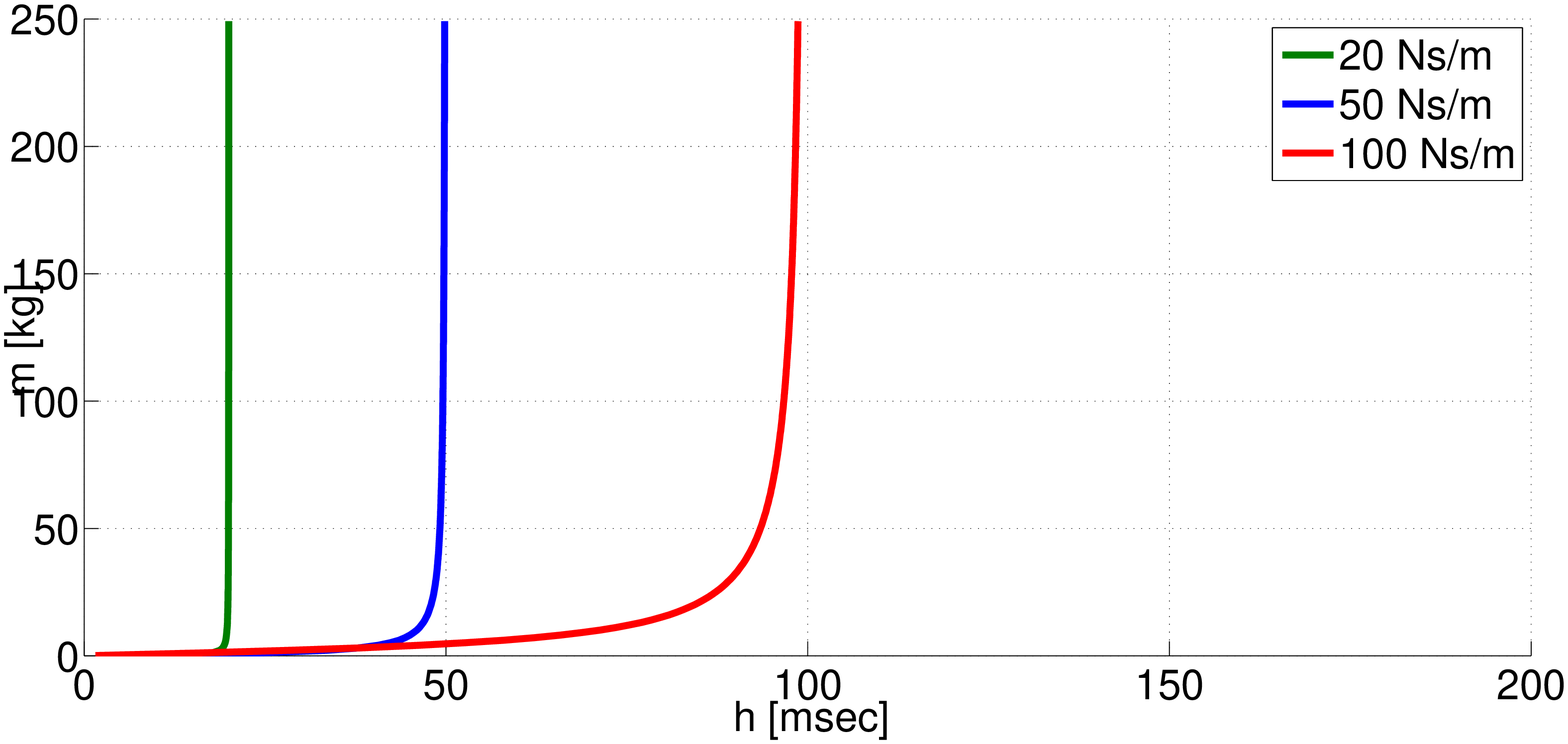}}
 \subfigure[$\mu$ vs $h$, $\beta=$50 N/m, varying stiffness]{\label{fig:f}\includegraphics[width=60mm, height=60mm]{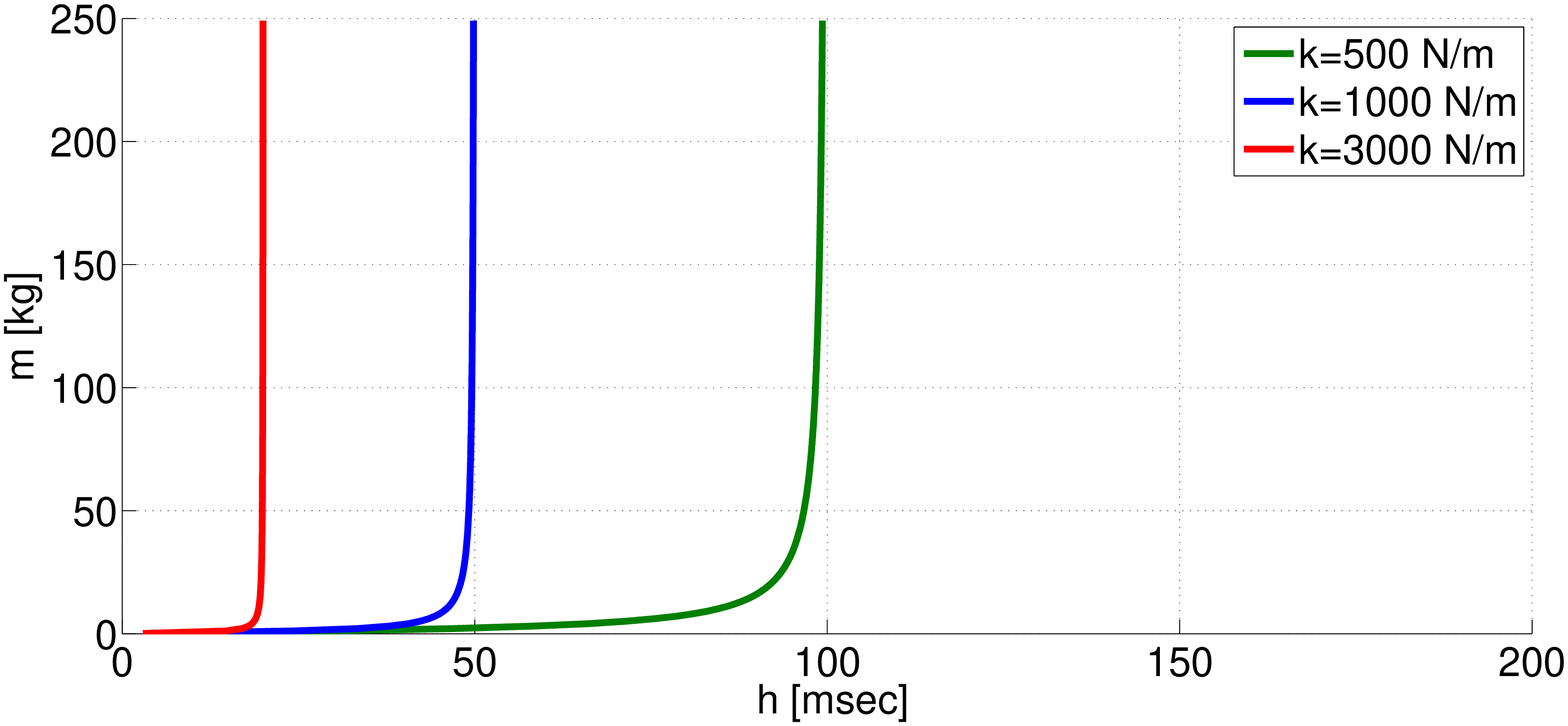}}
 \caption{Stability domains variations for an operational point $\mu$=60kg, $\kappa$=1000 N/m, $\beta$=50 Ns/m.}
   \label{f3}
\end{figure}

A numerical sensitivity investigation of the stability regions with respect the parameters $\mu$, $\kappa$, and $\beta$ was performed. The results are summarized in Figure~\ref{f3}. Figures~\ref{f3}a and \ref{f3}b depict the variations of Fig.~\ref{f2}a when $\kappa$ and $\mu$ are modified, respectively, while holding the second parameter constant. It appears that an increase in the stiffness $\kappa$ reduces the stability region (Fig.~\ref{f3}a), while an increase in the mass $\mu$ increases it (Fig.~\ref{f3}b). Henceforth, for a higher stiffness the HIL simulator will require more damping to guarantee stability. Notice that for small delays and damping values, the curve $\beta$ vs $h$ is approximately insensitive to the mass, as expected (see Eq.~\ref{s3eq16}). Figures \ref{f3}c-d illustrate the sensitivity of the curve $\kappa$ vs $h$ of Fig.~\ref{f2}b when $\mu$ and $\beta$ are varied, respectively. The sensitivity to changes in the mass is negligible: this is clearly seen from Eq.~\ref{s3eq16} where $\kappa$ decreases as $\beta/h$. On the other hand an increasing damping coefficient increases the stability region. Figures~\ref{f3}e-f depict the sensitivity of Fig.~\ref{f2}c to changes in $\beta$ and in $\kappa$. The increase in $\beta$ enlarges the domain of stability. It also shows that the maximum allowed delay becomes more mass-dependent for higher damping values. Notice that for the value of 20 Ns/m, and given a stiffness of 1000 N/m, the plot depicts a critical delay of 20 ms, which validates the approximation of Eq.~\eqref{s3eq15}. The increase in $\kappa$ has the inverse effect with a similar factor.

\subsection{Numerical Verification}

The objective of this subsection is to illustrate how well the proposed linear stability analysis performs when applied to the nonlinear system. For that purpose a numerical simulator of the nonlinear multibody dynamics of the two robots in contact was developed in 2D. It includes the nonlinear rigid body dynamics of the satellites, the linear contact dynamics model, and the robots pure delay models. The satellites' reduced mass $m$ is 60 kg, the probe length $a$ is 30 cm, and the target nozzle cone half-angle $\alfa$ is 30 deg. Including the chaser's inertia, the mass $m_a$ is 15.6 kg. The effective stiffness normal to the contact surface is 3000 N/m and the robots delay is 16 msec.

\begin{figure}[b!]
\centering
\subfigure[Linear stability analysis. The curve delineates the limit of stability. The dots show test points for varying damping values.]{\label{fig:a}\includegraphics[width=78mm]{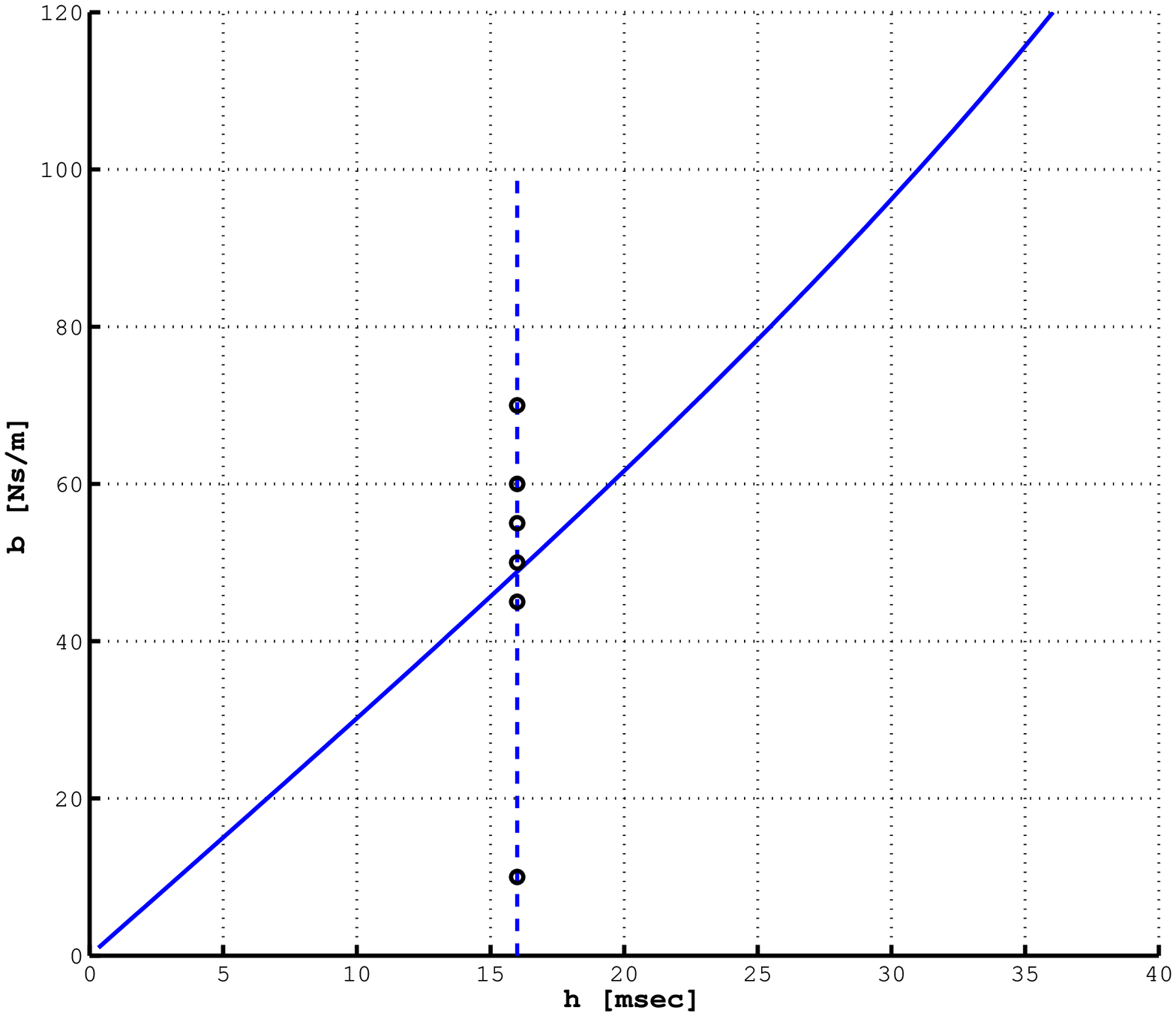}}
\subfigure[Nonlinear simulation. Penetration rate during the first contact]{\label{fig:b}\includegraphics[width=78mm]{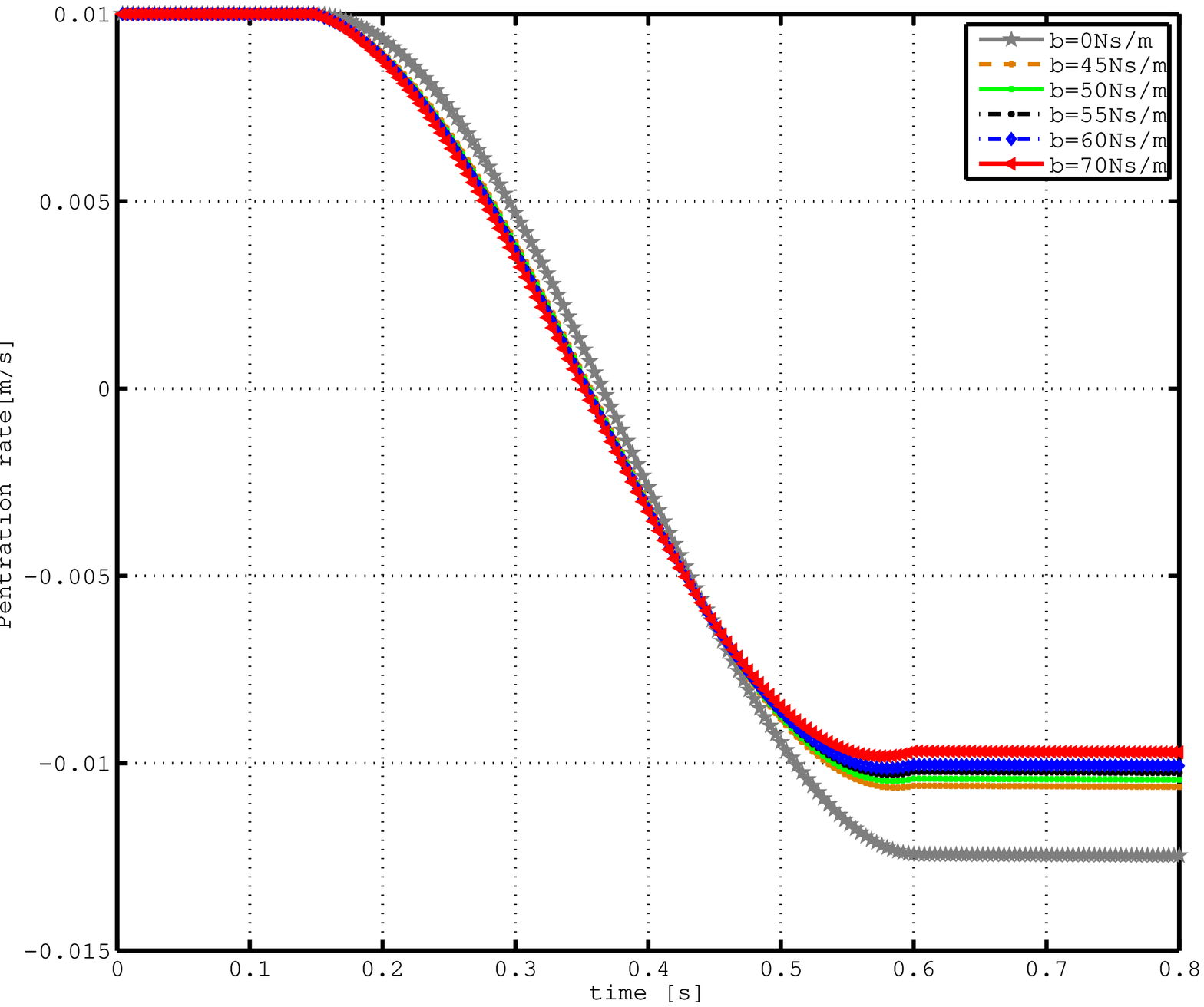}}
  \caption{Stability analysis validation. $m_a$ =15.6 kg, $h$ = 16 msec, $k$ = 3000 N/m}
  \label{fig:2Dpole}
\end{figure}

The numerical test consists in simulating contact for different values of the damping coefficient, $\beta$, and in comparing the observed stability limit for the nonlinear loop delay system with the predicted stability limits as shown in Fig~\ref{fig:2Dpole}. According to the linear analysis, the critical damping, $\beta_c$, is 50 Ns/m (as computed via Eqs.~\eqref{s3eq15}\eqref{eq:gainqu1}  and visualized on Fig.~\ref{fig:2Dpole}a. On the other hand, instability of the nonlinear system is cued via the coefficient of restitution~\cite{UyamaIROS}, denoted by $\epsilon$, which  is defined as follows:
\begin{align}
 \label{s3eq000}
 & \epsilon = \frac{v^{+}}{v^{-}} \
\end{align}
where $v^-$ and $v^+$ denote the penetration rate before and after the impact, respectively. The system is stable if $\epsilon < 1$, neutrally stable if $\epsilon= 1$, and unstable otherwise~\cite{UyamaIROS}. The results are summarized in Table~\ref{s3tab2} and Fig.~\ref{fig:2Dpole}.
\begin{table}[t]
 \centering
 \caption{Comparison of the linear and the nonlinear stability indices in  2D $b_c$=50 N/m.}
 \label{s3tab2}
\begin{tabular}{@{\extracolsep{10mm}}lcccccc}
 \hline\hline
$\beta$ [Ns/m] & 0      & 45  & 50  & 55 & 60 & 70 \\
\hline
$\beta_c/\beta$ & $\infty$  & 1.11 &  1  & 0.91 & 0.83 & 0.71 \\
$\epsilon$ & 1.6  &   1.14 &  1.09 & 1.03 & 1 & 0.82 \\
\hline\hline
\end{tabular}
\end{table}
Table~\ref{s3tab2} shows that the domains where the ratio $\beta_c/\beta$ and $\epsilon$ are smaller or greater than one are almost identical. Table~\ref{s3tab2} and ~\ref{fig:2Dpole}a indicate that the linear stability analysis is efficient in predicting the unstable behavior of the nonlinear system. There is however some discrepancy the linear analysis predicts a critical damping of 50 Ns/m while the nonlinear simulation produces a value of 60 Ns/m. Figure~\ref{fig:2Dpole}-a depicts the set of test points for the various values of $\beta$ along the vertical line corresponding to a 16 msec delay. Figure~\ref{fig:2Dpole}-b shows the time histories of the penetration rate during contact for each value of the damping coefficient. The plot with no damping clearly shows a significant increase in the relative velocity after impact which cues on an addition of energy by the robotics system due to the delay. For the case of $\beta$=0, the magnitude of the velocity profile after contact is greater than the initial velocity which yields an $\epsilon$=1.6 as show in in Tab~\ref{s3tab2}. The magnitude of the velocity profile is less than the initial velocity only when the virtual damping is more than {60 Ns/m} as see in Fig~\ref{fig:2Dpole}-b. The linear analysis indicates that it requires a virtual damping of {50 Ns/m} to remove all energy added to the system. However, the nonlinear simulator indicates that is is required {60 Ns/m} to stabilize the simulator.

\subsection{Stability analysis using passivity}

A ``passivity observer'' approach was introduced in~\cite{988969} with the purpose of monitoring the passivity property of a dynamical system from its input and output signals only. The passivity approach was applied and studied in previous experimental works~\cite{Rainer} since it easily lends itself to empirical procedures where input-output signals only are processed. As opposed to the pole location method, it is independent from particular model assumptions. It also has the advantage to provide an insight on the passivity of the system during the contact, as opposed to the coefficient of restitution approach where analysis uses data prior and after contact. To conclude, the energy-based approach provides a simple method for real-time monitoring of the passivity of the system. The consequences on the hybrid EPOS simulator operation are twofold: 1) it allows a real-time monitoring of its passivity property, which is necessary for the test to be faithful to nature, 2) it enables an output data driven adaptation mechanism to regulate the virtual damping in real-time according to specific requirements on the energy profile. A proper blending between this approach and the model-based approach for online tuning of the virtual damping is a promising direction for efficient and safe operation of the EPOS simulator. It consists in computing the following performance measure, a.k.a. observed or added energy:
\begin{align}
\label{eq:obstwo1}
\Delta E = \Delta t \sum_{i=1}^N [(f_{mx}v_{mx} - f_{inx} v_{rx})\\ \nonumber
+(f_{my}v_{my} - f_{iny} v_{ry}) \\ \nonumber
+(f_{mz}v_{mz} - f_{inz}v_{rz})\\ \nonumber
+(\tau_{mx}\omega_{mx} - \tau_{inx}\omega_{rx})\\ \nonumber
+(\tau_{my}\omega_{my} - \tau_{iny} \omega_{ry})\\ \nonumber
+(\tau_{mz}\omega_{mz} - \tau_{inz}\omega_{rz})] \nonumber
\end{align}
where $f_m$, $\tau_m$, $f_{in}$, $\tau_{in}$, $v_m$, $\omega_m$, $v_r$, $\omega_r$ are the sampled signals of the measured force and torque, the force and torque input to the hybrid simulator, the measured linear and angular velocity, and the command linear and angular velocity, respectively, and $\Delta t$ is the sample time (4 ms) with $N=1,2,\ldots$ denotes the number of samples. The hybrid simulator is passive if $\Delta E < 0$, lossless if $\Delta E = 0$, and active if $\Delta E >0$ at any particular time. Monitoring in real-time the value of $\Delta E$ thus gives a cue on the stability of the simulator: it is unstable if it becomes active.

\subsection{Concluding Remarks}
\par The stability analysis validation provided in this section shows encouraging results. The 4$^{th}$-order system which describes the linearized dynamics of the 2D system was decomposed, via a physically intuitive transformation, in two second-order delay systems. The stability of a standard second-order system, which was investigated using results developed in an earlier study~\cite{Melak2013} for a single-dimensional system, is straightforwardly extended to the case of the 4$^{th}$-order system. For this purpose, the formulas for the critical delay and frequency (Eqs.~\ref{s3eq14a}, \ref{s3eq15}) are applied by substituting $\mu = m_{\!a}$, $\beta = b$, and $\kappa= k$ (where $m_{\!a}$ is given in Eq.~\ref{s2eq59}) for the mode of the penetration depth, and $\mu = m$, $\beta = 2b$, and $\kappa = 2k$ for the other states dynamics. The critical delay for the 4$^{th}$-order system is the smaller of the two computed values. Notice that in the typical case where $\omega_c \beta << \kappa$, and Eq.~\eqref{s3eq16} is, thus, valid, the two values of the critical delay are identical, and equal to $\frac{b}{k}$.

Notice that the analysis of the 2D dynamics stability was enabled by the modeling of the penetration depth and rate. The extension to a 3D linear stability analysis, although not undertaken in this work, seems to be feasible along a similar approach.

Although the nonlinear simulation provided some validation of the closed-form design-model based linear stability analysis, cautious should be taken in applying these formulas for operational purposes. Margins should be taken in order to account for uncertainties and random effects in the loop delay system.

The pole location method is model-based and is thus sensitive to the uncertainty in the parameters knowledge. But it provides a simple and elegant framework in order to predict the system's behavior. On the other hand, an output-driven method (based on the coefficient of restitution or on a passivity approach) relies on incoming observations. It is thus robust to parameter uncertainty and it may be used for online adaptation. Indeed, the coefficient of restitution can be used as a control criteria to keep the system passive, or to perform a successful docking without back bouncing the Target satellite. Previous works~\cite{UyamaIROS} used it as a performance index in order to develop 1D control strategies for docking to uncooperative target satellites. Future works will use the passivity approach to monitor the stability of the hybrid simulator in 3D scenarios. But such approaches lack of a predictive feature. A blended methodology looks promising in order to benefit from the ``best of the both worlds''.

\section{Compliance Device Effective Stiffness}

\begin{figure}[t]
	\centering
	\includegraphics[width=0.8\textwidth]{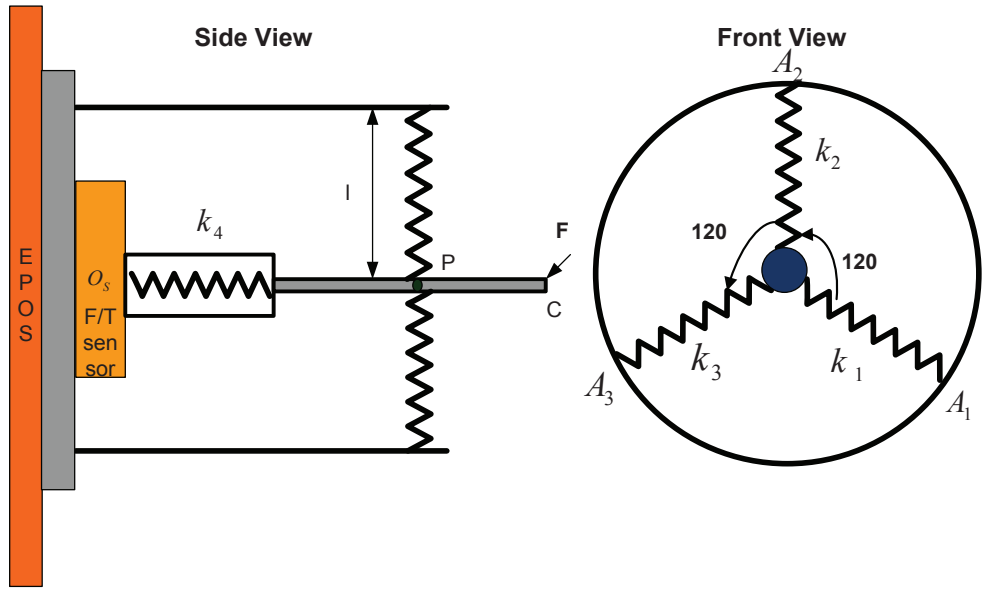}
	\caption{Drawing of the compliance device}
	\label{fig:comdev}
\end{figure}

The current section is concerned with the description of a compliance device, the presentation of the expression for its effective (scalar) stiffness along the penetration direction, and its relationship with the scalar $k$ introduced in the mathematical model [see Eq.~\eqref{s2eq02}]. This will eventually clarify how the resulting effective stiffness is implemented in the operation concept of the proposed hybrid docking simulator.  Figure~\ref{fig:comdev} depicts a drawing of the passive compliance device, which was designed, manufactured, and implemented for testing. It essentially consists of four springs and a shaft, ``the probe'', assembled in series with the force/torque (F/T) sensor, and rigidly attached to the chaser robot fixtures tool. The springs are linear with stiffness coefficients $k_i$, $i=1,2,3,4$, and the probe is rigid. The probe and the spring $k_4$ are clamped to the F/T sensor at its center $S$. The shaft is supported by the other three springs at point $P$. The attach point $P$ is free to slide along the shaft without friction. In the load-free conditions, neglecting gravity, the probe is perpendicular to the fixtures tool and the springs $k_i$ $i=1,2,3$ lie in a plane normal to the probe in a star configuration (see Fig. \ref{fig:comdev}). They are attached to a rigid cylinder at points $A_i$ $i=1,2,3$, which is clamped to the fixtures tool. These springs can freely rotate in a plane normal to the plane $A_1A_2A_3$.

The generalized expression for the stiffness tensor of this device is presented next. Assume that a force $\Fv$ is applied to the probe tip resulting in a differential displacement of the point $P$, $\delta \textbf{m}$, with respect to its load-free position. For $i=1,2,3,4$, let $\delta l_i$ denote the differential elongation of the spring $k_i$,
let $\lev_i$ denote the unit length vector along the direction $A_iP$, and let $\fv_i$ denote the force applied at $P$ by the spring $k_i$, then the expression for $\Fv$ is as follows:
\begin{align}
 \Fv & = \sum_{i=1}^4 \fv_i \\
 \label{s4eq02}
 & = - \underbrace{\left[ \sum_{i=1}^4 k_i (\leiv \leiv^T) \right]}_{\mathrm{K}} \delta \textbf{m}. \
\end{align}
where $\mathrm{K}$ denotes the generalized stiffness tensor.
It is assumed that the force and the motion of the point $P$ are along the direction normal to the nozzle wall, represented by the unit vector $\nev$. Therefore, the component of the displacement $\delta \textbf{m}$ perpendicular to $\nev$ is discarded and the component of the force $\Fv$ along $\nev$, denoted as $\fv_{\varphi}$, is considered. Its expression is provided next:
\begin{align}
 \label{s4eq07}
& \fv_{\varphi}  = - \underbrace{\left[ \sum_{i=1}^4 k_i (\leiv^T \nev)^2\right]}_{k_{\varphi}} \underbrace{(\nev^T \delta \textbf{m})}_d \nev \
\end{align}
where $k_{\varphi}$ denotes the effective stiffness along $\nev$ and $d$ denotes the penetration depth. To summarize, the contact force is expressed as follows
\begin{align}
\label{s4eq08}
& \fv_{\varphi} = f_{\varphi} \nev \\
\label{s4eq08a}
& f_{\varphi} = - k_{\varphi} d \
\end{align}
where the effective stiffness $k_{\varphi}$ is given as
\begin{align}
 \label{s4eq09}
& k_{\varphi} = \sum_{i=1}^4 k_i (\leiv^T \nev)^2 \
\end{align}
At contact, the compliance device produces an effective stiffness $k_\varphi$ such that the force is proportional to the penetration depth, $d(t)$. It is equivalent to the expression $d(t)$ as given in Eq.~\eqref{s2eq04}. Henceforth, the expression for the force, as given in Eq.~\eqref{s4eq08a} corresponds to the force magnitude $f(t)$ in the three-dimensional mathematical model, as given in Eq.~\eqref{s2eq15a}. The coefficient $k_\varphi$ can be adjusted by tuning the springs' coefficients $k_i$ and orientation vectors $\leiv$ $i=1,2,3,4$. The evaluation of $k_{\varphi}$ requires expressing the inner products $\lev^T\nev$, which are directly related to the orientation of the sensor frame $\Sfr$ with respect to the Nozzle frame $\Nfr$. The parameter $k_\varphi$ is thus state-dependent. Equations~\eqref{s4eq08}-\eqref{s4eq09} provide a mathematical model for the force feedback as sensed by the F/T sensor.
\subsection{Hybrid Contact Model}
 In addition to the measured force, $f_{\varphi}$, the  hybrid simulator concept of operations superposes a virtual force, $f_v$, at the input of the numerical simulation. The total input force magnitude is, thus, expressed as follows:
\begin{align}
\nonumber
f & =  f_{\varphi} + f_v \\
\label{s4eq10a}
& = -\underbrace{(k_\varphi + k_v)}_{k} d - \underbrace{b_v}_{b} \dot{d} \
\end{align}
where $k_v$ and $b_v$ are parameters that can be adjusted by the operator in order to provide the desired contact model properties without the need to physically change the contact interface. The expression for $f$ in Eq.~\eqref{s4eq10a} corresponds to Eq.~\eqref{s2eq15a} of the mathematical model.
This model is amenable to stability analysis along the approach presented  by replacing the time-varying $k_\varphi$ with a time-invariant upper bound. This was the approach adopted in this work. 
\section{Experimental Results}
\subsection{1D Test and Experimental Validation of the Stability Analysis}
\begin{figure}[h!]
	\centering
	\includegraphics[width=0.5\textwidth]{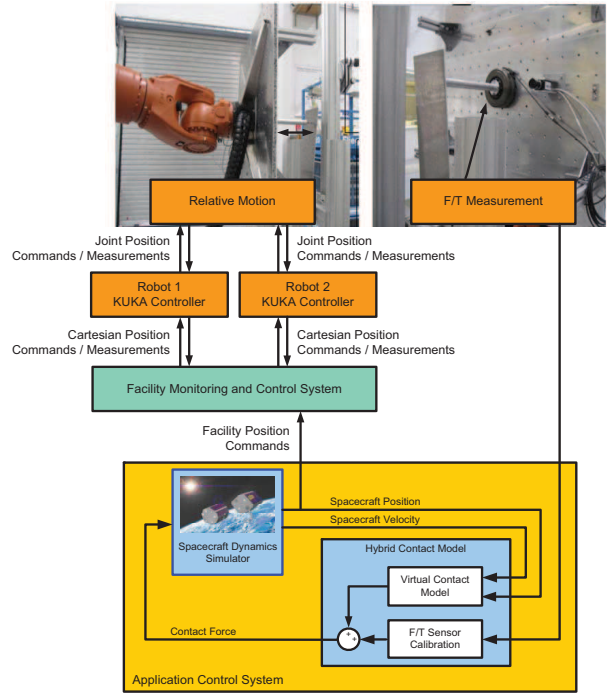}
	\caption{1D test setup on the DLR EPOS hybrid simulator}
	\label{fig:HILsetup}
\end{figure}

Figure~\ref{fig:HILsetup} conceptually pictures the experiment setup for the 1D case within the EPOS facility. The hardware module of the hybrid simulator consists of the chaser robot, its tracking controller, the target element, the force sensor, and a compliance device. The force sensor is attached to a tool plate that is fixed at the chaser's end-effector. The docking interface, rigidly attached to the tool plate, is equipped with a stiff shaft (the probe) with a pin-like head. The probe thus makes contact with the target element in a pin-pointed manner. The target element is a metal sheet at rest with respect to the room's referential. This was done for the sake of simplicity and does not limit the validity of the tests, since they are conducted in 1D only. The software module of the hybrid simulator includes the numerical simulation of the chaser and target satellites, an estimator of the current relative displacement of the target with respect to the chaser, the computation of a virtual contact force according to specified damping and stiffness coefficients, and the calibration of the force sensor. The robotics tracking system has a millimeter accuracy and operates at a frequency of 250 Hz. The force sensor output, after calibration, are corrupted with errors of order 0.25 N, and the force sampling frequency is 1000 Hz.
The tests were conducted with various values of the simulated reduced mass $m$ and of the virtual damping $b_v$. An account of the results for a mass of 63 kg is provided here. Additional results are proposed in~\cite{Melak2013}. The value for the delay used in the analytical formulas is 16 msec. Experimental values of the stiffness, $\kes$, were identified during each test. In each test, the chaser robot is moving towards the target at a constant speed of 20 mm/sec, makes contact, and bounces back. Several tests were performed where the damping coefficient, $b_v$, was gradually increased in the software.

\begin{figure}[h]
	\centering
	\includegraphics[width=0.6\textwidth]{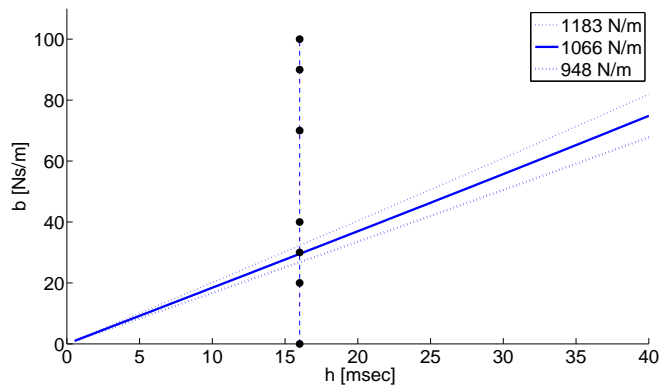}
	\caption{Experimental validation of the stability analysis. The $b$ vs $h$ curves stem from the analysis. The dots represent test points.}
       \label{f6}
\end{figure}

The results are summarized in Table~\ref{s6tab1} and Figure~\ref{f6}. Table~\ref{s6tab1} present seven test cases for various values of $b$. The relative velocity before and after impact was recorded and averaged over several seconds. These averages, by $v^{-}$ and $v^{+}$, are used in the computation of the coefficient of restitution (Eq.~\ref{s3eq000}). According to this criterion, the system is stable if $\epsilon < 1$, neutrally stable if $\epsilon= 1$, and unstable otherwise.
\begin{table}[t]
 \centering
 \caption{Tests results for varying values of the damping $b$}
 \label{s6tab1}
 \vspace{10mm}
\begin{tabular}{@{\extracolsep{10mm}}ccccc}
 \hline\hline\\
$b$     & $v^{-}$    & $v^{+}$    & $\epsilon$ & $\kes$   \\
$[Ns/m]$  &  [mm/s] & [mm/s]    &            & [N/m]   \\\\
\hline\\
$0$   & $21.0$  & $23.4$  & $1.11$  & $977$   \\
$20$   & $18.5$  & $20.0$  & $1.08$  & $1020$ \\
$\bf30$   & $18.0$  & $18.0$  & $\bf1.00$  & $975$   \\
$\bf40$   & $17.5$  & $17.0$  & $\bf0.97$  & $1050$ \\
$70$   & $20.0$  & $17.0$  & $0.85$  & $1030$  \\
$90$   & $20.0$  & $15.0$  & $0.75$  & $1040$ \\
$100$   & $21.0$  & $15.0$  & $0.71$  & $822$ \\\\
\hline\hline
\end{tabular}
\end{table}
When $b$ is zero, the system is, as expected, unstable, as evidenced by the fact that $\epsilon$ is greater than one. Incremental increases of the value of $b$, up to 30-40 Ns/m in the software, produce stronger damping forces, which results in a decrease of $\epsilon$ down to unity. This particular test (in bold in Table~\ref{s6tab1}) was repeated several times, consistently yielding values of $\epsilon$ between between $0.97$ and $1$. The system has thus become neutrally stable. Further increasing the coefficient $b$ to 70, 90, and 100 Ns/m, results in a consistent reduction of $\epsilon$. Comparison with the stability analytical results is done as follows. Using the values for the identified stiffness, $\kes$, as given in Table~\ref{s6tab1}, the sample average $\bar{k}$ and standard deviation $\sigma_k$ are computed, yielding 1066 N/m and 118 N/m, respectively. This is consistent with the levels of accuracy of 0.25 N and 1 mm in the force and position knowledge, respectively. This shows that the experiment was well calibrated. Using the values for the mass (63 kg), the delay (16 ms), and the three stiffness values $\bar{k}$, $\bar{k}\pm\sigma_{k}$, three curves of $b$ vs $h$ are plotted (see Fig.~\ref{f6}). These curves provide an envelope where one expects to find the experimental critical value for $b$, for a given delay $h$. The black dots represent the experimental data. It appears that the points corresponding to neutral stability (i.e. $b$ at 30 and 40 Ns/m) lie inside or are close to the critical envelope (in dotted lines). There is thus a good agreement between the tests and the analysis. These tests also provide a proof-of-concept in 1D of the EPOS hybrid simulator concept of operations.

\subsection{3D Test}
The objectives of the 3D tests are as follows: 1) to illustrate the feasibility of the concept of operations of the hybrid EPOS in 3D, 2) to illustrate a methodology for real-time monitoring of the simulator stability that is not model-based and is easily implementable in 3D.

\par
\begin{figure}[h]
\centering
\includegraphics[width=0.5\textwidth]{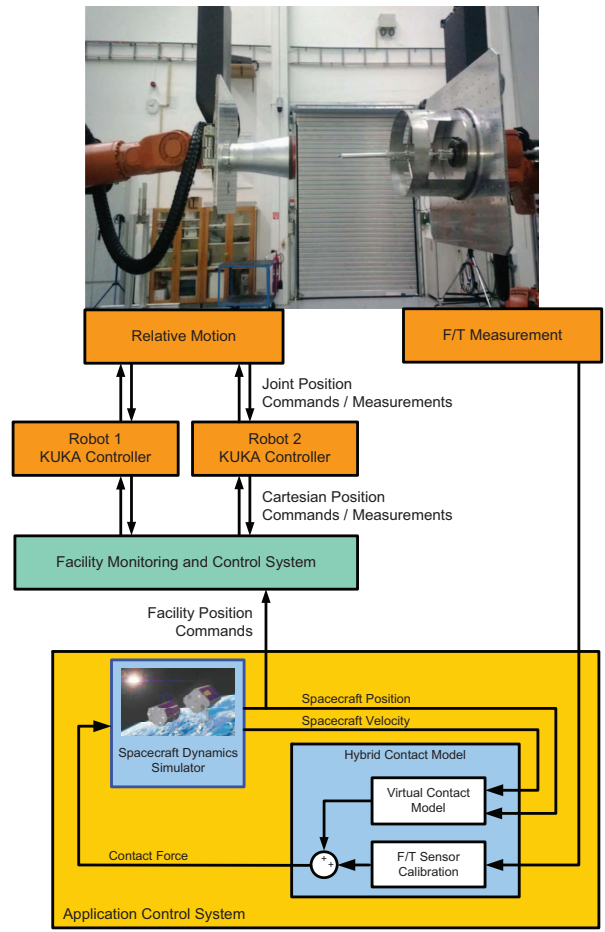}
\caption{3D test setup on the DLR EPOS hybrid simulator}
\label{fig:HILsetup1}
\end{figure}

 Figure~\ref{fig:HILsetup1} conceptually shows the EPOS experiment setup for the 3D case within the EPOS facility. The target hardware element is a conic shape metallic device, which has got the typical dimensions of a 10 Newton thruster nozzle. That is the type of orbit correction thruster to be found on geostationary satellites. The nozzle-like device is mounted on the fixtures tool of the target robot. Both the chaser and the target robots are set in motion in these tests. The software module of the hybrid simulator includes the numerical simulation of the chaser and target satellites, an estimator of the current relative displacement of the target with respect to the chaser, the computation of a virtual contact force according to specified damping and stiffness coefficients, and the calibration of the force sensor \cite{Melak2014}. The robotics tracking system has a millimeter accuracy and operates at a frequency of 250 Hz. The force sensor output, after calibration, are corrupted with errors of order 0.25 N for stationary chaser robot, and the force sampling frequency is 1000 Hz. 
The satellites have identical masses of {3000 kg}, and principal inertias of 500 kg-m$^2$ in each axis. The compliance device has got a stiffness of {4000 N/m} along the probe ($k_4$) and unknown stiffness orthogonal to the probe. Two tests were performed with two values of the damping coefficient $b_v$. In the first test, no damping was added to the physical force and torque measurements. In the second test, a virtual torque was added with a damping coefficient of {40 Ns/m} along the $z$-axis of the chaser body frame.
 
 \par
\begin{figure}[h]
\centering
\includegraphics[width=0.6\textwidth]{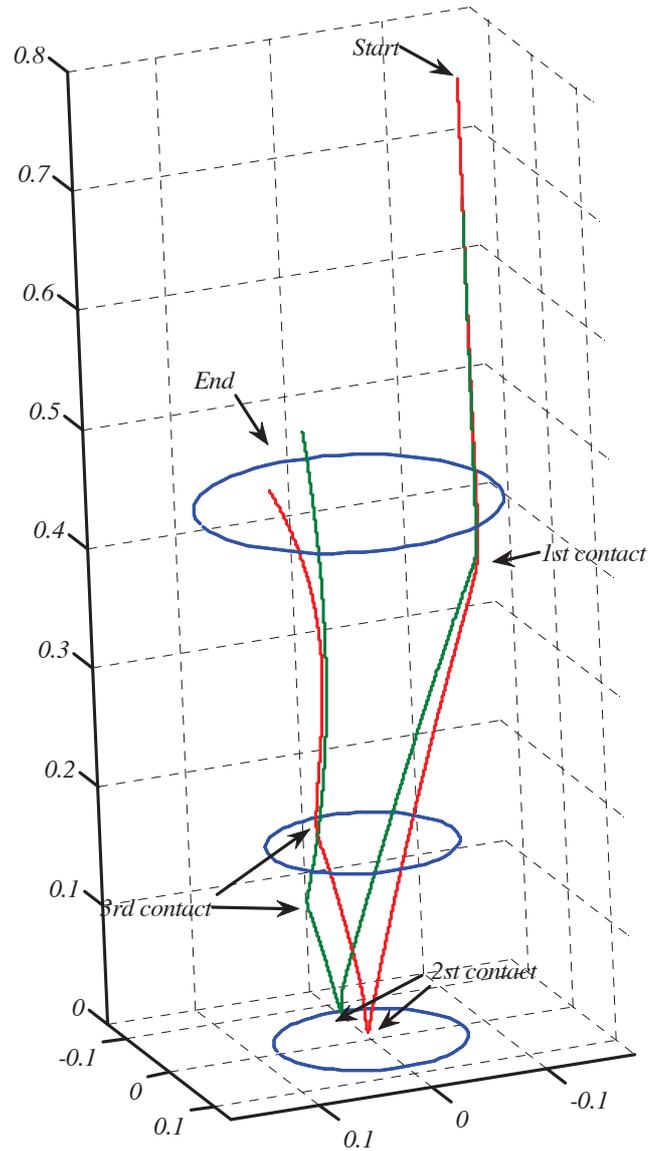}
\caption{Visualization of the probe tip trajectory as seen from the Nozzle frame $\Nfr$. Green line: no damping. Red plot: damping of the {z}-axis torque. The red line trajectory is less affected by the first shock than the green line trajectory.}
\label{fig:3Dprofile}
\end{figure}

 The trajectory of the probe tip with respect to the nozzle frame is visualized in Fig.~\ref{fig:3Dprofile} for both test cases. The dotted curve depict the trajectory in the first test (no damping) while the solid curve indicate the probe trajectory in the second test (40 Ns/m damping). A difference in the trajectory after the first contact can be observed: the virtual damping  compensated for the added energy due to the time delay of the controller that resulted in the probe tip change of motion after the first contact. All in all, three contacts were observed. The second one took place at the bottom of the nozzle before the third contact occurred and the back-bouncing probe left the nozzle's volume.
 
\FloatBarrier

\begin{figure}[h]
    \centering
	   \includegraphics[width=0.8\textwidth]{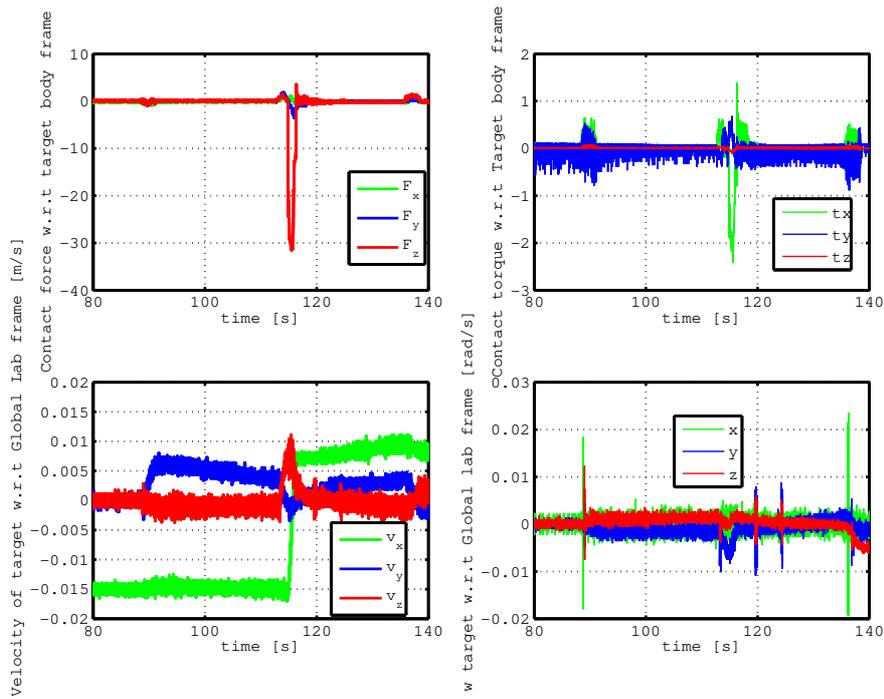}
        \caption{Time histories of the force and torque components in the Nozzle frame (upper graphs). Time histories of the relative velocity and position in the Global frame (lower graphs)}
       \label{fig:CDT}
\end{figure}

\begin{figure}[h]
\centering
\includegraphics[width=0.95\textwidth]{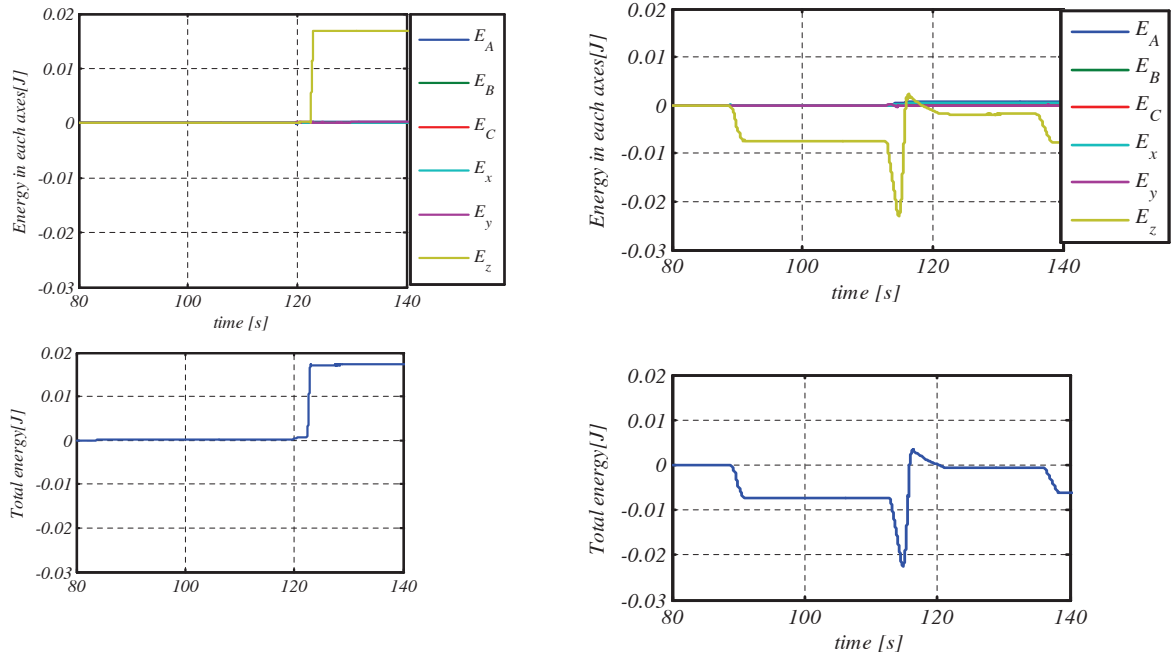}
\caption{The observed energy with (right plot) and without (left plot) virtual damping}
\label{fig:3Denergy}
\end{figure}

 The upper graphs in Fig.~\ref{fig:CDT} show the time histories of the components in the force and torque, as measured by the force/torque sensor, during the second test. The force/torque components are produced along the Nozzle frame $\Nfr$. The lower graphs depict the time histories of the components of the relative velocity and position vectors of the target and chaser robots, as measured by the robots tracking systems. The components are along the global frame $\Gfr$. The test started such that the probe would enter the interior of the nozzle, and hit the lateral side first. The initial relative linear velocity was 15 mm/sec, and the initial rotational velocity was zero.
\FloatBarrier
 Figure~\ref{fig:3Denergy} shows the observed energy during the docking test. The left-hand-side graphs correspond to the first test (no damping) and the right-hand-side graphs correspond to the second test (some damping). The upper graphs show the energy plots along each separate axes (three for translation and three for rotation), while the lower graphs depict the total energy values. The upper-left plot shows that the system is active (in rotation about the z-axis of the chaser frame): this is due to the delay and the absence of damping. The upper-right plot shows that  energy dissipation took place, as observed for the z-axis in rotation, as expected.
\FloatBarrier


\section{Conclusion}

This work presented a 2D and 3D analytical and experimental investigation of the stability of the DLR robotics-based docking simulator EPOS.
\par The gerneral  concept of hybrid docking simulator concept was presented. The hybrid simulator's concept of operations implements a virtual/software force (and torque) feedback aside the physical/hardware force (torque) feedback. A nonlinear state-space model was developed for the concept in 3D and particularized to 2D. A hardware compliance device was designed to be rigidly attached to the chaser robot flange. Its effective stiffness, along the penetration direction, was analytically expressed as a function of the springs stiffness and the relative orientation target-chaser. A time-invariant upper-bound of the resulting scalar expression is conveniently used in the proposed theoretical framework of the stability analysis.   The characteristic polynomial could be developed and the pole location method was applied for stability analysis for the 2D case. Closed form expressions relating the robotics tracking delay with the system's parameters - chaser mass and inertia, probe length, nozzle aperture angle, contact surface stiffness, and contact surface damping - were developed. The general stability results were illustrated by a numerical example. They could be validated by comparison with a nonlinear simulation stability performances. The latter were evaluated using a coefficient of restitution expressed from the penetration rate.  The proposed analysis aims at predicting the stability of the hybrid simulator and at tuning the required parameters for safe operations.

Experiments were conducted in 1D and 3D. The 1D test results exhibited a very good agreement with the model-based analysis: the pole location method could satisfactorily predict the domain of stability of the hybrid EPOS simulator. The 3D test illustrated the feasibility of the hybrid simulator concept of operation: the damping coefficient could be chosen in a selected axis in order to vary the system passivity. The 3D test also illustrated the energy-based approach, which is model-free, for real-time monitoring of the system passivity.

Future work will relax the target stationary assumption, revisit the sliding without friction assumption, incorporate uncertainty and random errors in the design model, develop a linear design model for the 3D case, look for adequate upper-bounds on the compliance device effective stiffness, exploit a combination of the model-based approach and the model-free energy approach for check/recover of the hybrid simulator stability, and design an active approach for online adaptation of the virtual damping for stable \emph{and} truthful EPOS operations.

\section*{Acknowledgments}
The authors acknowledge many fruitful discussions with Mr. Rainer Krenn from the DLR Robotics and Mechatronics Institute and thank him for his help during the 3D tests.

\bibliography{bibliography}
\bibliographystyle{elsarticle-num}

\end{document}